\documentclass[12pt]{article}
\usepackage{epsfig,amsfonts,amssymb}
\usepackage{hyperref}
\pdfoutput=1

\usepackage{cite}
\usepackage{cite}

\usepackage{comment}

\def\ZZZ{{\hbox{ Z\kern-1.6mm Z}}}
\def\RRR{{\hbox{ R\kern-2.4mm R}}}
\def\CCC{{\hbox{ C\kern-2.0mm C}}}
\def\zzz{{\hbox{z\kern-1mm z}}}

\newcommand{\qeq}{{\hbox{=\kern-2.3mm ? \kern.5mm }}}
\renewcommand{\qeq}{=}

\newcommand{\eps}{\epsilon}

\newcommand{\vp}{\varphi}
\newcommand{\ve}{\varepsilon}

\newcommand{\BB}{{\cal B}}

\newcommand{\BBB}{{\cal B}}
\newcommand{\II}{{\cal I}}
\newcommand{\AAA}{{\cal A}}

\newcommand{\JJ}{{\cal J}}

\newcommand{\CC}{{\cal C}}

\newcommand{\OO}{{\cal O}}

\newcommand{\wt}{\widetilde}
\newcommand{\wh}{\widehat}

\newcommand{\NN}{{\cal N}}

\newcommand{\be}{\begin{equation}}
\newcommand{\ee}{\end{equation}}
\newcommand{\ben}{\begin{eqnarray}\displaystyle}
\newcommand{\een}{\end{eqnarray}}

\newcommand{\refb}[1]{(\ref{#1})}
\newcommand{\p}{\partial}
\newcommand{\sectiono}[1]{\section{#1}\setcounter{equation}{0}}

\newcommand{\gsim}{\stackrel{>}{\sim}}

\def\one{{\hbox{ 1\kern-.8mm l}}}
\def\zero{{\hbox{ 0\kern-1.5mm 0}}}

\newcommand{\bea}[1]{\begin{eqnarray}\label{#1} }
\newcommand{\eea}{\end{eqnarray}}

\newcommand{\eqref}{\refb}

\usepackage{bm}
\usepackage[table]{xcolor}

\newcommand{\tom}{\tilde\omega}

\def\figimaginary{

\def\JPicScale{0.8}
\ifx\JPicScale\undefined\def\JPicScale{1}\fi
\unitlength \JPicScale mm
\begin{picture}(110,60)(0,0)
\linethickness{1mm}
\multiput(10,20)(0.12,0.12){167}{\line(1,0){0.12}}
\linethickness{1mm}
\multiput(30,40)(0.12,-0.12){167}{\line(1,0){0.12}}
\linethickness{1mm}
\put(30,40){\line(0,1){20}}
\linethickness{1mm}
\put(90,20){\line(0,1){20}}
\linethickness{1mm}
\multiput(70,60)(0.12,-0.12){167}{\line(1,0){0.12}}
\linethickness{1mm}
\multiput(90,40)(0.12,0.12){167}{\line(1,0){0.12}}
\put(30,60.5){\makebox(0,0)[cc]{$\times$}}

\put(70,60){\makebox(0,0)[cc]{$\times$}}

\put(110,60){\makebox(0,0)[cc]{$\times$}}

\put(30,10){\makebox(0,0)[cc]{(a)}}

\put(90,10){\makebox(0,0)[cc]{(b)}}

\end{picture}
}

\def\figone{

\def\JPicScale{0.8}
\ifx\JPicScale\undefined\def\JPicScale{1}\fi
\unitlength \JPicScale mm
\begin{picture}(110,70)(0,0)
\linethickness{1mm}
\put(20,50){\line(1,0){20}}
\linethickness{0.2mm}
\put(40,50){\line(1,0){20}}
\linethickness{0.2mm}
\multiput(70,70)(0.12,-0.12){167}{\line(1,0){0.12}}
\linethickness{0.2mm}
\multiput(90,50)(0.12,0.12){167}{\line(1,0){0.12}}
\linethickness{0.2mm}
\put(90,30){\line(0,1){20}}

\put(40,50){\makebox(0,0)[cc]{$\times$}}

\put(90,50){\makebox(0,0)[cc]{$\times$}}

\put(40,20){\makebox(0,0)[cc]{(a)}}

\put(90,20){\makebox(0,0)[cc]{(b)}}

\end{picture}

}

\def\figtwo{

\def\JPicScale{0.8}
\ifx\JPicScale\undefined\def\JPicScale{1}\fi
\unitlength \JPicScale mm
\begin{picture}(70,60)(0,0)
\linethickness{1mm}
\put(10,50){\line(1,0){20}}
\linethickness{0.2mm}
\put(30,50){\line(1,0){20}}
\linethickness{0.2mm}
\multiput(50,50)(0.24,0.12){83}{\line(1,0){0.24}}
\linethickness{0.2mm}
\multiput(50,50)(0.24,-0.12){83}{\line(1,0){0.24}}

\linethickness{1mm}
\put(90,50){\line(1,0){20}}
\linethickness{0.2mm}
\put(110,50){\line(1,0){20}}
\linethickness{0.2mm}
\put(110,35){\line(0,1){15}}

\put(30,50){\makebox(0,0)[cc]{$\times$}}

\put(50,50){\makebox(0,0)[cc]{$\times$}}

\put(110,50){\makebox(0,0)[cc]{$\times$}}

\put(40,30){\makebox(0,0)[cc]{(a)}}

\put(110,30){\makebox(0,0)[cc]{(b)}}

\end{picture}

}

\def\figthree{

\def\JPicScale{0.8}
\ifx\JPicScale\undefined\def\JPicScale{1}\fi
\unitlength \JPicScale mm
\begin{picture}(70,60)(0,0)
\linethickness{1mm}
\put(10,50){\line(1,0){20}}
\linethickness{0.2mm}
\put(30,50){\line(1,0){20}}
\linethickness{1mm}
\put(50,50){\line(1,0){20}}

\linethickness{1mm}
\put(90,50){\line(1,0){20}}
\linethickness{1mm}
\put(110,50){\line(1,0){20}}

\put(30,50){\makebox(0,0)[cc]{$\times$}}

\put(50,50){\makebox(0,0)[cc]{$\times$}}

\put(110,50){\makebox(0,0)[cc]{$\times$}}

\put(40,40){\makebox(0,0)[cc]{(a)}}

\put(110,40){\makebox(0,0)[cc]{(b)}}

\end{picture}

}

\def\figfour{

\def\JPicScale{0.8}
\ifx\JPicScale\undefined\def\JPicScale{1}\fi
\unitlength \JPicScale mm
\begin{picture}(70,60)(0,0)
\linethickness{1mm}
\put(10,50){\line(1,0){20}}
\linethickness{0.2mm}
\put(30,50){\line(1,0){20}}
\linethickness{1mm}
\put(50,50){\line(1,0){20}}

\linethickness{0.2mm}
\put(40,35){\line(0,1){15}}

\linethickness{1mm}
\put(90,50){\line(1,0){20}}
\linethickness{0.2mm}
\put(110,50){\line(1,0){20}}
\linethickness{1mm}
\put(130,50){\line(1,0){20}}

\linethickness{0.2mm}
\put(130,35){\line(0,1){15}}

\put(30,50){\makebox(0,0)[cc]{$\times$}}

\put(50,50){\makebox(0,0)[cc]{$\times$}}

\put(110,50){\makebox(0,0)[cc]{$\times$}}

\put(40,50){\makebox(0,0)[cc]{$\times$}}

\put(130,50){\makebox(0,0)[cc]{$\times$}}

\put(40,30){\makebox(0,0)[cc]{(a)}}

\put(110,30){\makebox(0,0)[cc]{(b)}}

\put(35,55){\makebox(0,0)[cc]{$q_1$}}

\put(45,55){\makebox(0,0)[cc]{$q_2$}}

\put(120,55){\makebox(0,0)[cc]{$q_1$}}

\end{picture}

}

\def\figfive{

\def\JPicScale{0.8}
\ifx\JPicScale\undefined\def\JPicScale{1}\fi
\unitlength \JPicScale mm
\begin{picture}(70,60)(0,0)
\linethickness{1mm}
\put(10,50){\line(1,0){20}}
\linethickness{0.2mm}
\put(30,50){\line(1,0){20}}
\linethickness{1mm}
\put(50,50){\line(1,0){20}}

\linethickness{0.2mm}
\put(30,35){\line(0,1){15}}

\linethickness{1mm}
\put(90,50){\line(1,0){20}}
\linethickness{1mm}
\put(110,50){\line(1,0){20}}

\linethickness{0.2mm}
\put(110,35){\line(0,1){15}}

\put(30,50){\makebox(0,0)[cc]{$\times$}}

\put(50,50){\makebox(0,0)[cc]{$\times$}}

\put(110,50){\makebox(0,0)[cc]{$\times$}}

\put(40,30){\makebox(0,0)[cc]{(c)}}

\put(110,30){\makebox(0,0)[cc]{(d)}}

\put(40,55){\makebox(0,0)[cc]{$q_2$}}

\end{picture}

}

\def\figsix{

\ifx\JPicScale\undefined\def\JPicScale{1}\fi
\unitlength \JPicScale mm
\begin{picture}(140,55.59)(0,0)
\linethickness{0.2mm}
\put(45.6,50){\circle{11.18}}

\linethickness{1mm}
\put(20,50){\line(1,0){10}}
\linethickness{0.2mm}
\put(30,50){\line(1,0){10}}

\linethickness{1mm}
\put(60,50){\line(1,0){10}}
\linethickness{0.2mm}
\put(70,50){\line(1,0){10}}

\linethickness{1mm}
\put(90,50){\line(1,0){10}}
\linethickness{0.2mm}
\put(105.5,50){\circle{11.18}}

\linethickness{1mm}
\put(125,50){\line(1,0){15}}

\put(30,50){\makebox(0,0)[cc]{$\times$}}

\put(70,50){\makebox(0,0)[cc]{$\times$}}

\put(80,50){\makebox(0,0)[cc]{$\otimes$}}

\put(100,50){\makebox(0,0)[cc]{$\times$}}

\put(140,50){\makebox(0,0)[cc]{$\otimes$}}

\put(35,35){\makebox(0,0)[cc]{(a)}}

\put(75,35){\makebox(0,0)[cc]{(b)}}

\put(105,35){\makebox(0,0)[cc]{(c)}}

\put(135,35){\makebox(0,0)[cc]{(d)}}

\put(40,50){\makebox(0,0)[cc]{$\times$}}

\put(35,53){\makebox(0,0)[cc]{$q_1$}}

\put(45,58){\makebox(0,0)[cc]{$q_2$}}

\put(75,53){\makebox(0,0)[cc]{$q_1$}}

\put(105,58){\makebox(0,0)[cc]{$q_2$}}

\end{picture}

}

\def\figpsione{

\def\JPicScale{1.2}
\ifx\JPicScale\undefined\def\JPicScale{1}\fi
\unitlength \JPicScale mm
\begin{picture}(140,55.59)(0,0)
\linethickness{0.2mm}
\put(45.6,50){\circle{11.18}}

\linethickness{1mm}
\put(20,50){\line(1,0){10}}
\linethickness{0.2mm}
\put(30,50){\line(1,0){10}}

\linethickness{1mm}
\put(90,50){\line(1,0){10}}
\linethickness{0.2mm}
\put(105.5,50){\circle{11.18}}

\put(30,50){\makebox(0,0)[cc]{$\times$}}

\put(100,50){\makebox(0,0)[cc]{$\times$}}

\put(35,35){\makebox(0,0)[cc]{(a)}}

\put(105,35){\makebox(0,0)[cc]{(b)}}

\put(40,50){\makebox(0,0)[cc]{$\times$}}

\put(55,50){\makebox(0,0)[cc]{$\psi^1$}}

\put(115,50){\makebox(0,0)[cc]{$\psi^1$}}

\end{picture}

}

\def\figseven{

\def\JPicScale{1.2}
\ifx\JPicScale\undefined\def\JPicScale{1}\fi
\unitlength \JPicScale mm
\begin{picture}(140,55.59)(0,0)
\linethickness{0.2mm}
\put(45.6,50){\circle{11.18}}

\linethickness{0.2mm}
\put(30,50){\line(1,0){10}}

\linethickness{0.2mm}
\put(70,50){\line(1,0){10}}

\put(40,50){\makebox(0,0)[cc]{$\times$}}

\put(80,50){\makebox(0,0)[cc]{$\otimes$}}

\put(45,35){\makebox(0,0)[cc]{(a)}}

\put(75,35){\makebox(0,0)[cc]{(b)}}

\end{picture}

}

\def\figeight{

\ifx\JPicScale\undefined\def\JPicScale{1}\fi
\unitlength \JPicScale mm
\begin{picture}(130,55)(0,0)
\linethickness{0.3mm}
\multiput(30,40)(0.36,-0.12){83}{\line(1,0){0.36}}
\linethickness{0.3mm}
\multiput(60,30)(0.36,0.12){83}{\line(1,0){0.36}}
\linethickness{0.3mm}
\put(90,30){\line(0,1){10}}
\linethickness{0.3mm}
\put(30,30){\line(0,1){10}}
\linethickness{0.3mm}
\put(-10,30){\line(1,0){140}}
\linethickness{0.3mm}
\put(-10,40){\line(1,0){40}}
\linethickness{0.3mm}
\put(90,40){\line(1,0){40}}
\linethickness{0.3mm}
\put(30,40){\line(1,0){60}}
\put(35,35){\makebox(0,0)[cc]{I}}

\linethickness{0.3mm}
\put(-10,55){\line(1,0){140}}

\put(85,35){\makebox(0,0)[cc]{I$'$}}

\linethickness{0.3mm}
\put(60,30){\line(0,1){10}}
\put(65,35){\makebox(0,0)[cc]{II}}

\put(55,35){\makebox(0,0)[cc]{II$'$}}

\put(10,35){\makebox(0,0)[cc]{III}}

\put(100,35){\makebox(0,0)[cc]{III$'$}}

\put(60,47){\makebox(0,0)[cc]{IV}}

\put(130,58){\makebox(0,0)[cc]{$y=1$}}

\put(130,43){\makebox(0,0)[cc]{$y\simeq 1/\wt\lambda^2$}}

\put(60,27){\makebox(0,0)[cc]{$x=0$}}

\put(90,26){\makebox(0,0)[cc]{$x\simeq 1/\wt\lambda$}}

\put(30,26){\makebox(0,0)[cc]{$x\simeq -1/\wt\lambda$}}

\end{picture}

}

\def\fignine{

\def\JPicScale{0.8}
\ifx\JPicScale\undefined\def\JPicScale{1}\fi
\unitlength \JPicScale mm
\begin{picture}(125,90)(0,0)
\linethickness{0.3mm}
\put(30,30){\line(1,0){90}}
\linethickness{0.3mm}
\put(30,30){\line(0,1){60}}
\linethickness{0.3mm}
\put(30,70){\line(1,0){90}}
\linethickness{0.3mm}
\put(50,30){\line(0,1){10}}
\linethickness{0.3mm}
\put(30,40){\line(1,0){90}}
\linethickness{0.3mm}
\qbezier(35,70)(37.58,59.59)(41.19,52.38)
\qbezier(41.19,52.38)(44.8,45.16)(50,40)
\put(35,35){\makebox(0,0)[cc]{(a)}}

\put(70,35){\makebox(0,0)[cc]{(b)}}

\put(35,50){\makebox(0,0)[cc]{(c)}}

\put(70,55){\makebox(0,0)[cc]{(d)}}

\put(129,30){\makebox(0,0)[cc]{$v\to$}}

\put(25,90){\makebox(0,0)[cc]{$x\uparrow$}}

\put(50,27){\makebox(0,0)[cc]{$v\simeq\alpha^{-2}$}}

\put(135,41){\makebox(0,0)[cc]{$x\simeq (2\pi\wt\lambda)^{-1}$}}

\put(130,70){\makebox(0,0)[cc]{$x= {1\over 4}$}}

\end{picture}

}

\def\figcooo{

\def\JPicScale{0.8}
\ifx\JPicScale\undefined\def\JPicScale{1}\fi
\unitlength \JPicScale mm
\begin{picture}(150,70)(0,0)
\linethickness{1mm}
\put(10,50){\line(1,0){15}}
\linethickness{0.2mm}
\put(25,50){\line(1,0){15}}
\linethickness{0.2mm}
\multiput(40,50)(0.12,-0.12){83}{\line(1,0){0.12}}
\linethickness{0.2mm}
\multiput(40,50)(0.12,0.12){83}{\line(1,0){0.12}}
\linethickness{0.2mm}
\multiput(50,60)(0.12,0.24){42}{\line(0,1){0.24}}
\linethickness{0.2mm}
\put(50,60){\line(1,0){10}}
\linethickness{1mm}
\put(70,50){\line(1,0){15}}
\linethickness{0.2mm}
\multiput(85,50)(0.12,0.12){83}{\line(1,0){0.12}}
\linethickness{0.2mm}
\multiput(85,50)(0.12,-0.12){83}{\line(1,0){0.12}}
\linethickness{0.2mm}
\multiput(95,60)(0.12,0.24){42}{\line(0,1){0.24}}
\linethickness{0.2mm}
\put(95,60){\line(1,0){10}}
\linethickness{1mm}
\put(120,50){\line(1,0){15}}
\linethickness{0.2mm}
\multiput(135,50)(0.12,0.12){83}{\line(1,0){0.12}}
\linethickness{0.2mm}
\put(135,50){\line(1,0){15}}
\linethickness{0.2mm}
\multiput(135,50)(0.12,-0.12){83}{\line(1,0){0.12}}
\put(25,30){\makebox(0,0)[cc]{(a)}}

\put(85,30){\makebox(0,0)[cc]{(b)}}

\put(135,30){\makebox(0,0)[cc]{(c)}}

\linethickness{1mm}
\put(160,50){\line(1,0){10}}
\linethickness{0.2mm}
\put(170,50){\line(1,0){10}}
\linethickness{0.2mm}
\multiput(180,50)(0.12,0.12){83}{\line(1,0){0.12}}
\linethickness{0.2mm}
\put(180,50){\line(1,0){10}}
\linethickness{0.2mm}
\multiput(180,50)(0.12,-0.12){83}{\line(1,0){0.12}}
\put(180,30){\makebox(0,0)[cc]{(d)}}

\put(32,53){\makebox(0,0)[cc]{$q_2$}}

\put(43,58){\makebox(0,0)[cc]{$q_1$}}

\put(177,53){\makebox(0,0)[cc]{$q_2$}}

\put(88,58){\makebox(0,0)[cc]{$q_1$}}

\end{picture}

}

\def\figoooo{

\def\JPicScale{0.7}
\ifx\JPicScale\undefined\def\JPicScale{1}\fi
\unitlength \JPicScale mm
\begin{picture}(120,60)(0,0)
\linethickness{0.2mm}
\multiput(10,60)(0.12,-0.12){167}{\line(1,0){0.12}}
\linethickness{0.2mm}
\multiput(10,20)(0.12,0.12){167}{\line(1,0){0.12}}
\linethickness{0.2mm}
\put(30,40){\line(1,0){20}}
\linethickness{0.2mm}
\multiput(50,40)(0.12,0.12){167}{\line(1,0){0.12}}
\linethickness{0.2mm}
\multiput(50,40)(0.12,-0.12){167}{\line(1,0){0.12}}
\linethickness{0.2mm}
\multiput(80,60)(0.12,-0.12){333}{\line(1,0){0.12}}
\linethickness{0.2mm}
\multiput(80,20)(0.12,0.12){333}{\line(1,0){0.12}}
\put(40,10){\makebox(0,0)[cc]{(a)}}

\put(100,10){\makebox(0,0)[cc]{(b)}}

\end{picture}

}

\begin{document}

\baselineskip 24pt

\begin{center}

{\Large \bf D-instantons, String Field Theory and Two Dimensional String Theory}


\end{center}

\vskip .6cm
\medskip

\vspace*{4.0ex}

\baselineskip=18pt

\centerline{\large \rm Ashoke Sen}

\vspace*{4.0ex}

\centerline{\large \it Harish-Chandra Research Institute, HBNI}
\centerline{\large \it  Chhatnag Road, Jhusi,
Allahabad 211019, India}


\vspace*{1.0ex}
\centerline{\small E-mail:  sen@hri.res.in}

\vspace*{5.0ex}

\centerline{\bf Abstract} \bigskip

In arXiv:1907.07688 Balthazar, Rodriguez and Yin (BRY) computed the one instanton 
contribution to the two point scattering amplitude in two dimensional string theory 
to first
subleading order in the string coupling. Their analysis left undetermined two constants
due to divergences in the integration over world-sheet variables, but they were fixed
by numerically comparing the result with that of the dual matrix model. 
If we consider n-point scattering amplitudes to the same order, there are actually four
undetermined constants in the world-sheet approach. We show  that using string
field theory we can get finite unambiguous
values of all of these constants, and we explicitly compute three
of these four constants. 
Two of the three constants determined this way 
agree with the numerical result of BRY within the accuracy of
numerical analysis, but the third 
constant seems to differ by  1/2. We also discuss a shortcut to determining the fourth
constant  if we assume the equality of the quantum corrected
D-instanton action and the action of the matrix model instanton. 
This also agrees with the numerical result of BRY.

\noindent
Note added: The apparent discrepancy in the third constant has been resolved in
arXiv:2210.11473.

\vfill \eject

\tableofcontents

\sectiono{Introduction and summary} \label{sintro}

D-instantons represent saddle points of the path integral in second quantized string theory
and give non-perturbative contribution to the string amplitudes. The usual world-sheet
approach to computing these corrections suffers from infra-red divergences from the
boundaries of the moduli spaces of Riemann surfaces. Furthermore, unlike the perturbative 
amplitudes in string theory, these infra-red divergences cannot be removed by analytic
continuation in the external momenta, since the open strings living on the D-instanton
carry strictly zero momentum. Even if the divergences cancel 
at the end\cite{9407031,9701093,0211250}, the fact that
they are present in the intermediate stages of the analysis and need to be tamed by
suitable regulator, makes the final result ambiguous.

A concrete example of such ambiguities arose in a recent analysis of two dimensional string
theory\cite{1907.07688}. The authors of \cite{1907.07688} computed 
the D-instanton contribution to the closed 
string scattering amplitude, but could determine the final result only up to two undetermined
constants. These constants were then determined by comparison with the results
in the dual matrix model\cite{dj,sw,gk,kr,9111035}. 
This comparison was carried out numerically, leading to the best
fit values $-1.399$ and .496 for these constants.\footnote{Ref.\cite{1912.07170}
extended this analysis to multi-instanton contribution. However since the analysis
was done only for the leading order terms in the expansion in powers of the string
coupling around the multi-instanton solution, 
the ambiguities of the type discussed in this paper did not arise in
\cite{1912.07170}.}

String field theory is well poised to address the issues related to infra-red divergences
arising in string theory\cite{1703.06410,1902.00263,1702.06489}. 
It does so by drawing  insights from quantum field 
theory.
Indeed, in an 
earlier analysis\cite{1908.02782}, 
string field theory was used to determine the first of these constants 
unambiguously, leading to the value $-\ln 4$, which is within 1\% of the numerical
result $-1.399$. In this paper, we shall use the systematic procedure for carrying
out perturbation theory in the presence of D-instantons, as developed in 
\cite{2002.04043}, 
to rederive this result in a more general set-up. We shall also see that to the order
to which the analysis of \cite{1907.07688,1908.02782} was carried out, there are
actually four independent constants that are apparently divergent / ambiguous in the
world-sheet description, but can be fixed in string field theory. 
The second constant $\simeq .496$ that appeared in the analysis of 
\cite{1907.07688} is a particular linear combination of these four constants.
We determine three of these four constants
explicitly using string field theory. 
Determination of the fourth constant is possible within string field theory,
but involves computations more elaborate than what will be performed in this paper.
All of these constants can also be determined independently by
(numerically) comparing the string theory results with the known results for higher
point amplitudes in the matrix model. 
This has now been done\cite{private}. We find that two of the three constants, 
determined from string field theory,
agree with the results of the matrix model. However, the third
constant differs by an additive constant close to 1/2.

Since most of the paper will be technical in nature, we shall now give an overview of
the main results. 
We begin by describing our convention. 
A naive amplitude with D-instanton boundary condition does not satisfy energy
conservation, since time translation invariance is broken by the Dirichlet boundary
condition on the time coordinate. As will be discussed later, energy conservation is 
eventually restored after integration over the collective modes of the D-instanton, but
it will be convenient for us to work with the amplitude before integration over the 
collective coordinates. Furthermore, we focus on amplitudes on connected world-sheets,
since the contribution from world-sheets with more than one connected component
can be found by taking the product of the contributions from the connected components.
We shall refer to these as the building blocks from which the final
amplitude will be constructed. Furthermore, every D-instanton amplitude is eventually
multiplied by a factor of $\NN e^{-1/g_s}$ where $g_s$ is the string coupling constant and
$\NN $ is a numerical constant. We shall drop this factor from
the building blocks. The final amplitude is to be constructed after taking sums
of products of appropriate building blocks and then including an overall factor of
$\NN \, e^{-1/g_s}$ and the energy conserving delta function. Finally, we shall use a 
convention in which all energies are counted as outgoing, i.e.\ an ingoing state
of energy $\omega$ will be described as a state of energy $-\omega$. In this 
convention the final energy conserving delta function that comes from integration over
the collective coordinate takes the form $2\pi\delta(\omega_{\rm tot})$ where 
$\omega_{\rm tot}$ is the total energy carried by all the external 
closed string states in an amplitude.

Ref.\cite{1907.07688} computed 
three building blocks in two dimensional string theory. One is the one point 
disk amplitude 
of a closed string tachyon of energy $\omega$. This is given by:
\be
2\, \sinh(\pi|\omega|)\, .
\ee
The powers of $g_s$ in this amplitude are explained as follows. The disk gives a
factor of $1/g_s$, and the closed string vertex operator is normalized so that it
carries a factor of $g_s$. Therefore the powers of $g_s$ cancel from this amplitude.
The second building block computed in 
\cite{1907.07688} is the two point function of a pair of
closed string tachyons of energy $\omega_1$ and $\omega_2$ on the disk. This takes
the form:
\be\label{eblockf}
4\, g_s\, \sinh(\pi|\omega_1|)\, \sinh(\pi|\omega_2|) \, f(\omega_1,\omega_2)\, ,
\ee
where
\ben \label{effull}
f(\omega_1,\omega_2) 
&=& 2^{-1/4}\pi^{1/2}
2^{(\omega_1^2+\omega_2^2)/2} \, {1\over \sinh(\pi|\omega_1|)\sinh(\pi|\omega_2|)}
\int_0^1 dy \nonumber \\ && 
y^{\, \omega_2^2/2} (1-y)^{1-\omega_1\omega_2} 
(1+y)^{1+\omega_1\omega_2}
\langle V_{|\omega_1|/2}(i) V_{|\omega_2|
/2}(i\, y)\rangle_{\rm UHP}\, .
\een
Here $\langle V_{|\omega_1|/2}(i) V_{|\omega_2|
/2}(i\, y)\rangle_{\rm UHP}$ denotes the
two point function of a pair of primaries in the
c=25 Liouville theory of momenta $|\omega_1|/2$ and $|\omega_2|/2$, inserted
at $i$ and $iy$ respectively in  the upper half plane.
The third building block computed in \cite{1907.07688} is the one point function of a
closed string tachyon of energy $\omega$  on the annulus. This takes
the form:
\be
2\, g_s\, \sinh(\pi|\omega|) \, g(\omega)\, ,
\ee
where
\be\label{egfull}
g(\omega)
= {2\, \pi^2} \,  {1\over \sinh(\pi|\omega|)} \int_0^\infty dt \int_0^{1/4} dx \,
\eta(it) \left( {2\pi \over \theta_1'(0|it)}\, \theta_1(2\, x|i\, t)\right)^{\omega^2/2}
\langle V_{|\omega|/2} (2\pi x) \rangle_A \, .
\ee
Here $\theta_1(z|\tau)$ is the odd Jacobi theta function and $\theta_1'(z|\tau)\equiv
\p_z \theta_1(z|\tau)$.
$\langle V_{|\omega|/2} (2\pi x) \rangle_A$ denotes the
one point function of the Liouville primary of momentum $|\omega|$ on an annulus
described by $0\le {\rm Re}(w)\le \pi$, $w\equiv w+2\pi \, i\, t$, with the vertex
operator inserted at ${\rm Re}(w)=2\pi x$. 
Explicit expressions for the correlation
functions in the Liouville theory can be found in \cite{1907.07688}.

The integrals defining $f(\omega_1,\omega_2)$ and $g(\omega)$ are divergent from the
$y\simeq 0$, $t\simeq \infty$ and $x\simeq 0$ regions.\footnote{There are also
divergences from the $y\simeq 1$ and $t\simeq 0$ regions for real energies. 
These are associated with
closed string degenerations and were avoided in \cite{1907.07688} 
by working with imaginary external
energies. In \cite{2003.12076} 
we showed how to use string field theory to get finite results for these
amplitudes for real energies.}
We now divide each of these functions into a finite
part and a divergent part:
\be \label{efdec}
f(\omega_1,\omega_2) = f_{\rm finite}(\omega_1, \omega_2) 
+ f_{\rm div}(\omega_1, \omega_2) \, ,
\ee
\be \label{egdec}
g(\omega) = g_{\rm finite}(\omega) + g_{\rm div}(\omega)\, ,
\ee
where,
\ben \label{effinite}
f_{\rm finite}(\omega_1, \omega_2)
&=& 2^{-1/4}\pi^{1/2}
2^{(\omega_1^2+\omega_2^2)/2} \, {1\over \sinh(\pi|\omega_1|)\sinh(\pi|\omega_2|)}
\int_0^1 dy \nonumber \\ && 
\Bigg[ y^{\, \omega_2^2/2} (1-y)^{1-\omega_1\omega_2} 
(1+y)^{1+\omega_1\omega_2}
\langle V_{|\omega_1|/2}(i) V_{|\omega_2|
/2}(i\, y)\rangle_{\rm UHP} \nonumber \\ &&
- 2^{-3/4} \pi^{-1/2} \, 2^{-(\omega_1^2+\omega_2^2)/2} \,
\sinh(\pi|\omega_1|)\sinh(\pi|\omega_2|)\, 
y^{-2} (1+2\omega_1\omega_2 y) 
\Bigg]\, ,
\een
\be \label{efdivexp}
f_{\rm div}(\omega_1, \omega_2)={1\over 2}
\int_0^1 dy \,
y^{-2} (1+2\omega_1\omega_2 y) \, ,
\ee
and,
\ben \label{egfinite}
g_{\rm finite}(\omega)
&=& {2\, \pi^2} \,  {1\over \sinh(\pi|\omega|)} \int_0^\infty dt \int_0^{1/4} dx 
\Bigg[ \eta(it) \left( {2\pi \over \theta_1'(0|it)}\, \theta_1(2\, x|i\, t)\right)^{\omega^2/2}
\langle V_{|\omega|/2} (2\pi x) \rangle_A \nonumber \\ && 
- {1\over \pi}\,  \sinh(\pi|\omega|) \, \left({e^{2\pi t}-1\over \sin^2(2\pi x)} + 2\, \omega^2
\right)\Bigg]\, ,
\een
\be\label{egdivexp}
g_{\rm div}(\omega)=
2\, \pi\,  \int_0^\infty dt \int_0^{1/4} dx \left({e^{2\pi t}-1\over \sin^2(2\pi x)} + 2\, \omega^2
\right) =  \int_0^1 dv \,  \int_0^{1/4} dx \left\{ {v^{-2} - v^{-1} \over \sin^2(2\pi x)}
+ 2\, \omega^2 \, v^{-1}\right\}\, .
\ee
In the second step we have made a change of variable from $t$ to $v$ via 
$v=e^{-2\pi t}$. \refb{efdivexp} and \refb{egdivexp} suggests that we
parametrize $f_{\rm div}(\omega_1, \omega_2)$ and $g_{\rm div}(\omega)$ as
\be \label{efgpar}
f_{\rm div}(\omega_1, \omega_2) = A_f + B_f \, \omega_1\omega_2, \quad
g_{\rm div}(\omega) = A_g+B_g\, \omega^2\, ,
\ee
where $A_f$, $B_f$, $A_g$ and $B_g$ are constants.

There is one more building block that was not analyzed in 
\cite{1907.07688}, but will be needed for our discussion below. This is the zero point
function on surfaces of Euler number $-1$, that includes a disk with two holes and 
a torus with one hole. Since there are no external momenta, we can parametrize it
as $C\, g_s$ for some constant $C$.

Ref.~\cite{1907.07688} used these building blocks to construct a D-instanton induced
two point closed string tachyon
amplitude to order $g_s$, but we shall present the result in a more general
form where we use the same building blocks to
construct a general $(n+1)$-point amplitude of closed string tachyons to this order.
The leading order contribution comes from the product of $n+1$ disk one
point functions. Our interest will be in the subleading order $g_s$ contribution which
comes from three sources: 1) the product of $(n-1)$ disk one point functions and
one disk two point function, 2) the product of $n$ disk one point functions and
one annulus one point function, and 3)  the product of $(n+1)$ disk one point functions and
one zero point function on a surface of Euler number $-1$. 
A fourth contribution, given by the product of a sphere three point function and $(n-2)$
disk one point functions, has a separate energy conserving delta-function from the sphere
three point function and does not contribute for generic energies of incoming and
outgoing particles satisfying overall energy conservation.
After restoring the factors of
$\NN e^{-1/g_s}$ and the $2\pi\, \delta(\sum_i\omega_i)$, the
D-instanton induced $(n+1)$-point amplitude  up to order $g_s$ takes the form:
\ben\label{emulti1int}
A_{n+1} &=& \NN\, e^{-1/g_s}\,  2\pi \delta(\omega_1+\omega_2+\cdots
\omega_n+\omega_{n+1}) 
\, \left[\prod_{i=1}^{n+1} \{2\, \sinh(\pi|\omega_i|)\} \right] \nonumber \\ &&
\left[
1 +  g_s\, \sum_{i,j=1\atop i<j}^{n+1} f(\omega_i, \omega_j) +  g_s\, \sum_{i=1}^{n+1} g(\omega_i) 
+ C\,  g_s\right]\, .
\een
Using \refb{efdec}, \refb{egdec} and \refb{efgpar} we now get,
\ben\label{emulti1intfinite}
A_{n+1} &=& 2^{n+1}\, \NN\, e^{-1/g_s}\,  2\pi \delta(\omega_1+\omega_2+\cdots
+\omega_{n+1}) 
\, \left\{\prod_{i=1}^{n+1} \sinh(\pi|\omega_i|) \right\} \nonumber \\ &&
\Bigg[
1 +  g_s\, \sum_{i,j=1\atop i<j}^{n+1} f_{\rm finite}(\omega_i, \omega_j) +  g_s\, 
\sum_{i=1}^{n+1} g_{\rm finite} (\omega_i) \nonumber \\ &&
+ \left\{ {1\over 2}\, 
n(n+1) A_f + (n+1)\, A_g +C\right\} \,  g_s + \left\{ B_g -{1\over 2}B_f\right\} 
g_s\, \sum_{i=1}^{n+1} \omega_i^2\Bigg]\, , 
\een
where in the last term we have used the energy conservation law to replace $\sum_{i<j}\omega_i\omega_j$ by $-\sum_i \omega_i^2/2$. This shows that the collection of the
amplitudes for different values of $n$
depends on 
the coefficients
$A_f$, $A_g$ and $C$ individually, but only on  the combination $2B_g - B_f$. 

The matrix model result for the scattering amplitude 
of a closed string tachyon of energy $-\omega_{n+1}$ into $n$ closed string
tachyons of energy $\omega_1,\cdots,\omega_n$ is known explicitly,
and is given by\cite{1907.07688,2003.12076}
\ben \label{ematrix}
&&   2^{n+1}\, \NN\, e^{-1/g_s}\,  2\pi \delta(\omega_1+\omega_2+\cdots
\omega_n+\omega_{n+1}) 
\, \left\{\prod_{i=1}^{n+1} \sinh(\pi|\omega_i|) \right\} \nonumber \\ && \hskip 1in
\left[1 - i\, g_s\, \sum_{j=1}^n \omega_j \, 
\left(1 - \sum_{i=1}^n \pi\omega_i \coth(\pi\omega_i)\right)\right]\, ,
\een
with $\NN=-1/(8\pi^2)$.
Ref.~\cite{1907.07688} compared \refb{emulti1intfinite} and \refb{ematrix} 
for $n=1$ numerically 
and found that the results match well if we choose:
\be \label{eABvalue}
A_f + 2 A_g + C \simeq -.496, \qquad 2B_g-B_f \simeq -1.399\, .
\ee
We emphasize however that we can numerically compare \refb{emulti1intfinite} 
and \refb{ematrix}
for different values of $n$ to extract the coefficients $A_f$, $A_g$, $C$ and
$2B_g-B_f$ separately. This has now been done, and the comparison yields\cite{private}:
\be \label{enumerical}
A_f\simeq -0.50, \qquad A_g\simeq 0.00, \qquad C\simeq 0.00, \qquad
2B_g-B_f\simeq -1.40\, ,
\ee
with an expected error of order $.01$ in each term.

In this paper we use string field theory to show that,
\ben\label{egfdiv}
g_{\rm div}(\omega)&=& -{1\over 2}+ {1\over 2}\, \omega^2 \, \ln{\lambda^2\over 4}\, ,
\nonumber \\
f_{\rm div}(\omega_1, \omega_2) &=& -{1\over 2} (1 - 2\omega_1
\omega_2 \ln\lambda^2) \, ,
\een
where $\lambda$ is an arbitrary parameter that enters the construction of string field
theory. Different values of $\lambda$ give apparently different string field theories, but
they can be shown to be related by field redefinition\cite{9301097}.
Comparison with \refb{efgpar} now gives:
\be\label{eactualvalue}
A_f = -{1\over 2}, \qquad B_f = \ln\, \lambda^2, \qquad A_g =  -{1\over 2}, 
\qquad B_g = {1\over 2}
\ln {\lambda^2\over 4}\, .
\ee
We see that $B_f$ and $B_g$ are individually ambiguous but the combination 
$2B_g-B_f$ that enters the physical amplitudes is unambiguous and is given by:
\be \label{ebgbf}
2B_g-B_f=-\ln 4 \simeq -1.386\, .
\ee
Even though we have not computed $C$ explicitly, we have argued in \S\ref{s5.4}
that if we make the additional assumption
that one can equate the renormalized D-instanton action with the renormalized action of the
matrix 
model instanton, we get
\be\label{eCzero}
C=0\, .
\ee 

The values of $A_f$, $2B_g-B_f$ and $C$ given in
\refb{eactualvalue}, \refb{ebgbf} and \refb{eCzero} agree with the numerical values
\refb{enumerical}
determined by comparison with the matrix model within an error of about $.01$.
However $A_g$ given in \refb{eactualvalue} seems to differ from the numerical
value of $A_g$
given in \refb{enumerical} by an additive constant $-1/2$. 
While this discrepancy could be removed by rescaling the closed string
tachyon field by $(1+{1\over 2} g_s)$, this option is not available since it would
generate corrections of order $g_s$ in the perturbative amplitudes that are not present
in the theory.
The origin of this discrepancy
is not clear to us. The most likely explanation is some missing contribution in the string field
theory analysis, but we have not been able to identify such a contribution. 
Furthermore, as discussed in \S\ref{sopen}, our results pass various consistency tests.
We discuss various other possible origins of this discrepancy in \S\ref{sopen}.

We shall now briefly mention the main subtleties that arise in our analysis. 
\begin{enumerate}
\item The divergent parts of the amplitude are handled by interpreting these divergences
in string field theory and following the standard procedure of quantum  field theories to
remove these divergences. This in particular requires us to not use the
Schwinger parameter
representation for the tachyon propagator that is commonly used in arriving at the 
world-sheet
description of the amplitude. Instead we use the string field theory result for the 
tachyon propagator given by the inverse of the mass$^2$ of the tachyon. This
removes all power law divergences.

\item We also need to remove the zero mode fields propagating in the internal lines, and
treat them as we would do in a quantum field theory. 
Once the contribution due to the zero mode fields is removed, all logarithmic divergences
also disappear. The effect of integration over the zero mode fields
will be discussed
separately below.
\item In the D-instanton background the Siegel gauge, that is normally used in string field
theory, breaks down and we need a different gauge fixing\cite{2002.04043}. 
In this new gauge there is
an additional field whose contribution to the amplitude is not captured in the usual
world-sheet formulation. We need to explicitly include the contribution from this field
propagating in the internal propagators in all the amplitudes.
\item The open string field theory has a zero mode field that represents the freedom 
of translating the D-instanton along the (euclidean) time direction. This can be related to the
collective coordinate of the D-instanton by a field redefinition, after which the integration over the collective coordinate produces the usual energy conserving delta function. 
However, the original string field theory integration measure
produces an additional Jacobian factor  during this change of variable. This gives
additional contributions to the amplitudes that need to be accounted for.
\item In the Siegel gauge the ghost sector contains a pair of zero mode fields, 
one of which is
removed in the new gauge that we use. However the other one remains, and  
can be interpreted as the ghost field associated with a particular 
gauge symmetry of string field theory.
Integration over this ghost field can be interpreted as division by the volume of the 
corresponding gauge group. Now this particular gauge symmetry is related, by a 
field dependent multiplicative factor, to 
the rigid $U(1)$ gauge
symmetry under which an open string stretched from the D-instanton under consideration
to any other D-brane picks up a phase. 
Since the rigid $U(1)$ gauge group has a constant volume, the effect of division by this
volume can be absorbed into a redefinition of the overall normalization constant 
$\NN$.
However the field dependent multiplicative factor that takes us from the
string field theory gauge transformation parameter 
to the rigid $U(1)$ gauge transformation parameter
gives rise to an additional Jacobian factor in the path integral measure. This also
needs to be taken into account while computing amplitudes.
\end{enumerate}
String field theory that we use has two ad hoc parameters $\lambda$ and $\wt\lambda$ and a
function $f(\beta)$.
The contribution to the coefficients \refb{eactualvalue} from the different effects mentioned
above -- the two jacobians, the effect of integration over the out of Siegel gauge modes,
and the original contribution after removal of the zero mode contribution and the string
field theory treatment of the tachyonic modes, all depend on these ad hoc parameters and
the function $f(\beta)$ in
a non-trivial way. 
An illustration of this can be found in \refb{egdivtotal}, 
which lists contribution to $g_{\rm div}(\omega)$
from various sources.
However after summing over all the contributions, the dependence on these
parameters and the function $f(\beta)$ 
disappears and we get the unambiguous results for the coefficients 
$A_f, A_g$ and $2B_g-B_f$ as given in \refb{eactualvalue}. This is consistent with the
fact that different string field theories corresponding to different values of these ad hoc
parameters and the function $f(\beta)$ are related by field redefinition\cite{9301097}.

The rest of the paper is organized as follows. In \S\ref{srev} we review 
some aspects of string field theory and the main results
of \cite{2002.04043} that we shall be using to tame the infrared divergences in 
the computation of 
string amplitudes in a D-instanton background. 
One of the ingredients in this computation is the determination of the relationship between
the standard coordinates of the moduli space of Riemann surfaces and the coordinates
that arise from the string field theory description of the amplitudes. However it is often
sufficient to determine approximate relations between these coordinates instead
of the exact relations. 
In \S\ref{serror} we estimate the errors that we make in using these approximations,
and determine how much error we are allowed to make so as not to affect the final
results. In \S\ref{s1} we introduce the interaction vertices of string field theory
by choosing appropriate local coordinates at the punctures of various Riemann surfaces.
The latter are 
needed for describing off-shell amplitudes. Using these data, we find the (approximate)
relations between the coordinates of the moduli space of Riemann surfaces and the
parameters
arising from string field theory description 
for various amplitudes. In \S\ref{s2} we use these results
to compute the D-instanton contribution to the two point disk amplitude 
of closed string
tachyons in the two dimensional string theory, avoiding the infrared divergences by using 
the procedure described in \S\ref{srev}. This determines the function
$f_{\rm div}(\omega_1,\omega_2)$ leading to the result
given in the second line of \refb{egfdiv}.
In \S\ref{s7} we use the results of \S\ref{s1}
to compute the D-instanton contribution to the one point function of closed string tachyon
on the
annulus. This determines 
$g_{\rm div}(\omega)$,  given in the first line of \refb{egfdiv}.
In \S\ref{s5.4} we discuss the steps needed to compute the two loop correction 
$C$ to the D-instanton action. However we stop short of actually doing this
computation, which involves an open string two loop amplitude and a closed
string one loop amplitude. Nevertheless we give an indirect determination of this
constant $C$ by assuming the equality of the quantum corrected actions of the 
D-instanton and the matrix model instanton.
We end in \S\ref{sopen} by listing some open questions.
The appendices describe the details of some of the computations whose results were
used in the text.  

\sectiono{Background} \label{srev}

In this section we shall 
give a brief introduction to some aspects of string field 
theory\cite{zwiebach_open,9206084,9705241} 
that we shall use, and
also
review the algorithm developed in  \cite{2002.04043} 
for carrying out 
systematic computation of closed string amplitudes in the presence
of D-instantons. We shall focus on the contribution associated with a single D-instanton.
Furthermore, we shall assume that the background does not have any other D-branes,
so that all the boundaries of the world-sheet have D-instanton boundary conditions.

\subsection{String field theory}

The usual world-sheet description of string amplitudes involves integration over 
moduli spaces of Riemann surfaces with punctures, with the external closed
string vertex operators 
inserted at the punctures in the bulk of the world-sheet and the external
open string vertex operators inserted at the punctures on the boundary of the
world-sheet. The integrand is computed from appropriate correlation functions
of the vertex operators and world-sheet 
ghost fields on the Riemann surface. In the two dimensional string theory that we
shall consider, the world-sheet theory consists of a free scalar field $X$ describing the
time coordinate, a Liouville field theory with central charge 25, and the usual $b$, $c$
ghost system.

At a superficial
level, string field theory may be regarded as a regular quantum field theory, whose
perturbation expansion generates the same result at the one in the world-sheet 
approach. Each Feynman diagram of string field theory generates an integral 
similar to what we would get in the world-sheet approach, but the integration over the
moduli runs over a subspace of the full moduli space. The sum over all Feynman
diagrams gives back integration over the full moduli space. We shall see however,
that this correspondence is only formal, -- often the world-sheet expression is divergent
while the corresponding string field theory expression yields finite result.

Although the relevant field theory that describes the interaction of closed strings 
and
open  strings on the D-instanton is the open-closed string field theory,
since our goal will be to understand how to integrate over the open string fields,
we shall regard the closed string fields as fixed background fields.
In general there will be infinite number of closed string fields, but for most of our 
analysis in the 
two dimensional
string theory, we shall only need the coupling of the
on-shell closed string tachyon to the open string fields.  So we shall only write down the
coupling of the open string fields to the closed string tachyon field
carrying energy $\omega$ and Liouville momentum $P$, which we shall denote
by $\Phi_C(\omega,P)$. Despite its name, it actually describes a massless field in two
dimensions. In the convention of \cite{1912.07170}, the on-shell condition takes the
form $P=|\omega|/2$.

If $H$ denotes the vector space of states of the open string, including
matter and ghost excitations, then the off-shell open string field is taken to be an arbitrary
element $|\Psi\rangle$ of $H$. This means that if $\{|\phi_r\rangle\}$ is the set of basis
states in $H$, and if we expand $|\Psi\rangle\in H$ as
\be\label{esftexpansion}
|\Psi\rangle =\sum_r \chi^r |\phi_r\rangle\, ,
\ee
then $\{\chi^r\}$ are the dynamical degrees of freedom on which the path integral
is to be performed (after suitable gauge fixing). For string field theory on a D$p$-brane
the sum over $r$ will also include integration over $(p+1)$ continuous variables 
labelling the energy and momenta of the state, but on the D-instanton the open strings
do not carry any continuous momenta and the sum over $r$ is a discrete sum.
Therefore we can regard this as a field theory in zero space-time dimensions. 
The grassmann parity of $\chi^r$ is opposite to that of the state $|\phi_r\rangle$.
Therefore
the string field $\Psi$ is grassmann odd.

The string field theory action takes the form:
\be\label{esftaction}
S={1\over 2} \langle \Psi|Q_B|\Psi\rangle +\cdots \, 
\ee
where $Q_B$ is the world-sheet BRST operator and $\cdots$ denotes interaction
terms. Note that we have not explicitly written down the 
kinetic term for the closed string fields since
they will be treated as background fields and the kinetic term will not be needed for
our analysis. The interaction term takes the form
\ben \label{einteractionvertex}
&& \sum _{m\ge 0} \sum_{n\ge 0} {1\over m!n!}
\int {d\omega_1}\, dP_1\cdots {d\omega_m}\, dP_m\, 
\Phi_C(\omega_1, P_1)\cdots \Phi_C(\omega_m, P_m) \nonumber \\
&& \hskip 1in \times \, \chi^{r_1}\cdots \chi^{r_n}\, 
F^{(m,n)}_{r_1\cdots r_n}(\omega_1,P_1,\cdots, \omega_m, P_m)\, ,
\een
where the functions $F^{(m,n)}_{r_1\cdots r_n}(\omega_1,P_1,\cdots, \omega_m,P_m)$ 
are constructed
from appropriate correlation functions of $m$ closed string tachyon vertex operators
with energies $\omega_1,\cdots,\omega_m$,
$n$ open string vertex operators $\phi_{r_1},\cdots,\phi_{r_n}$ 
and appropriate set of $b$-ghost insertions,
integrated over some appropriate subspaces of moduli space of Riemann surfaces
with punctures. In general the functions $F^{(m,n)}_{r_1\cdots r_n}$ depend 
on the choice
of local coordinate systems at the punctures, but since we shall need these functions
only for on-shell closed string states, the dependence on the local coordinates at the
punctures in the bulk of the world-sheet, where closed string vertex operators are
inserted, drops out. 
For on-shell open string states the
correlation function is also independent of the choice of local coordinate system at the
punctures on the boundaries of the world-sheet, but this
is not the case for off-shell open string states. 
Different choices of local coordinates
lead to apparently different string field theories, but they can be shown to be related
to each other by field redefinition\cite{9301097}.

The action \refb{esftaction} is to be regarded as the quantum master action in the
Batalin-Vilkovisky formalism\cite{bv1,bv2}. Let us assign ghost number to the states in 
$H$ by declaring $c,\bar c$ to have ghost number 1, $b,\bar b$ to have ghost number 
$-1$, matter fields to have ghost number 0 and the $SL(2,R)$ invariant vacuum to have ghost
number 0. In the BV formalism
we declare the open string fields
multiplying states of ghost number $\le 1$ as fields and the fields multiplying
states of ghost number $\ge 2$ as antifields. 
To be more specific, 
let us introduce basis states $|\vp_r\rangle$ in ghost number
$\le 1$ subspace and $|\vp^r\rangle$ in the ghost number $\ge 2$ subspace such that
\be
\langle \vp^r |\vp_s\rangle=\delta^r_s=\langle\vp_s|\vp^r\rangle, \qquad 
\langle\vp^r|\vp^s\rangle
=0, \qquad \langle \vp_r|\vp_s\rangle=0\, ,
\ee
and expand the string field as,
\be \label{edefantifield}
|\Psi\rangle = \sum_r( \psi^r |\vp_r\rangle + \psi_r |\vp^r\rangle)\, .
\ee
Then we declare $\psi^r$ as a field and $\psi_r$ as the conjugate anti-field (up to a sign).

String field theory constructed this way has gauge symmetries where the gauge
transformation parameters are in also in one to one correspondence with states of the 
open string. Therefore we can represent a gauge transformation parameter as:
\be\label{esftgauge}
|\Lambda\rangle =\sum_r \theta^r |\phi_r\rangle\, ,
\ee
with $\theta^r$'s being the parameters of infinitesimal gauge transformation. 
Even though $\theta^r$ and $\chi^r$ appear in similar expansions 
\refb{esftexpansion} and
\refb{esftgauge}, their grassmann
parities are opposite of each other. The significance of this will become clear later
The gauge
transformation of a field $\chi^r$ under a gauge transformation parameter $\theta^s$
is given (up to a sign) by the double derivative of the action with respect to $\chi_r^*$ and
$\chi^s$, where $\chi_r^*$ is the anti-field conjugate to $\chi^r$, computed via
\refb{edefantifield}\cite{henn}. More general gauge transformations where the
transformation parameters themselves are functions of fields have been discussed in
\cite{9309027}, but we shall not discuss them here.

The actual path integral of string field theory
is carried out over fields in a Lagrangian submanifold -- this is the analog of
gauge fixing in the Fadeev-Popov formalism. We shall not give a general definition
of Lagrangian submanifolds, but it suffices to say that one set of choices correspond
to setting, for each pair $(\psi^r, \psi_r)$, either $\psi^r$ to 0 or $\psi_r$ to 0. This can
be done independently for each $r$. The result of the path integral for physical
quantities can be shown to be
formally independent of the choice of the Lagrangian submanifold.

A particular choice of Lagrangian submanifold corresponds to setting all the anti-fields
to 0. Therefore $|\Psi\rangle$ has ghost number $\le 1$. It then follows from ghost
number conservation that the action depends on only the ghost number 1 sector
fields. These are called classical fields since the physical open string states of the string 
belong to this sector. In this case the path integral decomposes into an integration over
the classical fields on which the action depends and an integration over the fields
of ghost number $\le 0$ on which the action does not depend. Since the string fields 
$\chi^r$ have opposite grassmann parity compared to the corresponding 
gauge transformation parameters $\theta^r$, the integration over the string fields
of ghost number $\le 0$ can be interpreted as division by the volume of the gauge group
generated by gauge transformation parameter $\Lambda$ carrying ghost number
$\le 0$. This connects the BV formalism to the usual path integral formulation of gauge
theories, in which we start with the path integral over the classical fields weighted by
the exponential of the classical action and divide this by the volume of the gauge group.
While this is a sensible starting point when the gauge group has a finite volume, as in
lattice gauge theories on a finite lattice and with a 
compact gauge group, this is
usually not suitable for perturbation theory, since the kinetic operator of the classical
action has flat directions associated with the gauge deformations. This makes
the kinetic operator non-invertible, requiring us to gauge fix the theory. In the BV
formalism this will correspond to choosing a different Lagrangian submanifold.

The Lagrangian submanifold commonly used for carrying out perturbation theory is
called the Siegel gauge. Here we restrict the path integral to a subspace of string fields
of the form:
\be \label{esiegelgauge}
b_0|\Psi\rangle=0\, .
\ee
More precisely this means the following. We can choose the basis states $|\vp^r\rangle$
and $|\vp_r\rangle$ in \refb{edefantifield} in such a way that one of the following holds:
\begin{enumerate}
\item $b_0|\vp^r\rangle=0$, $c_0|\vp_r\rangle=0$. In this case we choose $\psi^r=0$.
\item $c_0|\vp^r\rangle=0$, $b_0|\vp_r\rangle=0$. In this case we choose $\psi_r=0$.
\end{enumerate}
Since for each $r$ we set to zero either the field or the anti-field, this is clearly a
Lagrangian submanifold. The corresponding $|\Psi\rangle$ given in \refb{edefantifield}
satisfies \refb{esiegelgauge}.

In the Siegel gauge,  the kinetic operator $Q_B$ appearing in \refb{esftaction}
reduces to $c_0L_0$ and the open string propagator is proportional to
\be\label{eprop}
b_0 L_0^{-1} = b_0 \int_0^1 dq\, q^{L_0-1}\, ,
\ee
where $L_n$'s denote the total Virasoro generators of the world-sheet conformal
field theory of matter and ghosts.
As we shall discuss later, equality of the two sides of \refb{eprop} 
is somewhat formal and breaks down for
states with $L_0\le 0$, but let us proceed with this for now.
In this case there is a simple algorithm to determine which region of the moduli
space is covered by a given Feynman diagram. The local coordinate
system at the punctures, needed for defining the interaction term of string field
theory for off-shell external states, play an important role in this analysis.
If we contract the external state
$\chi^r$ from one interaction vertex with the external state $\chi^s$ from 
another (or the same) 
interaction vertex\footnote{Throughout this paper an interaction vertex of string field
theory will have the
same meaning as in the Feynman diagrams of a quantum field theory, and should not
be confused with vertex operators that refer to operators in the world-sheet theory.}
by a propagator,  then geometrically this corresponds to 
sewing the Riemann surfaces associated with the two interaction vertices
by identifying the corresponding local coordinates $w$ and $w'$ via the
relation
\be 
w\, w'=-q\, , \qquad 0\le q\le 1\, .
\ee
For this reason, we shall call 
$q$ the sewing parameter associated with the open string
propagator. 
Following this procedure we can determine the regions in the moduli space of
punctured Riemann surfaces covered by each Feynman diagram. 
This not only generates the family of Riemann surfaces corresponding to the
Feynman diagram, but also generates the local coordinates at the punctures of these
Riemann surfaces in terms of the local coordinates at the punctures of the 
Riemann surfaces associated with the interaction vertices.
The interaction
vertices are chosen such that the region corresponding to the sum of all Feynman
diagrams covers the full moduli space. We shall discuss several examples of this
in \S\ref{s1}. This condition is needed for string field theory to reproduce the usual
world-sheet results for the amplitudes, but at a more fundamental level, this condition
is also necessary for the action to satisfy the BV master equation.

We shall use a convention in which the path integral measure 
is weighted by $e^S$. In this 
convention the contribution to an amplitude from a Feynman diagram containing
a single interaction 
vertex is given directly
by the corresponding function $F^{(m,n)}_{r_1\cdots r_n}(\omega_1,P_1,\cdots,\omega_m,
P_m)$ 
without any additional sign or numerical factors.
However we shall keep the overall normalization of $S$ arbitrary, by choosing an
arbitrary normalization of the vacuum in the world-sheet theory (see \refb{econvention}).
This will have the advantage that we can use our result both in the Euclidean and
in the Lorentzian theory (in which the standard measure is $e^{iS}$) by changing the
overall normalization of $S$.

\subsection{Dealing with the open string tachyons}

We shall now review the divergences that arise in the world-sheet description
of the amplitudes in the presence of D-instantons. They are of two kinds -- those 
associated with the open string tachyons and those associated with the 
open string zero mode fields. In this subsection we shall describe some general 
aspects of D-instanton perturbation theory, and review the procedure 
discussed in  \cite{2002.04043} for dealing with the open string tachyons. The effect
of open string zero modes will be discussed in \S\ref{s2.4}.
\begin{enumerate}

\item In the presence of D-instantons we need to sum over contributions to string 
amplitudes from Riemann surfaces with boundaries, with the world-sheet fields
satisfying boundary condition associated with the D-instantons at the boundaries. 
We need to include disconnected world-sheets in the computation, with the rule that
each connected component must have either at least one external closed string
insertions or at least one boundary with D-instanton boundary condition, and at least
one of the connected components must have both, an external closed string insertion 
and a boundary.
\item The integrals over the moduli spaces of Riemann surfaces 
suffer from infra-red divergences from the boundaries of the moduli space. These
divergences are most problematic for open string degenerations since the open
strings living on the D-instanton do not carry any momentum and therefore the 
usual approach of dealing with the divergences by analytic continuation in the external
momenta does not work. However, as we shall review below, 
string field theory can be used to deal with these
divergences. 
\item The infrared divergences
associated with open string degeneration arise from the $q\simeq 0$ region on the right
hand side of \refb{eprop}. 
As is obvious from \refb{eprop}, these divergences  arise from $L_0\le 0$ states.
The divergences associated with $L_0<0$ states can be easily tackled by using
the left hand side of \refb{eprop} instead of the right hand side. 
 \label{p3}
This leads to the replacement rule:
\be\label{erp0}
\int_0^1 dq\, q^{\alpha-1} \quad \to \quad {1\over \alpha} \qquad
\hbox{for $\alpha\ne 0$}\, .
\ee
For $\alpha>0$ this is an identity but for $\alpha<0$ this is a prescription that
ensures that the world-sheet computations reproduce the string field theory results.

Since the $L_0<0$ states represent tachyonic modes, 
the reader may worry about treating them this way when they propagate
in the loops. Indeed, this will not be a sensible way to treat tachyons in a quantum field
theory. However, for D-instantons the tachyons do not carry any momentum and 
the path integral over the tachyons represent ordinary integrals, with the integrand having
a local minimum in the direction of the tachyon field instead of a local maximum. This means
that the steepest descent contour runs along the imaginary axis of the tachyon field. Once we
take
the integration contour over the tachyonic mode along this steepest descent contour, as
is customary for evaluating contributions from saddle points of a path integral, there is no
problem in carrying out systematic perturbation expansion to evaluate the contribution
to the integral.
\item For treating the divergences associated with the $L_0=0$ states we need to
introduce the notion of Wilsonian effective action that we shall discuss next.
\end{enumerate}

\subsection{Wilsonian effective action} \label{swilson}

In string field theory, it is possible to integrate out the set of fields corresponding to 
a given set of $L_0$ eigenvalues and write an effective action for the other fields.
The resulting effective action shares all the properties of the original master 
action, both in the classical theory\cite{0112228,0306332,2006.16270,2012.09521} and in the 
quantum theory\cite{1609.00459}.
We can use this to integrate out all the open string 
fields with $L_0\ne 0$, including the open string
tachyon, leaving us with the master action of the $L_0=0$ fields only. We shall call
this the Wilsonian effective action.
The interaction
vertices of this effective action are obtained by summing over all the Feynman diagrams
of the original string field theory, with only the $L_0=0$ open string states (and
closed string states) as external states, and by removing, from each internal open
string propagator,
the contribution from the $L_0=0$ states. From the world-sheet perspective, this corresponds to subtracting all contributions of the form $\int_0^1 dq\, q^{-1}$ in the
integrand. Combining this with \refb{erp0} we arrive at the modified replacement
rule:
\be\label{erp1}
\int_0^1 dq\, q^{\alpha-1} \quad \to \quad {1\over \alpha} \left\{ 1 - \delta_{\alpha,0}\right\}\, .
\ee
The $ \left\{ 1 - \delta_{\alpha,0}\right\}$ reflects that for $\alpha=0$, we need to remove the
contribution from the region $0\le q\le 1$ altogether. Since this removes all
divergent terms, computation of the Wilsonian effective action does not suffer from any
divergences.

\subsection{An alternative replacement rule} \label{salternative}

We shall now describe an alternative to the replacement rule \refb{erp1}. 
We shall consider the cases $\alpha=0$ and $\alpha=-1$ only, since these are the relevant
cases for our analysis.
If we replace the
lower limit of $q$ integration by some small number $\delta$, then \refb{erp1} can also
be stated as,
\be
\int_\delta^1 dq\, q^{\alpha-1} \quad \to \quad  \cases{-1 \quad 
\hbox{for $\alpha=-1$},\cr
0  \, \quad \hskip .1in \hbox{for $\alpha=0$}\, .}
\ee
On the other hand if we explicitly evaluate the left hand side, we get
\be \label{erep213}
\int_\delta^1 dq\, q^{\alpha-1} = \cases{\delta^{-1}-1 \quad \hbox{for $\alpha=-1$},\cr
-\ln\delta  \, \quad  \hskip .1in \hbox{for $\alpha=0$}\, .}
\ee
Therefore the replacement rule simply corresponds to dropping the boundary terms
proportional to $\delta^{-1}$ and $\ln\delta$.

This procedure is particularly useful if the integrand is a total derivative -- which is indeed
the case in many of the expressions that we shall encounter. In this case we can express 
this as an  integral over the various boundary segments labeled by $q_i=\delta_i$ 
following the procedure described
in \cite{1902.00263}, and then drop terms proportional to $\delta_i^{-1}$ and
$\ln\delta_i$ in the final expression. 
Although we shall not describe this analysis in this paper, we have checked that
following this procedure we get the same results as the ones reported here.

\subsection{Dealing with the zero mode fields} \label{s2.4}

At the level of the Wilsonian effective action, the open string fields 
associated with the $L_0=0$ states, which we shall collectively call
the zero modes, remain unintegrated. However in computing physical amplitudes
we have to eventually integrate over them. Doing this
requires physical understanding of
the origin of these fields.

\begin{enumerate}

\item First let us consider the zero modes arising from the pure ghost sector. 
We begin by writing down the full expansion of the string field in the $L_0=0$ sector,
involving pure ghost excitations:
\be\label{efieldexpandfirst}
\psi^1 c_0|0\rangle + \psi^2|0\rangle + \psi_1 \, c_{-1} c_1|0\rangle + \psi_2 \, c_{-1}
c_0 c_1|0\rangle \, .
\ee 
The analog of the field $\psi^1$ in heterotic string theory has been identified 
in \cite{1912.05463}
with the zero mode of the Nakanishi-Lautrup field\cite{NL1,NL2}, while the
fields $\psi^2$ and $\psi_1$ can be identified with the zero mode of the ghost and
the anti-ghost fields associated with the $U(1)$ gauge symmetry. Siegel gauge
condition would set $\psi^1$ and $\psi_2$ to zero, but the resulting kinetic operator
$c_0L_0$ will vanish while acting on the remaining $L_0=0$ states, leading to
divergent propagator of the fields $\psi^2$ and $\psi_1$. 
For this reason, in the $L_0=0$ sector
we choose a new gauge\cite{2002.04043}
in which we set all the anti-fields to zero, and express the
string field involving pure ghost excitations as:
\be\label{efieldexpand}
\psi^1 c_0|0\rangle + \psi^2|0\rangle +\cdots \, .
\ee 

\item We are supposed to
integrate over the fields $\psi^1$ and $\psi^2$. First consider the effect of $\psi^1$.
One can show that the field $\psi^1$
associated with the state $c_0|0\rangle$ has finite kinetic term and propagator, and can
be treated perturbatively. This means however that besides the usual world-sheet
contribution, with the replacement rule \refb{erp1}, we must also separately include the
contribution from the propagator of this new field $\psi^1$ in every propagator of a 
Feynman diagram. A classical version of a similar result 
can be found in 
\cite{1912.05463}.

\item
The field $\psi^2$ associated with the
state $|0\rangle$ multiplies a state of ghost number zero. 
Using the same argument as for the full string field theory, one can show that it
decouples from the action due to ghost number
conservation. The interpretation of integration over $\psi^2$ follows from the
same analysis that was used for the full string field theory -- since
the string field $\psi^2$
associated with the state $|0\rangle$ is grassmann odd, 
the integration over $\psi^2$ corresponds to dividing by the volume of
gauge transformation in string field theory 
generated by $\theta|0\rangle$, where $\theta$ in the infinitesimal
gauge transformation parameter. 
This gauge transformation, in turn, is related to the rigid 
U(1) gauge transformation with
parameter
$\tilde\theta$ after a suitable redefinition that relates $\theta$ to $\tilde\theta$.
For example, 
under the rigid gauge transformation any open string that stretches from the original
D-instanton to a second D-instanton picks up a phase $e^{i\wt\theta}\simeq (1+i
\wt\theta)$, 
but in general
the transformation law under the gauge transformation generated by $\theta|0\rangle$
in open string field theory is
more complicated.
We can find the relation between $\theta$ and $\tilde\theta$ by comparing these
transformation laws. If $\theta$ is related to $\tilde\theta$ as $\theta=J \tilde \theta$ where $J$
is some function of the fields, then $d\theta$ is given by $J d\tilde \theta$ and dividing by
$\int d\theta$ is equivalent to dividing by $J\int d\tilde \theta$. Since $\int d\tilde\theta$
is the volume of the rigid U(1) gauge group,
this contributes an overall constant to the path integral that can be absorbed into
the normalization constant $\NN$ in \refb{emulti1int}. Therefore the net effect of
dividing by $\int d\theta$ is to include a factor of $J^{-1}=\exp[-\ln J]$ in the path integral.
This can give additional contribution to the amplitudes
that needs to be taken into account in our analysis.

\item Finally, let us turn to the zero modes coming from the matter sector in the
ghost number $1$ sector. Typically for single  instanton, these arise 
from the collective excitations of
the D-instanton that translate the D-instanton along transverse directions along
which we have translational symmetry. 
For the two dimensional string theory that we shall analyze, there is
only one such zero mode described by the vertex operator $c\p X$ in the world-sheet
theory. We shall denote this field by $\phi$. 
The integration over this mode is carried out at the end of the 
computation. In general, $\phi$ itself is not the collective coordinate,
but it is expected to be related to the collective 
coordinate $\wt\phi$
describing time translation of the D-instanton by a field redefinition\cite{1702.06489}. 
We can
find this explicit field redefinition by noting that the dependence
of the amplitude on the collective mode 
$\wt\phi$ must be proportional to $e^{-i\wt\phi\omega}$ 
where $\omega$ 
denotes the total outgoing energy of all the external closed string states in the amplitude.
Therefore we first make the change of variable of integration in the path integral from
$\phi$ to $\wt\phi$ so that the amplitude has the correct dependence on $\wt\phi$, and then
perform the $\wt\phi$ integration to pick up the energy conserving delta function. 
Any Jacobian that arises from this change of variable must be taken into account in
the computation of the amplitudes.

\end{enumerate}

This discussion can be summarized by saying that after evaluating the amplitudes
using the replacement rule \refb{erp1} that renders the world-sheet integrals finite, we
need to include the following extra contributions:
\begin{enumerate}
\item We need to take into account the contribution from the $\psi^1$ exchange in each
internal propagator of the Feynman diagrams.
\item We need to take into account the additional contribution to the effective action 
from the Jacobian arising from change of variables from the open string matter sector
zero modes to the collective coordinates. 
\item We also need to take into account the additional contribution to the effective action 
from the Jacobian arising from change of variables from the open string gauge 
transformation parameter $\theta$ multiplying the vacuum state $|0\rangle$ to the
rigid $U(1)$ gauge transformation parameter.
Both these Jacobians may give additional contribution
to the amplitudes.
\item The amplitudes on the D-instanton do not conserve energy (and momentum). 
This is achieved at
the very end when we integrate over the collective mode(s) $\wt\phi$ after changing
variable from the open string matter sector zero mode(s) to the collective coordinate(s).
\end{enumerate}

\sectiono{Approximations} \label{serror}

In order to implement the procedure described in \S\ref{srev},  we need to
know the precise relation between the
integration parameters entering the expression for a world-sheet amplitude and the
variables $q_i$ that are used to express the propagators via \refb{eprop}. We can then
express the integral as integration over the variables $q_i$'s near the $q_i=0$ 
region, expand the integrand in a power series expansion in the $q_i$'s and make the
replacement \refb{erp1}. Our goal in this section will be to analyze how much error
we are allowed to make in identifying the variables $q_i$ without actually changing the
value of the integral.

In order to convince the reader that this is not an empty exercise, we shall first 
demonstrate, with an example, that identifying the correct variables $q_i$ is
important before applying the replacement rule \refb{erp1}. 
Let us consider an integral
\be\label{example1}
\int_0^1 dq\, q^{-2} \to -1\, ,
\ee
where we have used the replacement rule \refb{erp1}. Now suppose that instead
of analyzing \refb{example1} directly, we first make a change of variable:
\be
q' = {q\over 1 - c\, q}\, ,
\ee
with $c<1$, so that the change of variable is well defined in the range $0\le q\le 1$.
It is easy to check that $dq\, q^{-2}=dq'\, q^{\prime -2}$. Using this we can express
\refb{example1} as
\be\label{example2}
\int_0^{1/(1-c)} \, dq'\, q^{\prime -2} =\int_0^{1} \, dq'\, q^{\prime -2} +
\int_1^{1/(1-c)} \, dq'\, q^{\prime -2} \to -1 + \{ 1-(1-c)\} = - (1-c)  \, ,
\ee
where in the last but one 
step we have used the replacement rule \refb{erp1} for the
$q'$ integral from 0 to 1, and have explicitly evaluated the finite integral from
1 to $1/(1-c)$. We now see that the final result in \refb{example2} differs
from \refb{example1}. This demonstrates that it is important to identify the variable
$q$ correctly using string field theory before applying \refb{erp1}.

Before proceeding with the analysis, we shall make a change in the notation.
For the type of string field theory we shall be considering, the 
integration parameters entering the expression for a world-sheet amplitude will have natural
relation not with the $q_i$'s but with $u_i=\eps_i q_i$'s where $\eps_i$ are small parameters.
In other words, the integrand, when expanded in powers of the $u_i$'s, will have finite
coefficients of expansion.
In this case it is more convenient to trade in the variables $q_i$'s for the $u_i$'s, and
express the replacement rule \refb{erp1} by
\be\label{erp2}
\int_0^{\eps_i} du_i\, u_i^{\alpha-1} \quad \to \quad {\eps_i^\alpha\over \alpha}
\,  \left\{ 1 - \delta_{\alpha,0}\right\}\, .
\ee

If we can determine the exact relation between the moduli parameters of the Riemann
surface in terms of which we compute the integrand, 
and the $u_i$'s, then we can carry out the procedure described above.
However this is not strictly necessary -- we shall show that
even if we make some error in finding the
relation between the moduli parameters and the $u_i$'s, as long as the error goes to
zero sufficiently fast in the $u_i\to 0$ limit, we can still get the
result for the integral exactly. 
Our goal in this section will be to study how fast the error should vanish as $u_i\to 0$
so as to not make an error in the final value of the integral.
This
question is important since during our analysis we often will need to make
approximations in relating the natural moduli space variables to the string field theory variables
$q_i$, or equivalently, $u_i$.

We shall first consider the case where we have one integration variable $u$, and
the integrand is $u^{-2}$.
In this case 
the rule \refb{erp2} can be stated more explicitly as:
\be\label{erp3}
\II\equiv \int_0^a du\, u^{-2} =  \int_0^\eps du\, u^{-2}  +  \int_\eps^a du\, u^{-2} 
 \quad \to \quad I\equiv - {1\over \eps}
+ \int_{\eps}^a du\, u^{-2}\, ,
\ee
for any positive constant $a$. Note that $\II$ is a formal integral since the integral
diverges from $u=0$, but $I$ is well defined. 
It can be easily verified that
$I$ defined in \refb{erp3} is independent of $\eps$.
Now suppose that instead of using the variable $u$, we erroneously use the variable
$v$, related to $u$ via
\be\label{erp2.5}
u = f(v)=a_0 v + a_1 v^2 + a_2 v^3+\cdots\, ,
\ee
and use the replacement rule \refb{erp2} with the variable $u$ replaced by $v$. 
Then the formal integral $\II$, when converted into an integration
over the variable $v$, reduces to another formal integral:
\be
\II'
= \int_0^b dv f'(v) f(v)^{-2} = \int_0^{\eps'}  dv f'(v) f(v)^{-2} 
+ \int_{\eps'}^{b} dv f'(v) f(v)^{-2}, 
\quad f(b)=a, \quad f(\eps')=\eps\, .
\ee
Since
\be
f(v)^{-2} f'(v) = a_0^{-1} v^{-2} + \OO(1)\, ,
\ee
we can express $\II'$ as
\be
\II' =  {1\over a_0} \int_0^{\eps'} \, dv\, v^{-2}  + \int_{0}^{\eps'} dv\, \left\{f(v)^{-2} 
f'(v)
- a_0^{-1} v^{-2} \right\}+    \int_{\eps'}^{b}  dv\, f(v)^{-2} f'(v)\, .
\ee
Note that only the first integral on the right hand side is divergent.
If we now use the replacement rule \refb{erp2} with $u$ replaced by $v$, we can
replace the formal integral $\II'$ by:
\be\label{erp3new}
\II'\quad \to\quad I' \equiv -{1\over a_0\, \eps'} + \int_{0}^{\eps'} dv\, \left\{f(v)^{-2} f'(v) 
- a_0^{-1} v^{-2} \right\}+   \int_{\eps'}^{b}  dv\, f(v)^{-2} f'(v)\, .
\ee
It can be verified that the finite integral $I'$ is independent of $\eps'$.
Using
\be 
\int_{\eps'}^{b} f(v)^{-2} f'(v) dv = \int_\eps^a {du\over u^2}\, ,
\ee
we get
\be \label{erp3p}
 I' = -{1\over a_0\, \eps'} + \int_{0}^{\eps'} dv\,  \left\{f(v)^{-1} f'(v) 
- a_0^{-1} v^{-2} \right\}+   \int_\eps^a {du\over u^2}\, .
\ee
We can now use 
the $\eps$ independence of $I$ and $\eps'$ independence of
$I'$ to evaluate both $I$ and $I'$ in the limit $\eps,\eps'\to 0$. 
Since \refb{erp2.5} gives 
\be
\eps = a_0\eps' + a_1\eps^{\prime 2}+\cdots\, ,
\ee
we get, from \refb{erp3}, \refb{erp3p},
\be \label{eapr1}
I' = I -{a_1\over a_0^2} \, .
\ee
This shows that if instead of
applying the replacement rule \refb{erp2} on the original integral $\II$ expressed as
integration over $u$, we first make a change of variables from $u$ to $v$ as given
in \refb{erp2.5}, and then
use analogous replacement rule on the integral, the result differs by $-a_1/a_0^2$.
The
example described at the beginning of this section is a special case of this 
general result.

Next we consider the case $\alpha=0$. In this case the replacement rule can be
expressed as:
\be 
\JJ=\int_0^a {du\over u} \quad \to \quad J\equiv \int_\eps^a  {du\over u}\, 
\ee
This time $J$ does depend on the cut-off $\eps$ since the effect of this cut-off is to
leave unintegrated the zero mode fields whose definition implicitly 
depends on the cut-off $\eps$. Under a change of variable of the form 
\refb{erp2.5}, the formal integral $\JJ$ may be expressed as,
\ben\label{eap12}
&& \JJ' = \int_0^{b} f(v)^{-1} f'(v) dv \nonumber \\
&=& \int_0^{\eps'} v^{-1} dv 
+ \int_0^{\eps'} \{f(v)^{-1} f'(v)-v^{-1}\} \, dv
+  \int_{\eps'}^{b}  f(v)^{-1} f'(v) dv \, .
\een
Only the first term is divergent. Using the replacement rule \refb{erp2} with $u$ replaced
by $v$, we can now replace $\JJ'$ by:
\ben
\JJ' \to J' &\equiv& \int_{\eps}^{\eps'} v^{-1} dv 
+ \int_0^{\eps'} \{f(v)^{-1} f'(v)-v^{-1}\} \, dv + \int_{\eps'}^{b}  f(v)^{-1} f'(v) dv \, ,
\een
where we have used the fact that in a given variable, the rule for subtraction in
a logarithmically divergent integral is to
replace the lower limit of integration by $\eps$ instead of 0.
Using
\be
\int_\eps^a  {du\over u}=\int_{\eps'}^{b}  f(v)^{-1} f'(v) dv \, ,
\ee
we get
\be\label{eapr2}
J'= J - \ln\, a_0 \, .
 \ee

It is worth noting that the origins of the errors \refb{eapr1} and \refb{eapr2} are
somewhat different. The error \refb{eapr1} associated with linearly divergent integral
arises due to the fact that under a change of variable from $u$ to $f(u)$, the 
replacement rule \refb{erp2} does not remain invariant but gets shifted by a constant.
However, since the right hand sides of \refb{erp3} and \refb{erp3new} are independent of the cut-offs $\eps$ and $\eps'$, the error is 
not related to a change in the cut-off. On the other hand the
error \refb{eapr2} arises from the fact that for logarithmically divergent integral the
integration region $u\le \eps$ is removed by hand, making the integral dependent on
the cut-off $\eps$. Therefore if we make a change of variable from $u$ to
$v$ and use the same cut-off $\eps$ for the $v$ variable, then it 
corresponds to a different cut-off in the original variable $u$ and result changes.

Eqs.\refb{eapr1} and \refb{eapr2} show that if we do not want to make an error in
the evaluation of an amplitude, then for a linearly divergent integral we are allowed to
make a fractional error of order $u^2$, while for a logarithmically divergent integral we
are allowed to make a fractional error of order $u$.
Intuitively these results may be understood as follows. When there is a tachyon
in the intermediate state, it can produce a contribution of order $1/\eps$ in the
integral according to \refb{erp2}. 
The $1/\eps$ term is expected to cancel at the end, leaving behind a finite
contribution. However a fractional error of order $\eps$ can combine with the $1/\eps$
term to give a finite contribution that does not cancel.
In contrast, divergences associated with massless particle poles
are logarithmic. Therefore a fractional error of order $\eps$ is not important, but the
leading order identification of variables must be correct. 
 
Even though we have not carried out a detailed analysis for the many variable case, we
shall adopt this guideline for determining what are allowed approximations. Regarding
the $\eps_i$'s associated with different degenerations as independent, we can see that
in general we cannot ignore fractional errors of order $\eps_i$, but can ignore higher order
corrections. Thus, for example, for two variables we shall not ignore terms of order
$\eps_1$, $\eps_2$ or $\eps_1\eps_2$ but will ignore terms of order $\eps_1^2$ and
$\eps_2^2$. One might worry that $1/\eps_1\eps_2$ contribution from the tachyon
may combine with the $\eps_1^2$ term to give a finite contribution $\eps_1/\eps_2$.
However since we know that the final results are independent of $\eps_1$ and $\eps_2$,
such terms must cancel at the end. 
If in some cases $\eps_1$ and $\eps_2$ cannot be taken to be independent
-- since they are determined by the underlying string field theory -- then we need to
be a bit more careful and keep terms up to sufficient order in order not 
to make an error.

In our analysis, the $\eps_i$'s will be determined in terms of two string field theory
parameters $\wt\lambda$ and $\alpha$, which we shall take to be large. The contributions
from different Feynman diagrams will have the structure of Laurent series expansion
in $\alpha^{-1}$ and $\wt\lambda^{-1}$. We shall drop terms with positive powers of
$\alpha^{-1}$ and / or $\wt\lambda^{-1}$ by taking $\alpha$ and $\wt\lambda$ to be large.
Since the final result is expected to be independent of $\alpha$ and $\wt\lambda$, we could
in principle also drop terms with positive powers of $\alpha$ and / or $\wt\lambda$,
keeping only the terms that are independent of $\alpha$ and $\wt\lambda$. However we
keep these terms, since their cancellation in the final result provides a consistency test
of our results.

Finally, we note that the results of this section are also valid if we use the 
alternative procedure described in \S\ref{salternative}. To see this, recall that the 
procedure of \S\ref{salternative} involves dropping terms proportional to $\delta^{-1}$ 
and $\ln\delta$ from the boundary at $q=\delta$ before taking the $\delta\to 0$ limit. 
Now if we make a change of variable
from $q$ to $\tilde q=b_0 q+ c_0 q^2 +\OO(q^3)$, then the boundary at
$q=\delta$ corresponds to 
$\tilde q=\tilde\delta$
were $\tilde\delta= b_0\delta + c_0\delta^2+\OO(\delta^3)$. Therefore, if we were using
the variable $\tilde q$ as the sewing parameter, then we would drop terms proportional
to $\tilde \delta^{-1}$ and $\ln\tilde\delta$. It is easy to see that $\tilde\delta^{-1}
=b_0^{-1} \delta^{-1} - c_0 b_0^{-2} +\OO(\delta)$ and $\ln\tilde \delta =\ln\delta + \ln b_0
+\OO(\delta)$. Therefore as long as $b_0=1$ and $c_0=0$, we shall not make any error
if we interpret $\tilde q$ as the sewing parameter instead of $q$. This argument easily
generalizes to the case of multiple sewing parameters.

\sectiono{Interaction vertices and covering of the moduli space} \label{s1}

The string field theory that we shall use falls in the general framework discussed in
\cite{zwiebach_open,9206084,9705241}, but the actual interaction vertices that we use
in our analysis will be based on the choice of local coordinates that are related to the
global upper half plane coordinates by $SL(2,R)$ transformation. 
In this section we shall describe our choice of local coordinates for defining various interaction vertices
of string field theory, and also describe how the contributions from different Feynman diagrams cover
different regions of the moduli space of Riemann surfaces with punctures. We shall denote an external open string
by O and an external closed string by C. 
Since our goal will be to understand how to integrate over the open string fields in fixed
background closed string fields, we shall only consider Feynman diagrams with internal
open string propagators and not those with internal closed string propagators.

There are more elegant choice of interaction vertices based on half 
string overlap\cite{wittensft}, the minimal area metric\cite{9206084,9705241} 
or the hyperbolic 
metric\cite{1703.10563,1706.07366,1708.04977,1909.00033,1912.00030}, 
but for practical calculations the vertices based on $SL(2,R)$ 
transformations
seem to be more useful.\footnote{See \cite{1704.01210} for the use of 
$SL(2,C)$ vertex for the 
computation of one point function on the torus.}
Since string field theories associated with different choices of
interaction vertices are related to each other by field redefinition\cite{9301097},
all physical results will
be independent of this choice.

\subsection{C-O amplitude on the disk} 

\begin{figure}
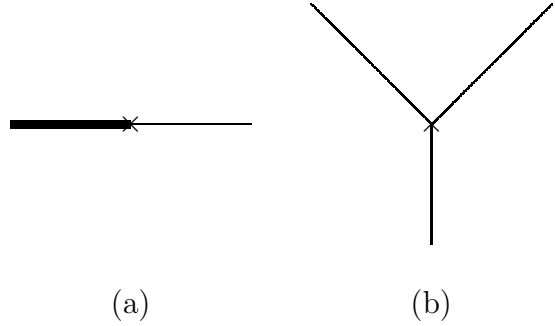

\begin{center}

\figone

\vskip -.5in

\caption{This figure shows two elementary interaction vertices with zero dimensional moduli space. The thick
lines denote closed strings and the thin lines denote open strings. A $\times$
represents disk amplitude.
Fig.~(a) represent open-closed disk amplitude while Fig.~(b) represents open-open-open disk amplitude.
\label{figone}
}

\end{center}
\end{figure}

The C-O amplitude on the disk has zero dimensional moduli space, and therefore the entire contribution to the
amplitude comes from the elementary C-O interaction vertex shown in 
Fig.~\ref{figone}(a). This can be described as
a unit disk parametrized by the complex coordinate $\xi $,
with a closed string puncture at the center $\xi =0$ and an open string puncture at the boundary (which we can take
to be at $\xi =1$). 
We choose the local coordinates
at the open string puncture to be
\be\label{e1}
w = i\, \lambda\, {1-\xi \over 1+\xi } \, ,
\ee
where $\lambda$ is a large positive number.
We can map the unit disk to the upper half
plane labelled by $z$ via
\be\label{e2}
z = i\, {1-\xi \over 1+\xi } \, ,
\ee
so that \refb{e1} takes the form
\be \label{e1alt}
w = \lambda\, z\, .
\ee
In the $z$-plane the closed string puncture is located at $i$.

On the unit disk, instead of choosing the open string puncture at $\xi=1$, we could choose it at another point 
$\xi=\xi_0$ on the unit circle. 
Rotational invariance suggests the associated choice of local coordinate to be
\be
w = i\, \lambda\, {\xi_0-\xi \over \xi_0+\xi } \, .
\ee
In the $z$ coordinate system this translates to
\be
w = \lambda \, {z -z_0\over 1+z\, z_0}, \qquad z_0\equiv i\, {1-\xi_0\over 1+\xi_0}\, .
\ee

\subsection{O-O-O amplitude on the disk}

The O-O-O amplitude on the disk also has zero dimensional moduli space, and therefore the entire contribution to the
amplitude comes from the elementary O-O-O interaction vertex shown in 
Fig.~\ref{figone}(b). In the coordinate $z$, labelling
the upper half plane, the punctures are taken to be located at 0,1 and $\infty$ with local coordinates:
\be\label{eooo}
w_1 = \alpha {2 z\over 2-z}, \quad w_2 = -2\, \alpha\, {1-z\over 1+z} , \quad w_3 = \alpha\, {2\over 1-2z}\, ,
\ee
where $\alpha$ is a large positive number. These choices of local coordinates get cyclically permuted under the transformation
$z\to 1/(1-z)$ that cyclically permutes 0, 1 and $\infty$.

\subsection{C-O-O amplitude on the disk} \label{scoo}

\begin{figure}
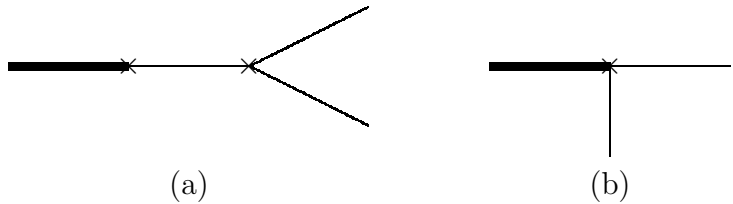

\begin{center}

\hbox{~\hskip 1in \figtwo}

\vskip -1in

\caption{This figure shows two Feynman diagrams contributing to the disk amplitude with one external
closed string and two external open strings.
\label{figtwo}
}
\end{center}
\end{figure}

The moduli space of the C-O-O amplitude on the disk 
is one dimensional, and can be represented by an upper half plane with the
closed string puncture at $i$ and open string punctures at $\pm\beta$ with $0\le\beta\le\infty$. This amplitude receives
contribution from two sources, the Feynman diagram obtained
by joining the C-O interaction vertex to the O-O-O interaction vertex by an 
open string propagator, as shown in Fig.~\ref{figtwo}(a), and
the elementary C-O-O interaction vertex shown in  Fig.~\ref{figtwo}(b).
Let us first consider the contribution from   Fig.~\ref{figtwo}(a). 
For this we denote by $z$ the 
coordinate on the upper half plane associated with the C-O interaction vertex, and by $\tilde z$ the 
coordinate on the upper half plane associated with the O-O-O interaction vertex. 
The Feynman rules of string field theory translate to the statement that the 
local coordinate $w$ around the open string puncture of the C-O interaction vertex, given in 
\refb{e1alt}, should be identified with the local coordinate around one of the open string
punctures of the O-O-O interaction vertex (say $w_1$ given in \refb{eooo} with $z$ replaced
by $\tilde z$) via the identification:
\be
w\, w_1 = -q,  \qquad 0\le q\le 1\, .
\ee
This gives:
\be\label{e5b}
\wt\lambda \, z \, {2 \tilde z\over 2 -\tilde z} = -q\, ,  \qquad \wt\lambda \equiv \lambda\, \alpha\, , \qquad 0\le q\le 1\, .
\ee
This gives an upper half plane labelled by $z$, with a closed string puncture located at $z=i$ and a pair
of open string punctures located at $\tilde z=1,\infty$. In the $z$ plane, these translate to:
\be \label{e4.9a}
z_1=-{q\over 2\wt\lambda},   \qquad z_2 = {q\over 2\wt\lambda}\, .
\ee 
Therefore this Feynman diagram covers part of
the moduli space corresponding to $\beta\le (2\wt\lambda)^{-1}$. Furthermore, the local
coordinates at the open string punctures are given by $w_2$ and $w_3$ in \refb{eooo}, 
which
we shall rename as $w_1$ and $w_2$ respectively:
\be \label{e5a}
w_1 = -2\, \alpha\,  {1-\tilde z\over 1+\tilde z}=-2\, \alpha\,  {z-z_1\over z+3\, z_1},  
\qquad w_2 =\alpha\, {2\over 1-2\tilde z}= 2\, \alpha \, {z-z_2\over z + 3 z_2}\, .
\ee
These can be written in a compact form as
\be  \label{e5c}
w_a = {4\alpha\wt\lambda\over 3\, q} \, {z-z_a\over 1 + (3 z_a)^{-1} z}, \qquad a=1,2\, .
\ee

The C-O-O interaction vertex shown in Fig.~\ref{figtwo}(b)
must be defined so that it covers the moduli space corresponding to 
$\beta>(2\wt\lambda)^{-1}$, and
at $\beta=(2\wt\lambda)^{-1}$ the choice of local coordinates at the open string punctures match the results given in
\refb{e5c} at $q=1$. We choose the local coordinates at the punctures to be
\be\label{e5}
w_a = {\alpha\wt\lambda} \, {4\wt\lambda^2+1\over 4\wt\lambda^2} \,
{z-z_a\over (1+z_a\, z) + \wt\lambda\, f(z_a) (z-z_a)}, \qquad a=1,2, \qquad z_1=-\beta, \quad z_2=\beta\, ,
\ee
where
$f(z_a)$ interpolates between
\be \label{e4.13new}
f\left(\pm {1\over2\wt\lambda}\right)=\pm  {4\wt\lambda^2-3\over 8\wt\lambda^2}
 \quad \hbox{to}
\quad f(\pm 1)=0\, .
\ee
This ensures that at $\beta=(2\wt\lambda)^{-1}$ the local coordinate of the C-O-O interaction vertex matches the local
coordinate of the Feynman diagram Fig.~\ref{figtwo}(a) 
given in \refb{e5c}.
The significance of the second condition in \refb{e4.13new} will be discussed later.
We shall further simplify the choice by taking 
\be
f(-\beta)=-f(\beta)\, .
\ee

We restrict $\beta$ to lie in the range $\beta\le 1$ since a $z\to -1/z$ transformation leaves the closed string
puncture at $i\infty$ fixed and exchanges the range $0 <\beta< 1$ with $1< \beta<\infty$. This also exchanges
the positions of the two open string punctures and their local coordinates. 
Therefore once we restrict $\beta$ to the range $\beta<1$, we also need to
explicitly add contributions that are related by the exchange of two 
open string punctures. Alternatively, we could extend the range of $\beta$ all the way
to $2\wt\lambda$, defining $f(\beta)= f(-1/\beta)=-f(1/\beta)$ for $\beta\ge 1$.
In this case we do not need to separately add contributions where we exchange the
two open string punctures. The condition $f(\pm 1)=0$ in \refb{e4.13new} has 
been chosen to
ensure compatibility with the $f(\beta)=-f(1/\beta)$ condition at $\beta=\pm 1$.

Note that we could have multiplied the choice of $w_a$ given in \refb{e5} by an arbitrary function $g(\beta)$ of
$\beta$ that takes value 1 at $\beta=(2\wt\lambda)^{-1}$ and satisfies the condition
$g(\beta)=g(-\beta)=g(1/\beta)$,
but is otherwise arbitrary. This will make the later analysis a bit more complicated. For this reason,
we shall proceed with the choice $g(\beta)=1$. We shall check in appendix \ref{se} that
inclusion of such multiplicative function does not change the final results.

\subsection{C-C amplitude on the disk}

\begin{figure}
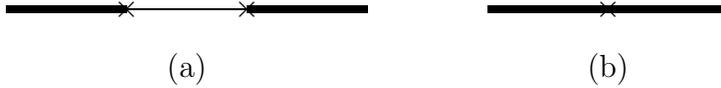

\begin{center}

\hbox{~\hskip 1in \figthree}

\vskip -1in

\caption{This figure shows two Feynman diagrams contributing to the disk amplitude with two external
closed strings.
\label{figthree}
}
\end{center}
\end{figure}

The moduli space of the C-C amplitude on the disk 
is one dimensional, and can be represented by an upper half plane with the first
closed string puncture at $i$ and the second closed  string puncture at $i\, y$ with $0 < y <1$. 
This amplitude receives
contribution from two sources, the Feynman diagram obtained
by joining the C-O interaction vertex to the C-O interaction vertex by an open string propagator, as shown in Fig.~\ref{figthree}(a), and
the elementary C-C interaction vertex shown in  Fig.~\ref{figthree}(b).
Let us first consider the contribution from   Fig.~\ref{figthree}(a). If we denote by
$z$ and $\hat z$ the coordinates on the upper half planes associated with 
the two
C-O vertices, then as a consequence of the
local coordinate choice \refb{e1alt}, the two disks are glued via the relation
\be 
\lambda z\, \lambda \hat z = -q\, , \qquad 0\le q\le 1\, .
\ee
This maps $\hat z=i$ to $z=i y$ with \be\label{eyqrel}
y=q/\lambda^2\, .
\ee 
The condition $q\le 1$ corresponds to the following
region in the $y$-space:
\be \label{edefA}
{\rm A}: \qquad y\le \lambda^{-2}\, .
\ee
Therefore the C-C interaction vertex shown in Fig.~\ref{figthree}(b) must cover the region 
\be\label{edefB}
{\rm B}: \qquad 1\ge y\ge \lambda^{-2}\, .
\ee
Note that the $y\ge 1$ region can be 
brought back to the $y\le 1$ region by $z\to -1/z$ transformation.

The region B actually contains a boundary of the moduli space associated with
closed string degeneration, since as $y\to 1$ the two closed string punctures approach
each other. In the full open-closed string field theory, we need to represent the contribution 
from this region as coming from a new Feynman diagram in which a C-C-C vertex is
connected to a C vertex (one point function of the closed string on the disk) 
by an internal closed string propagator. This will be necessary
for evaluating the full contribution to $f(\omega_1,\omega_2)$ given in 
\refb{effull} that
has a singular contribution from this region\cite{2003.12076}. 
However, since our goal is to evaluate
$f_{\rm div}$ given in \refb{efdivexp} which does not have any divergence from the
$y\to 1$ region, we do not need this special treatment of the $y\simeq 1$ region.

\subsection{C-C-O amplitude on the disk} \label{scconew}

\begin{figure}
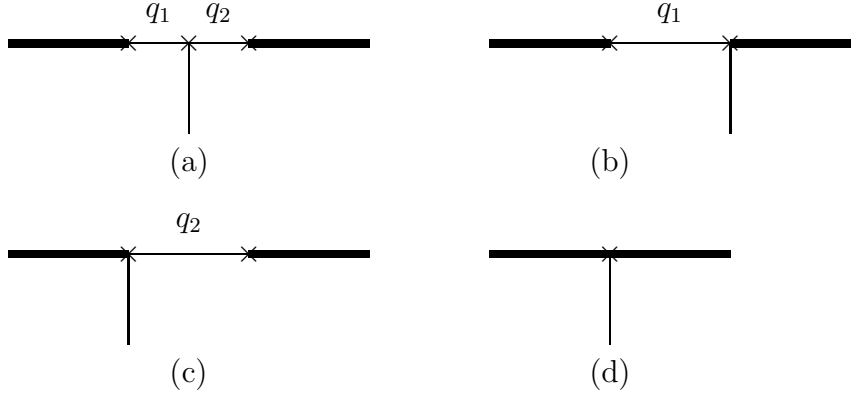

\begin{center}

\hbox{~\hskip 1in \figfour}

\vskip -.8in

\hbox{~\hskip 1in \figfive}

\vskip -1in

\caption{This figure shows four Feynman diagrams contributing to the disk amplitude with two external
closed strings and one external open string. $q_i$'s denote the sewing parameters of the
corresponding open string propagators. They have been chosen such that when a propagator
in Fig.(a) collapses, the parametrization agrees with the parametrization given in Fig.~(b)
or (c). With this
parametrization, the $q_1=\delta_1$ boundary segments of Fig.~(a) and Fig.~(b)
will describe a continuous curve, and the $q_2=\delta_2$ boundary 
segments of Fig.~(a) and Fig.~(c)
will describe a continuous curve.
This will play no role in the analysis presented in this paper, 
but is useful when we follow the
alternative procedure described in \S\ref{salternative} and \cite{1902.00263}. 
\label{figfour}
}
\end{center}
\end{figure}

The moduli space of the C-C-O amplitude on the disk 
is two dimensional, and can be represented by an upper half plane with the first
closed string puncture at $i$, the second closed  string puncture at $i y$ with $0 < y <1$, and the open string
puncture at a point $x$ on the real axis. 
This amplitude receives
contribution from the four Feynman diagrams shown in Fig.~\ref{figfour}.

Let us begin by analyzing the Feynman diagram  shown in Fig.~\ref{figfour}(a) comprising
the C-O, O-O-O, C-O vertices connected by a pair of open string propagators. 
We denote by
$z$ the coordinate on the upper half plane representing the C-O interaction vertex on the left, 
by $\tilde z$ the 
coordinate of the upper half plane representing the O-O-O interaction vertex, and by $\hat z$
 the coordinate on the upper half plane representing the C-O interaction vertex on the right.
Using \refb{e1alt} and \refb{eooo} we get the sewing relations to be,
\be
\wt\lambda \, z\, {2 \tilde z\over 2-\tilde z} = -q_1\, , \qquad \wt\lambda \, \hat z\, {2\over 1-2\tilde z}
= -q_2\, ,  \qquad \wt\lambda\equiv \lambda \, \alpha\, , \qquad 0\le q_1,q_2\le 1\, .
\ee
In the $z$ plane, the location of the first closed string puncture at $z=i$, the 
second closed string puncture at $\hat z=i$ and the open string puncture
at $\tilde z=1$ are mapped to:
\be
z_c^{(1)}=i, \qquad z_c^{(2)} =-u_1 \, {3\, u_2 - 2\, i\over 2(u_2 + 2\, i)} , \qquad z_o=-{u_1\over 2}\, , \qquad
u_i = q_i/\wt\lambda \quad \hbox{for $i=1,2$}\, .
\ee
We shall now make a coordinate transformation
\be
z = {z'+ a\over -a \, z'+1} \qquad \Leftrightarrow \qquad z' =  {z - a\over a\, z+1}\, ,
\ee
and adjust the real constant $a$
so that in the $z'$ plane the closed string punctures are located at $i$ and $iy$ for 
some real number $y$.
This gives
\be
iy = {z_c^{(2)} -a\over a\, z_c^{(2)}  + 1}\, .
\ee
The vanishing of the real part of the right hand side fixes $a$, which can then be substituted into the same
equation to determine $y$. We shall carry out the analysis for small $u_1$, $u_2$. This gives:
\be
a\simeq {1\over 2} u_1 \left\{1- u_2^2+ 
\OO(u_1^4, u_2^4, u_1^2 u_2^2)\right\}\, ,
\ee
and
\be \label{ech1p}
y = u_1 u_2 \left\{1 -{u_1^2\over 4} - {u_2^2\over 4} + 
\OO(u_1^4, u_2^4, u_1^2 u_2^2)\right\} \, .
\ee
Also in the $z'$  plane the open string puncture is located at $x$ with
\be \label{ech2p}
x = -u_1 \left\{1 +{u_1^2\over 4} - {u_2^2\over 2} + \OO(u_1^4, u_2^4, u_1^2 u_2^2)
\right\} \, .
\ee
The region of the moduli space covered by this diagram corresponds to 
\be \label{erangea}
0\le u_1\le \wt\lambda^{-1}, \qquad 0\le u_2\le \wt\lambda^{-1}\, .
\ee
The range \refb{erangea} translates to a region in the $x$-$y$ plane bounded by
the following curves:
\be\label{erangeb}
{\rm I}: \qquad x=0, \quad {y\over x}=0, \quad x=-\wt\lambda^{-1} \, \left\{ 1 +
{1\over 4}\, \wt\lambda^{-2} -{1\over 2} \, {y^2\over x^2}\right\}, \quad 
{y\over x}= -\wt\lambda^{-1} \, \left\{1 -{x^2\over 2} + {1\over 4}\,
\wt\lambda^{-2}\right\}
\, ,
\ee
where we have neglected fractional errors of order higher than 
$\wt\lambda^{-2}$. This can be
justified as follows. We shall see in \S\ref{scco2} that the maximally divergent part of the
integrand goes as $dx\, dy\, y^{-2} \, (x-i\, y)^{-1}$. Using the fact that $y\sim 
u_1 u_2\sim \wt\lambda^{-2}$ and $x\sim u_1\sim\wt\lambda^{-1}$, we see that the
maximally divergent part of the integral goes as $\wt\lambda^{2}$. Therefore, it is
sufficient to keep fractional corrections to order $\wt\lambda^{-2}$ to extract all the
finite contributions.

There is another contribution to the diagram containing C-O, O-O-O and C-O vertices, where we reverse the
cyclic order of the O-O-O interaction vertex. This is equivalent to exchanging the two
punctures of the O-O-O interaction vertex to which the left and the right C-O interaction
vertices are sewed.  The new sewing relations take the form:
\be
\wt\lambda \, z\,  {2\over 1-2\tilde z} = -q_1\, , \qquad \wt\lambda \, \hat z\,{2 \tilde z\over 2-\tilde z}
= -q_2\, ,  \qquad 0\le q_1,q_2\le 1\, .
\ee
In the $z$ plane, the locations of the first closed string puncture at $z=i$, the 
second closed string puncture at $\hat z=i$ and the open string puncture
at $\tilde z=1$ are given by
\be
z_c^{(1)}=i, \qquad z_c^{(2)} =u_1 \, {3\, u_2 + 2\, i\over 2(u_2 - 2\, i)} , \qquad z_o={u_1\over 2}\, , \qquad
u_i = q_i/\wt\lambda \quad \hbox{for $i=1,2$}\, .
\ee
We now make a change of coordinate
$z' =  {(z - a) /  (a\, z+1)}$
so that in the $z'$ plane the closed string punctures are located at $i$ and $i\, y$ for some real number $y$.
This gives
\be
a\simeq -{1\over 2} u_1 \left\{1- u_2^2 + \OO(u_1^4, u_2^4, u_1^2 u_2^2)\right\}\, ,
\ee
and
\be \label{ech3p}
y \simeq u_1 u_2 \left\{1 - {u_1^2\over 4} - {u_2^2\over 4} 
+ \OO(u_1^4, u_2^4, u_1^2 u_2^2)\right\} \, .
\ee
Also in the $z'$  plane the open string puncture is located at $x$ with
\be \label{ech4p}
x = u_1 \left\{1 +{u_1^2\over 4} - {u_2^2\over 2}+ \OO(u_1^4, u_2^4, u_1^2 u_2^2)
\right\}\, .
\ee
The region of the moduli space covered by this diagrams corresponds to 
\be \label{erangeag}
0\le u_1\le \wt\lambda^{-1}, \qquad 0\le u_2\le \wt\lambda^{-1}\, .
\ee
This translates to the region in the $x$-$y$ plane bounded by the curves
\be \label{e4.35}
{\rm I}': \qquad x=0, \quad {y\over x}=0, \quad x=\wt\lambda^{-1} \, \left\{ 1 +
{1\over 4}\, \wt\lambda^{-2} -{1\over 2} \, {y^2\over x^2}\right\}, \quad 
{y\over x}= \wt\lambda^{-1} \, \left\{1 -{x^2\over 2} + {1\over 4}\,
\wt\lambda^{-2}\right\}
\, .
\ee

Next we consider the  diagram shown in Fig.~\ref{figfour}(b). If we denote by $z$ the coordinate on the
upper half plane associated with the C-O interaction vertex, and by $\hat z$ 
the upper half plane coordinate associated with the C-O-O interaction vertex, then, using \refb{e5} and \refb{e1alt}, the
sewing relation takes the form:
\be \label{ef1}
 {\alpha\wt\lambda} \, {4\wt\lambda^2+1\over 4\wt\lambda^2} \,
{\hat z+\beta\over (1-\beta\, \hat z) + \wt\lambda\, 
f(-\beta) (\hat z+\beta)} \, \lambda \,  z = -q_1\, .
\ee
We also need to separately consider
\be \label{ef2}
 {\alpha\wt\lambda} \, {4\wt\lambda^2+1\over 4\wt\lambda^2} \,
{\hat z-\beta\over (1+\beta\, \hat z) + \wt\lambda\,
f(\beta) (\hat z-\beta)} \, \lambda \,  z = -q_1\, .
\ee

First consider the case of sewing via \refb{ef1}.
In this case the second closed string puncture, located at $\hat z=i$, is mapped in the $z$-plane to
\be
z_c = u\, \{ -f(-\beta)+i \, \wt\lambda^{-1}\}\, , \qquad u\equiv {4\, \wt\lambda\, q_1\over 4\wt\lambda^2+1}, \qquad 0\le u\le  {4\, \wt\lambda\over 4\wt\lambda^2+1}\, ,
\ee
where we have used $\wt\lambda=\alpha\, \lambda$.
We need to bring this on the imaginary axis via the change of coordinate
$z' = {(z-a)/(az + 1)}$
that maps the location of the first closed string puncture at $z=i$ to $z'=i$ and the
second closed string puncture at $z=z_c$ to $z'_c=(z_c-a)/(1+ a\, z_c)$.
$a$ is determined as before
by demanding that the real part of $z'_c$ vanishes. 
This gives, using $f(-\beta)=-f(\beta)$: 
\be 
a= {u\,  f(\beta)}\,
\{1 + u^2 \wt\lambda^{-2} + \OO(u^4 \wt\lambda^{-4})\} \, ,
\ee
and,
\be
z'_c = i\, u\, \wt\lambda^{-1} \{1 - u^2 f(\beta)^2 + \OO(u^4,u^2\wt\lambda^{-2})\}
\, .
\ee

We also need to find the location of the external open string puncture in the $z'$ coordinate.  In the $\hat z$
coordinate it is located at $\hat z = \beta$. Using \refb{ef1}, we get
\be 
z_o = {u}\,  \left\{ f(\beta)-{1-\beta^2\over 2\, \beta\, \wt\lambda}\right\} \, . 
\ee
In the $z'$ coordinate this is located at
\be \label{eoriginal1}
z'_o =-{u\over \wt\lambda} \, 
{1-\beta^2 \over 2\, \beta} 
\left\{1 - u^2 f(\beta)^2 + u^2 \,\wt\lambda^{-1}
\, f(\beta)\, {(1+\beta^2)^2\over 2\beta (1-\beta^2)
}+ 
\OO(u^2\wt\lambda^{-1})\right\}
\, .
\ee
Note that due to the presence of the
$\wt\lambda^{-1}$ factor in the third term inside the curly bracket, 
this term can contribute only for 
$\beta\sim 1/\wt\lambda$. Therefore we can replace the $\beta$ dependent factor 
multiplying this by
any function of $\beta$ that approaches $1/(2\beta)$ for small $\beta$.
We shall choose this to be $(1-\beta^2)/ (2\beta)$ for convenience, but we have 
confirmed using Mathematica that the final result remains unchanged even if we use
the original form given in \refb{eoriginal1}.
Labelling $z_c'$ by $iy$ and $z'_o$ by $x$, and ignoring fractional corrections
of order higher than $\wt\lambda^{-2}$, we can write the  relations as,
\be \label{em2}
y= {u\over \wt\lambda}\,   \{1 - u^2\, f(\beta)^2\}, 
\quad x= -{u\over \wt\lambda}\,   {1-\beta^2 \over 2\, \beta} \,
\left\{1 - u^2 f(\beta)^2 + u^2 \, f(\beta)\, {1-\beta^2\over 2\beta\wt\lambda}
\right\}\, .
\ee
We now see from \refb{em2} that the range
\be\label{em4a}
0\le u\le  {4\, \wt\lambda\over 4\wt\lambda^2+1}, 
\quad 1\ge \beta\ge {1\over 2\wt\lambda}\, ,
\ee
corresponds to the region bounded by the curves:
\be\label{em5}
{\rm II}' : y=0, \quad y=\wt\lambda^{-2}\,
\left\{1-F(x)^2 \wt\lambda^{-2} - {1\over 4} 
\wt\lambda^{-2}
\right\}, \quad {x\over y}=0, \quad {x\over y} =  
-\wt\lambda \, 
\left\{1- {1\over 4}  \wt\lambda^{-2}+{1\over 2} x^2\right\}\, ,
\ee
where $F(x)$ is a function of $x$ defined as the result of elimination of $\beta$
from the equations:
\ben\label{edefFx1}
&& x = -{1\over \wt\lambda^2}\,   {1-\beta^2 \over 2\, \beta} \,
\left\{1 - {1\over 4} \wt\lambda^{-2} - \wt\lambda^{-2} f(\beta)^2 + \wt\lambda^{-2} 
\, f(\beta)\, {1-\beta^2\over 2\beta\wt\lambda}
\right\},  \qquad F(x) = f(\beta), \nonumber \\
&&\hskip 1in  {1\over 2\wt\lambda} \le\beta\le 1 \quad \Leftrightarrow 
\quad -\wt\lambda^{-1} \left(1 - {1\over 4\wt\lambda^2}\right)
 \le x \le 0\, .
\een

The case of sewing via \refb{ef2} can be analyzed similarly, and yields the result:
\be\label{em4}
y= {u\over \wt\lambda}\,   \{1 - u^2\, f(\beta)^2\}, 
\quad x= {u\over \wt\lambda}\,   {1-\beta^2 \over 2\, \beta} \,
\left\{1 - u^2 f(\beta)^2 + u^2 \, f(\beta)\, {1-\beta^2\over 2\beta\wt\lambda}
\right\}\, ,
\ee
The range of $x$ and $y$ covered by this diagram is bounded by:
\ben\label{em5pre}
{\rm II} &:&y=0, \quad y=\wt\lambda^{-2}\,
\left\{1-F(x)^2 \wt\lambda^{-2} - {1\over 4} 
\wt\lambda^{-2}
\right\}, \quad {x\over y}=0, \quad {x\over y} =  
\wt\lambda \, 
\left\{1- {1\over 4}  \wt\lambda^{-2}+{1\over 2} x^2\right\}\, ,
\nonumber \\
\een
where, $F(x)$ is now defined via:
\ben\label{edefFx2}
&& x = {1\over \wt\lambda^2}\,   {1-\beta^2 \over 2\, \beta} \,
\left\{1 - {1\over 4} \wt\lambda^{-2} - \wt\lambda^{-2} f(\beta)^2 + \wt\lambda^{-2} 
\, f(\beta)\, {1-\beta^2\over 2\beta\wt\lambda}
\right\},  \qquad F(x) = f(\beta), \nonumber \\
&&\hskip 1in  {1\over 2\wt\lambda} \le\beta\le 1 \quad \Leftrightarrow 
\quad 0\le x \le \wt\lambda^{-1} \left(1 - {1\over 4\wt\lambda^2}\right)
 \, .
\een

Next we consider the diagram shown in Fig.~\ref{figfour}(c). 
This is related to the contribution from Fig.~\ref{figfour}(b)
by the exchange of the two closed string punctures. In the $z'$ plane this can be 
achieved by the transformation $z'\to -y/z'$, taking the point $i$ to $iy$ and the point $iy$ to $i$. On the
open string punctures located at $z'=x$, this has the effect of sending $x$ to $-y/x$. Therefore the
equations analogous to \refb{em2} and \refb{em4} take the form:
\be\label{em7}
y= {u\over \wt\lambda}\,   \{1 - u^2\, f(\beta)^2\}, 
\qquad x=  {2\, \beta\over 1-\beta^2 } \,
\left\{1 - u^2 \, f(\beta)\, {1-\beta^2\over 2\beta\wt\lambda}
\right\}\, ,
\ee
and 
\be \label{em6}
y= {u\over \wt\lambda}\,   \{1 - u^2\, f(\beta)^2\}, 
\qquad x=  -{2\, \beta\over 1-\beta^2 } \,
\left\{1 - u^2 \, f(\beta)\, {1-\beta^2\over 2\beta\wt\lambda}
\right\}\, .
\ee
The range of $u$ and $\beta$ given in \refb{em4a} 
correspond to the regions in the $x$-$y$ plane bounded
by the curves:
\be\label{em18}
{\rm III}' \ :\quad  y=0, \quad y=\wt\lambda^{-2}\,
\left\{1-F(x)^2 \wt\lambda^{-2} - {1\over 4\wt\lambda^{2}
} 
\right\}, \quad x=\infty, \quad x =  
\wt\lambda^{-1} \, 
\left\{1+ {1\over 4\wt\lambda^{2}}  -{1\over 2} {y^2\over x^2}\right\}\, ,
\ee
and
\be \label{em17}
{\rm III} \ :\quad y=0, \quad y=\wt\lambda^{-2}\,
\left\{1-F(x)^2 \wt\lambda^{-2} - {1\over 4\wt\lambda^{2}
} 
\right\}, \quad {x}=-\infty, \quad {x} =  -
\wt\lambda^{-1} \, 
\left\{1+ {1\over 4 \wt\lambda^{2}} -{1\over 2} {y^2\over x^2}\right\}\, ,
\ee
where $ F(x)$ is defined via the equation:
\ben \label{edefFx3}
&& x=  \pm {2\, \beta\over 1-\beta^2 } \,
\left\{1 - \wt\lambda^{-2} \, f(\beta)\, {1-\beta^2\over 2\beta\wt\lambda}
\right\}, \qquad F(x) =f(\beta)\, , \nonumber \\ &&
\hskip 1in  {1\over 2\wt\lambda} \le\beta\le 1 \quad \Leftrightarrow 
\quad  \wt\lambda^{-1} \left(1 - {1\over 4\wt\lambda^2}\right) \le |x|<\infty
 \, .
\een
Eqs.\refb{edefFx1}, \refb{edefFx2}, \refb{edefFx3} define the function $F(x)$ over the
entire real $x$-axis.

\begin{figure}
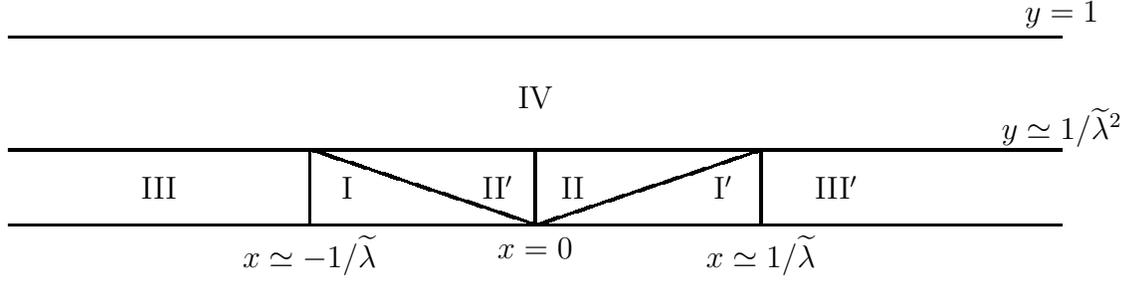

\begin{center}

\figeight

\vskip -.5in

\caption{This figure shows the covering of the moduli space of C-C-O amplitude on the disk by various 
Feynman diagrams. The horizontal axis labels $x$ -- the position of the open string puncture on the real
axis and the vertical axis labels $y$, the position of the second closed string puncture on the imaginary axis.
The first closed string puncture is situated at $i$. We use only the leading order results
for large $\wt\lambda$ to plot the boundaries of various regions; the full expressions may
be found in the text.
\label{figeight}
}

\end{center}
\end{figure}

A property of the function $F(x)$ that will be useful later is that to leading order in the
expansion in powers of $\wt\lambda^{-1}$, $F(x)$ satisfies the equation:
\be\label{eFid}
F(\wt\lambda^{-2}/x) = F(x).
\ee
Since $F(x)=f(\beta)$ for all values of $x$, to prove 
\refb{eFid} we need to verify that the leading order
relation between $x$ and $\beta$ remains unchanged under $x\to \wt\lambda^{-2}/x$.
Now, ignoring fractional corrections of order $\wt\lambda^{-2}$, we see that
for $|x|>\wt\lambda^{-1}$, the
relation between $x$ and $\beta$ given in \refb{edefFx3} takes the from
$|x| = 2\beta/(1-\beta^2)$, whereas for $|x|\le\wt\lambda^{-1}$, the 
relation between $x$ and $\beta$ given in \refb{edefFx1}, \refb{edefFx2} 
takes the from
$|x|= \wt\lambda^{-2} (1-\beta^2) / (2\beta)$. Since these are related by $x\to 
\wt\lambda^{-2}/x$ transformation, this establishes \refb{eFid}.

The rest of the moduli space must be covered by the C-C-O interaction vertex shown in Fig.~\ref{figfour}(d). 
This corresponds to the region:
\be \label{eivrange}
{\rm IV}: \qquad 1\ge y\ge\wt\lambda^{-2} \left\{1-F(x)^2 
\wt\lambda^{-2} 
- {1\over 4} 
\wt\lambda^{-2}
\right\}, 
\quad -\infty < x < \infty
\, .
\ee
Fig.~\ref{figeight} shows the regions in the $x$-$y$ plane given in eqs.\refb{erangeb},
\refb{e4.35}, \refb{em5}, \refb{em5pre}, \refb{em18}, \refb{em17} and \refb{eivrange}.

\subsection{O amplitude on the annulus}

Next we consider the one point amplitude of an open string 
on the annulus. 
We shall label the annulus by the complex coordinate $w$ with the
identification
\be\label{ecd1}
0\le {\rm Re}~w\le \pi, \qquad w \equiv w - i\, \ln v\, ,
\ee
where $v$ is a positive real number taking value in the range $0<v<1$.
The location of the open string puncture can be fixed at any arbitrary point on the
boundary using the shift symmetry ${\rm Im}(w)\to {\rm Im}(w)+\hbox{constant}$. 
Therefore the moduli space is one dimensional, parametrized by $v$.

\begin{figure}
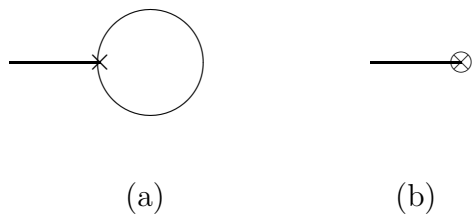

\begin{center}

\hbox{\figseven}

\vskip -1.5in

\caption{This figure shows the Feynman diagrams contributing to the annulus amplitude with one external
open string. The interaction vertex with a $\times$ represents 
a disk amplitude, while a interaction vertex with $\otimes$
represents part of an annulus amplitude.
\label{figsix}
}
\end{center}
\end{figure}

The amplitude receives contribution from two Feynman diagrams, shown in Fig.~\ref{figsix}. Let us first
analyze the
contribution from Fig.~\ref{figsix}(a). 
We denote by $ z$ the coordinate associated with the O-O-O interaction vertex. Then, using the local coordinates of
the O-O-O interaction vertex given in \refb{eooo}, we get the following identification in the $z$ plane:
\be
-2\, \alpha\,
{1- z\over 1+ z} \equiv -q\, {1 - 2 z\over 2\, \alpha},
\qquad 0\le q\le 1\, .
\ee
This can be expressed as
\be\label{et1}
 z \equiv {a z + b\over c z + d}\, ,
\ee
where
\be
\pmatrix{a & b\cr c & d} = \pmatrix{{1\over 2} u^{1/2} & u^{-1/2} - {1\over 4} u^{1/2}\cr
-{1\over 2} u^{1/2} & u^{-1/2} +{1\over 4} u^{1/2}}\, , 
\qquad u\equiv q/\alpha^2, \qquad 0\le u\le \alpha^{-2}\, .
\ee
We can `diagonalize' this matrix by defining new coordinate $\hat z$ via
\be\label{e4.61a}
z = {(4+3\, u +\OO(u^2))  \hat z -4 +{3} u\, +\OO(u^2)\over (4+u+\OO(u^2))   \hat z
- 2\, u+\OO(u^3)}, \qquad w = {1\over i} \ln \hat z, \qquad
0\le {\rm Re}~w\le \pi\, .
\ee
Then the identification \refb{et1} takes the form
\be
\hat z = u^{-1}\left\{1-{1\over 2}u +\OO(u^2)\right\}  \hat z, \qquad w \equiv w - i \ln \left[u
\left\{1-{1\over 2}u +\OO(u^2)\right\}^{-1}\right] \, .
\ee
Comparison with \refb{ecd1} gives 
\be
v = u\left\{1-{1\over 2}u +\OO(u^2)\right\}^{-1}\, .
\ee
This agrees with the result of \cite{1704.01210}.
Since according to the result  of \S\ref{serror}, fractional errors of order $u^2$ can be
absorbed into a redefinition of $u$ without changing the final result, we shall use the
relation
\be
v = u \left\{1-{1\over 2}u\right\}^{-1}\, .
\ee 
Therefore the range of $v$ spanned by Fig.\ref{figsix}(a) corresponds to
\be 
0\le v \le \left\{\alpha^2-{1\over 2}\right\}^{-1}\, .
\ee

In this coordinate system the open string puncture located at $z=0$ is mapped to,
\be\label{estick}
\hat z = 1 -{3\over 2} u +\OO(u^2)\simeq 1, \qquad w = i\left\{{3\over 2}\, u +\OO(u^2)
\right\} \, .
\ee
The local coordinate at the open string puncture is given
by
\be \label{elocal}
w_o = \alpha  \, {2\, z\over 2-z} =2\, \alpha  {(4+3\ u) \hat z -4 + 3\, u+\OO(u^2)\over
(4-u)\hat z + 4 - 7\, u+\OO(u^2)}\, .
\ee
Since translational invariance along the imaginary $w$ axis allows us to change
this to any arbitrary imaginary
value, the location of the open string puncture is not a modulus. 
Nevertheless, for definiteness, we shall stick to \refb{estick}, \refb{elocal}.

The contribution from Fig.~\ref{figsix}(b), corresponding to the open string one point interaction
vertex on the annulus, will have to span the range
\be\label{erange}
 \left\{\alpha^{2}-{1\over 2}\right\}^{-1} <v <1\, .
\ee
We can choose the local coordinate at the open string puncture to be the same as the one 
given in
\refb{elocal} for $u=\alpha^{-2}$, i.e.\
\be \label{elocal2}
w_o = 2\, \alpha  {(4+3\ \alpha^{-2}) \hat z -4 + 3\, \alpha^{-2} \over
(4-\alpha^{-2}) \hat z + 4 - 7\, \alpha^{-2}}\, .
\ee
We could make more general choice of $w_o$, {\it e.g.} include in the definition of
$w_o$ an arbitrary multiplicative function $h(v)$ that takes value 1 at 
$v=1/(\alpha^2-{1\over 2})$ but is otherwise arbitrary. In our analysis below, this will have the
effect of replacing $u$ by $u/h(v)$ in \refb{e2.81pre}, 
and multiplying the upper limit of $x$ in 
\refb{e4.91} and the lower limit of $x$ in \refb{e4104} by $1/h(v)$. 
This in turn will affect the integrals
$I_{(b)}$ and $I_{(d)}$ in \refb{eca4} and \refb{eca2pre} respectively. 
It is easy to verify that these 
changes cancel each other and do not affect the final result.

\subsection{C amplitude on the annulus} \label{scamp}

Next we consider the one point amplitude of a closed string 
on the annulus. The moduli space for this amplitude is two dimensional.
We shall label the annulus by the coordinate $w$ with the
identification
\be\label{ecc1}
0\le {\rm Re}~w\le \pi, \qquad w \equiv w - i\, \ln v\, ,
\ee
and the closed string puncture will be taken to be located at $w_c$ with
\be\label{ecc2}
{\rm Re}~ w_c = 2\pi x\, .
\ee
The pair $(v,x)$ label the moduli on the Riemann surface, with the full range given by:
\be
0<v< 1, \qquad 0\le x\le {1\over 4}\, .
\ee 
The range $x>{1\over 4}$ is related to the region described above by $w\to \pi -w$ transformation.
The imaginary part of $w_c$ can be shifted using the
translation invariance of the annulus along the periodically identified imaginary $w$ axis, and therefore does not
represent a modulus. 

\begin{figure}
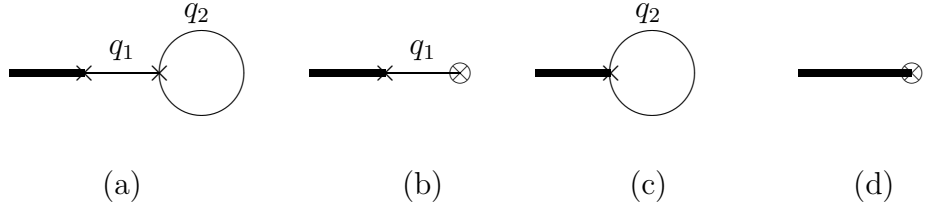

\begin{center}

\hbox{\figsix}

\vskip -1in

\caption{This figure shows the Feynman diagrams contributing to the annulus amplitude with one external
closed string. The interaction vertex with a $\times$ represents part of a disk amplitude, while the interaction vertex with $\otimes$
represents annulus amplitude. As in Fig.~\ref{figfour}, 
the $q_i$'s represent the sewing parameters
of the corresponding propagators.
\label{figfive}
}
\end{center}
\end{figure}

Fig.~\ref{figfive} shows the different Feynman diagrams contributing to this amplitude.
Our goal will be to find out the relation between the sewing parameters $\{q_i\}$
associated with these diagrams and the variables $v,x$ introduced above. Since we
shall derive approximate relations between these quantities instead of exact relations,
we need to make sure that we do not throw away terms that could give finite contribution
to the amplitude. For this we note the following facts:
\begin{enumerate}
\item The relations we are after depend on two string field theory parameters $\wt\lambda$
and $\alpha$ which are both taken to be large, 
and the amplitude, to be evaluated in \S\ref{s2.2}, 
will be given as a Laurent series expansion in
$\alpha^{-2}$ and $\wt\lambda^{-1}$. In this expansion we shall keep all terms that contain
non-negative power of $\alpha^2$ and $\wt\lambda$, but will throw away terms containing
negative powers of $\alpha^{2}$ {\em or} $\wt\lambda$, even if they contain 
positive power of the other variable.  The reason for this is that the $\alpha$ and 
$\wt\lambda$ dependent terms are expected to 
cancel at the end anyway, and once a term has got
negative power of one large variable, there is no way it can give a $\alpha$ and
$\wt\lambda$ independent term at higher order in the expansion.

\item The most divergent term in the integrand
that we shall encounter during our analysis in \S\ref{s2.2},
where these results will be used, 
goes as $dv\, dx\, x^{-2}
v^{-2}$ for small $v,x$. 
\end{enumerate}

\begin{figure}
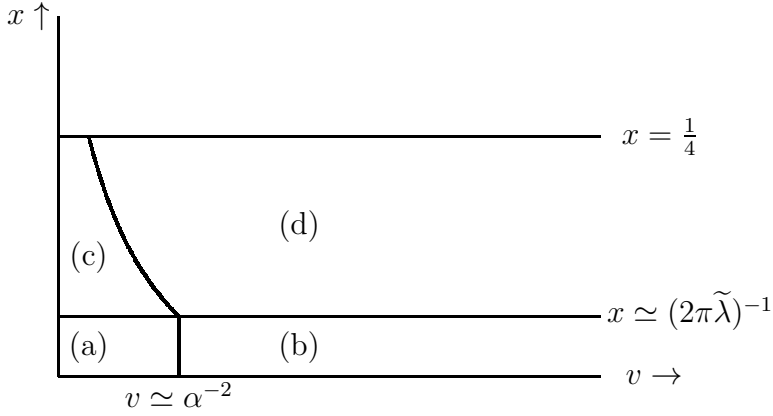

\begin{center}

\hbox{\fignine}

\vskip -.8in

\caption{This figure shows the different regions of the moduli space
covered by different Feynman diagrams of Fig.~\ref{figfive}. The regions are marked by keeping
only the leading order term in the expansion in powers of $\alpha^{-2}$ and $\wt\lambda^{-2}$; the subleading
corrections can be found in the text.
\label{fignine}
}
\end{center}
\end{figure}

We begin by analyzing the contribution from Fig.~\ref{figfive}(a).
We denote by
$z$ the coordinate on the C-O interaction vertex, and by $\tilde z$ the coordinate on the O-O-O 
interaction vertex.
Then, using \refb{e1alt} and \refb{eooo},  the sewing relation takes the form:
\be \label{e4.74}
\wt\lambda \, z \, {2\tilde z\over 2-\tilde z} = -q_1, \quad -2 \alpha
{1-\tilde z\over 1+\tilde z} \equiv -q_2\, {1 - 2\tilde z\over 2\, \alpha},
\quad 0\le q_1, q_2\le 1\, .
\ee
We shall define:
\be 
u_1 \equiv {q_1/ \wt\lambda}, \quad  u_2 = q_2/\alpha^2, \qquad
0\le u_1\le \wt\lambda^{-1}, \quad 0\le u_2\le \alpha^{-2}\, .
\ee
The second equation of \refb{e4.74} now gives
\be\label{et1rep}
\tilde z \equiv {a\tilde z + b\over c\tilde z + d}\, ,
\ee
where,
\be
\pmatrix{a & b\cr c & d} = \pmatrix{{1\over 2} u_2^{1/2} & u_2^{-1/2} - {1\over 4} u_2^{1/2}\cr
-{1\over 2} u_2^{1/2} & u_2^{-1/2} +{1\over 4} u_2^{1/2}}
\, .
\ee
As in \refb{e4.61a}, if we define new coordinate $\hat z$, $w$ via
\be
\tilde z= {(4+3\, u_2)  \hat z -4 +{3} u_2\, +\OO(u_2^2)\over 
(4+u_2)   \hat z- 2\, u_2+\OO(u_2^2)}, \qquad 
w = {1\over i} \ln \hat z, \qquad
0\le {\rm Re}~w\le \pi\, ,
\ee
then the identification \refb{et1rep} takes the form
\be\label{e4.78}
\hat z = u_2^{-1}\left\{1-{1\over 2}u_2 +\OO(u_2^2)\right\}  \hat z, \qquad w \equiv w -
i \ln \left[u_2
\left\{1+{1\over 2}u_2 +\OO(u_2^2)\right\}\right] \, .
\ee
In the $\tilde z$ plane, the closed string puncture at $z=i$ is at
\be\label{et0}
\tilde z_c = {2\, i\, u_1\over 2+i\, u_1}\, .
\ee
In the $w$ coordinate system this maps to
\be\label{e4.80}
w_c = {u_1} \left\{1-  u_2 + \OO(u_2^2)\right\} +\OO(u_1^3) \, ,
\ee
up to addition of imaginary terms which can be removed using translational invariance along
the imaginary $w$ direction.
Comparison of \refb{e4.78}, \refb{e4.80} with \refb{ecc1}, \refb{ecc2} gives 
\be\label{e171pre}
v = u_2\left\{1 - {1\over 2} u_2 +\OO(u_2^2)\right\}^{-1}, 
\qquad x = {1\over 2\pi} {u_1} \left\{1-u_2+ \OO(u_2^2)\right\} +\OO(u_1^3)\, .
\ee
\refb{e171pre} shows that for small $v,x$ the most divergent part $dx\, dv\, 
x^{-2}v^{-2}$
of the integrand
goes as $du_1\, du_2\, 
u_1^{-2} u_2^{-2}$. It then follows from the analysis of \S\ref{serror} that
it will be sufficient to  keep terms up to fractional errors of order
$u_1\sim\wt\lambda^{-1}$ and $u_2\sim\alpha^{-2}$ but we can ignore 
fractional errors of order 
$u_1^2\sim \wt\lambda^{-2}$ and / or
$u_2^2\sim \alpha^{-4}$. With this understanding,  we can replace 
\refb{e171pre} by:
\be\label{e171}
v = u_2\left\{1 - {1\over 2} u_2\right\}^{-1}, 
\qquad x = {1\over 2\pi} {u_1} (1-u_2) \, .
\ee
Therefore Fig.~\ref{figfive}(a) covers the range
\be
0\le v\le \left(\alpha^{2}-{1\over 2}\right)^{-1}, \quad 0\le 2\pi x \le \wt\lambda^{-1}\, {2-v\over 2+v}
\, ,
\ee
in the moduli space. This has been shown as region (a) in Fig.~\ref{fignine}.

We now turn to the contribution from Fig.~\ref{figfive}(b). We denote by
$\hat z=e^{iw}$ the coordinate associated with the interaction vertex for the
open string one point function on 
the annulus, and by $z$ the coordinate associated with the C-O interaction vertex. Then, according to
\refb{erange}, we have the identification:
\be
\hat z = v^{-1} \hat z, \quad \left\{\alpha^{2}-{1\over 2}\right\}^{-1}<v< 1\, .
\ee
Also, using \refb{e1alt}, \refb{elocal2}, the sewing relation takes the form:
\be
2\, \alpha  {(4+3\ \alpha^{-2}) \hat z -4 + 3\, \alpha^{-2} \over
(4-\alpha^{-2}) \hat z + 4 - 7\, \alpha^{-2}} \, \lambda z = -q_1, \quad 0\le q_1\le 1\, .
\ee
Using this we find that 
the closed string puncture at $z=i$ is located in the $w={1\over i} \ln\hat z$ plane at
\be
w_c={1\over i}\ln\hat z_c \simeq u\left\{1 - \alpha^{-2}+\OO(\alpha^{-4})\right\} +\OO(u^3), 
\qquad u ={q_1\over \wt\lambda}, 
\qquad 0\le u\le \wt\lambda^{-1} \, .
\ee
Comparison with \refb{ecc2} gives:
\be\label{e2.81pre}
2\pi x = u\left\{1 - \alpha^{-2}+\OO(\alpha^{-4})\right\} +\OO(u^3)\, .
\ee
Since $dx\, dv\, x^{-2} v^{-2}$ grows as $du\, dv\,
u^{-2} v^{-2}$ for small $u, v$, and since $v$ is
bounded from below by $\alpha^{-2}$, it follows from 
the analysis of \S\ref{serror} that we need to keep terms
up to fractional errors of order
$u\sim\wt\lambda^{-1}$ and $\alpha^{-2}$.
Therefore we can approximate \refb{e2.81pre} as:
\be \label{e2.81}
2\pi x = u\left\{1 - \alpha^{-2}\right\} \, , \qquad 0<2\pi x<  \wt\lambda^{-1} 
\left\{1 - \alpha^{-2}\right\}\, .
\ee
Therefore Fig.~\ref{figfive}(b)
covers the region:
\be \label{e4.91}
\left\{\alpha^{2}-{1\over 2}\right\}^{-1}<v<1, \qquad 0<2\pi x<  \wt\lambda^{-1}
\left\{1 - \alpha^{-2}\right\}\, .
 \ee
This has been shown as region (b) in Fig.~\ref{fignine}.

Next we turn to the contribution from Fig.~\ref{figfive}(c). We denote by 
$z$  the coordinate on the C-O-O interaction vertex with the closed string puncture located at 
$z=i$.
Then the sewing relation gives the identification
\be \label{eida}
w_1 =-q_2/w_2\, ,
\ee
where $w_1$ and $w_2$ have been defined in  \refb{e5}. Denoting $f(\beta)$ by
$f$ for notational simplicity and using the relation $f(-\beta)=-f(\beta)$,
we can express \refb{eida} as
\be \label{etrsz}
z \equiv {a  z + b\over c z + d}\, ,
\ee
where
\ben
&& a = \wt\lambda\, (1+\beta^2)^{-1}  \left[\beta \, u^{-1/2}+u^{1/2} \, \wt\lambda^{-2} \, \left\{1 - \wt\lambda f\beta\right\} 
\left\{ \wt\lambda f +\beta\right\}\right]  \, , \nonumber \\ &&
b = \wt\lambda\, (1+\beta^2)^{-1} \left[\beta^2 u^{-1/2}-  u^{1/2} \, \wt\lambda^{-2} \, \left\{1 - 
\wt\lambda f\beta\right\}^2
\right]\, ,
\nonumber \\ &&
c = \wt\lambda\, (1+\beta^2)^{-1} \left[u^{-1/2}-u^{1/2} \, \wt\lambda^{-2} \,  \left\{\wt\lambda 
f +\beta\right\}^2\right]   \, , \nonumber \\ 
&&
d = \wt\lambda\, 
(1+\beta^2)^{-1} \left[\beta u^{-1/2}+u^{1/2}\, \wt\lambda^{-2} \,  \left\{ \wt\lambda f 
+\beta\right\} 
\left\{ 1 - \wt\lambda f\beta\right\} \right]
\, , \nonumber \\
&& \hskip 2in f\equiv f(\beta)\, ,
\een
and
\be \label{e495u2}
u = q_2\, \alpha^{-2} \, \left\{ 1 + {1\over 4\wt\lambda^2}\right\}^{-2}, 
\qquad 0\le u\le \alpha^{-2} \left\{ 1 + {1\over 4\wt\lambda^2}\right\}^{-2}
\, .
\ee
The transformation \refb{etrsz} can be `diagonalized'
using new coordinate $\tilde z$, defined via
\ben
&& z = {A\tilde z + B\over C\tilde z + D}, \quad
{A\over C} = \beta\left[ 1 + u \, \wt\lambda^{-2} \,  \left\{ {1+\beta^2\over\beta} 
\wt\lambda f - {1-\beta^4\over 2\beta^2} 
\right\}+\OO(u^2)
\right], \nonumber \\ &&
{B\over D} = -\beta\left[ 1 + u \, \wt\lambda^{-2} \,  \left\{ {1+\beta^2\over\beta} 
\wt\lambda f - {1-\beta^4\over 2\beta^2}
\right\}+\OO(u^2)
\right]\, .
\een
We also define
\be
w ={1\over i} \, \ln \tilde z\, .
\ee
In these variables \refb{etrsz} takes the form:
\ben \label{e4.98}
\tilde z &\equiv& \left\{{4\beta^2 \wt\lambda^2 
\over (1+\beta^2)^2 \, u} + 8\, \beta {(\beta+\wt\lambda f) 
(1-\beta \wt\lambda f)\over (1+\beta^2)^2} -2 +\OO(u)
\right\} \, \tilde z, \nonumber \\
w &\equiv& w + i \ln \left\{{4\beta^2 \wt\lambda^2 
\over (1+\beta^2)^2 \, u} + 8\, \beta {(\beta+\wt\lambda f) 
(1-\beta \wt\lambda f)\over (1+\beta^2)^2} -2 +\OO(u)
\right\}\, .
\een
Also, in this variable
the closed string puncture at $z=i$ is located at
\be\label{e4.99}
\tilde z_c = - {D+iB\over C+iA}, \qquad
w_c = \pi +\tan^{-1} {B\over D} -\tan^{-1}{A\over C} + \ \hbox{imaginary}\, .
\ee
Comparison of \refb{e4.98}, \refb{e4.99}
with \refb{ecc1}, \refb{ecc2}, and use of the $x\to {1\over 2}-x$ symmetry, gives
\ben\label{e2.90pre}
&&\hskip -.3in  
2\pi x = 2\tan^{-1} \beta
+ {2\, u\, f} \, \wt\lambda^{-1} \,  - {1\over \beta} (1-\beta^2) \, \wt\lambda^{-2} \, u 
+\OO(u^2)\, , 
\nonumber \\ && 
\hskip -.3in v = u \,  {(1+\beta^2)^2\over 4\beta^2 \wt\lambda^{2}} 
\left\{ 1 +u\, \wt\lambda^{-2}\, {1\over 2\beta^2} (1-\beta^2 - 2\beta\wt\lambda f)^2 +\OO(u^2)\right\}\,.
\een
Since $dv\, dx\, v^{-2}x^{-2}\sim du\, d\beta\, u^{-2} 
\wt\lambda^{2}$ for small $u$, we can ignore fractional
errors of order $u^2$, since  they will
generate corrections containing inverse powers of  $\alpha^{2}$.  
Therefore we can write \refb{e2.90pre} as
\ben\label{e2.90}
&&\hskip -.3in  
2\pi x = 2\tan^{-1} \beta
+ {2\, u\, f} \, \wt\lambda^{-1} \,  - {1\over \beta} (1-\beta^2) \, \wt\lambda^{-2} \, u
\, , 
\nonumber \\ && 
\hskip -.3in v = u \,  {(1+\beta^2)^2\over 4\beta^2 \wt\lambda^{2}} 
\left\{ 1 + u\, \wt\lambda^{-2}\, {1\over 2\beta^2} (1-\beta^2 - 2\beta\wt\lambda f)^2\right\}
\,.
\een
The range
\be
{1\over 2\wt\lambda} \le \beta \le 1, \quad 0\le u\le \alpha^{-2} 
\left\{ 1 + {1\over 4\wt\lambda^2}\right\}^{-2}\, ,
\ee
translates to the region bounded by the curves:
\ben \label{e4.103}
&&\hskip -.2in  x = {1\over \pi} \, \tan^{-1} \left[{1\over 2\wt\lambda} (1- v)\right] , 
\quad x={1\over 4}, \quad v=0, \nonumber \\
&& \hskip -.2in
v= {1\over \alpha^2 \wt\lambda^2\sin^2(2\pi x)}
\left(1 + {1\over 4\wt\lambda^2}\right)^{-2} \Bigg[1- 2 \left\{ \cot^2(2\pi x)- \wt\lambda^2 f^2\right\} \alpha^{-2}\wt\lambda^{-2}
\left(1 + {1\over 4\wt\lambda^2}\right)^{-2}\Bigg]\, , \nonumber \\
\een
where now $f\equiv f(\tan(\pi x))$ and we have used
\be
f\left({1\over 2\wt\lambda}\right) = {1\over 2} \left\{1 - {3\over 4\, \wt\lambda^2}
\right\}, \qquad f(1)=0\, .
\ee
The region \refb{e4.103} has been shown as region (c) in Fig.~\ref{fignine}.

The contribution from
Fig.~\ref{figfive}(d) should cover the left-over region in the $x$-$v$ space:
\ben  \label{e4104}
&& {\pi\over 2}\ge 2\pi x\ge  \wt\lambda^{-1} (1-\alpha^{-2}), 
\nonumber \\
&& {1\over \alpha^2 \wt\lambda^2\sin^2(2\pi x)}
\left(1 + {1\over 4\wt\lambda^2}\right)^{-2} \Bigg[1- 2 \left\{ \cot^2(2\pi x)- \wt\lambda^2 f^2\right\} \alpha^{-2}\wt\lambda^{-2}
\left(1 + {1\over 4\wt\lambda^2}\right)^{-2}\Bigg]\le v<1\, , \nonumber \\
&& \hskip 1in f\equiv f(\tan(\pi x))\, .
\een
This has been shown as region (d) in 
Fig.~\ref{fignine}.
\refb{e4104} is an approximate description of the left-over region, but the arguments given
earlier ensure that the error due to this approximation affects the final integral only by
terms containing inverse powers of $\alpha^2$ or $\wt\lambda$.

The regions (b) and (d) in Fig.~\ref{fignine}  contain the region $v=1$. This is actually
a boundary of the moduli space associated with
closed string degeneration, where the annulus becomes an infinitely long cylinder. 
In the full open-closed string field theory, we need to represent the contribution 
from this region as coming from new Feynman diagrams containing internal closed string
propagators.
This will be necessary
for evaluating the full contribution to $g(\omega)$ given in 
\refb{egfull} that
has a singular contribution from this region\cite{2003.12076}. 
However, since our goal is to evaluate
$g_{\rm div}(\omega)$ given in \refb{egdivexp} which does not have
 any divergence from the
$v\to 1$ region, we do not need this special treatment of the $v\simeq 1$ region.

\sectiono{Disk two point function in two dimensional string theory} \label{s2}

Our analysis so far applies to any string theory D-instanton. 
Now we shall apply the results to the specific
case of D-instantons in
two dimensional string theory, whose world-sheet matter theory contains a free scalar 
describing 
the time coordinate and the Liouville theory with $c=25$. 
As discussed in \S\ref{sintro}, we shall first analyze the building blocks of the
amplitude, leaving out the energy conserving delta function and the 
$\NN\, e^{-1/g_s}$ factor that are common to all amplitudes.

In this section we shall 
analyze the building block corresponding to the disk two point function of a pair
of closed string tachyons of energies $\omega_1$ and $\omega_2$.
As described in \S\ref{sintro}, this has the form:
\be
4\,  g_s\, \sinh(\pi |\omega_1|)\, \sinh(\pi |\omega_2|) \, f(\omega_1,\omega_2)\, ,
\ee
where $f(\omega_1,\omega_2)$ is a sum of a finite piece and an apparently
divergent piece:
\be \label{e5.4}
f(\omega_1,\omega_2) = f_{\rm finite}(\omega_1, \omega_2) 
+ f_{\rm div}(\omega_1, \omega_2) \, .
\ee
$f_{\rm finite}$ has been given in \refb{effinite}, but since 
we shall not need it, we do not
reproduce it here. $f_{\rm div}(\omega_1,\omega_2)$ is given by:
\be \label{efdivor}
f_{\rm div}(\omega_1, \omega_2)={1\over 2}
\int_0^1 dy \,
y^{-2} (1+2\omega_1\omega_2 y) \, .
\ee

We shall follow the general strategy introduced in \S\ref{srev} 
to evaluate $f_{\rm div}$ (and also
$g_{\rm div}$ described in \S\ref{s7}). 
Let us suppose
that we have some on-shell amplitude, expressed as an integration over the moduli 
$\vec m\equiv
\{m_1,\cdots , m_n\}$ of the 
world-sheet with punctures. The issue that we have to deal with is that the integrand
diverges at various corners of the moduli space and we have to make sense
of these divergences. For this 
we express the amplitude as sum over Feynman diagrams of string field theory.
Let us suppose that we have a Feynman diagram with
$k$ propagators and that $q_1,\cdots q_k$ are the sewing parameters associated with these propagators,
each taking value in the range $0\le q_i\le 1$.
The analysis of the type performed in \S\ref{s1} determines the relation between the moduli $\vec m$ and the
sewing parameters $q_1,\cdots, q_k$ and $(n-k)$ other variables $\{\tilde m_1,\cdots \tilde m_{n-k}\}$ 
arising from the interaction vertices of the
Feynman diagram. We now express the integrand
as a function of the $q_i$'s and $\tilde m_j$'s and expand it in
a power series in $q_i$'s, with the coefficients of the power series being functions of the 
$\tilde m_j$'s. 
Next we apply the replacement rule 
\refb{erp1} to extract a finite answer. We must also include, for each propagator, the
contribution from the propagation of the $\psi^1$ field introduced in 
\refb{efieldexpandfirst} separately. The effect of the Jacobian factors discussed in
\S\ref{s2.4} contribute at the same order as one loop amplitudes, and will affect the
analysis of $g_{\rm div}$ in \S\ref{s7}. However they will not affect the analysis of
$f_{\rm div}$ that will be carried out in this section, since this, being a disk two point
function, represents a
tree amplitude.

To implement the strategy described above for the integral \refb{efdivor}, 
we first need the world-sheet interpretation
of $y$ appearing in \refb{efdivor}. There is a simple answer to this: $i$ and $iy$ are
the positions of the vertex operators carrying energies $\omega_1$ and $\omega_2$
in the upper half plane\cite{1907.07688}. 
Therefore we can use the relation between $y$ and $q$
given in \refb{eyqrel} for analyzing \refb{efdivor}.
We first divide the range of integration over $y$ into
the regions A and B described in \refb{edefA}, \refb{edefB}. 
The contribution from the region B does not suffer from any divergences and
simply gives back the same integral with integration range replaced by 
$\lambda^{-2}\le y\le 1$. The
contribution from the 
region A is evaluated by replacing $y$ by $q/\lambda^2$ according to 
\refb{eyqrel} and expressing the integral
as
\be\label{ereg}
{1\over 2} 
\int_0^1 dq\, q^{-2} (\lambda^2 + 2\, \omega_1 \omega_2 q)\quad \to \quad
- {1\over 2} \,
\lambda^2\, ,
\ee
where in the second step we have used the replacement rule \refb{erp1}. Therefore the net result is
\be \label{enetcc}
{1\over 2} \, 
\int_{\lambda^{-2}}^1 dy\, y^{-2} (1+2\, \omega_1 \omega_2 y)
- {1\over 2} \,
\lambda^2
= -{1\over 2} \left\{ 1 - 2\, \omega_1\, 
\omega_2 \ln \lambda^2\right\}\, .
\ee

We also have to add the contribution where in Fig.~\ref{figthree}(a) we replace the
internal open string propagator by the $\psi^1$ propagator. This contribution however
vanishes, since the open-closed string interaction vertex, with a closed string tachyon and
an open string state associated with the vertex operator $\p c$, vanishes for the choice
of local coordinate system given in \refb{e1alt}. To see this, note that if $c\bar c V_T$ 
denotes the closed string tachyon vertex operator, then this two point
function is proportional to
\be\label{eocvanish}
\langle \bar c\, c\,  V_T(i) \p c(0)\rangle \propto \left[\p_z \left\{(z-i)(z+i)\right\}
\right]_{z=0}=0\, .
\ee

Therefore \refb{enetcc} represents the full contribution to $f_{\rm div}$:\footnote{We 
shall follow the
convention
that $f_{\rm div}$ stands for any contribution to the disk two point function that is
not already included in $f_{\rm finite}$ defined in \refb{effinite}.
A similar convention will be followed for $g_{\rm div}$ in \S\ref{s7}.}
\be\label{efdiv}
f_{\rm div}(\omega_1, \omega_2) = -{1\over 2} (1 - 2\omega_1
\omega_2 \ln\lambda^2) \, .
\ee
Comparing this with \refb{efgpar} we get
\be
A_f = -{1\over 2}, \quad B_f = \ln \, \lambda^2\, .
\ee

Note that the vanishing of the $\psi^1$ exchange amplitude 
is a specific property of the choice of local coordinate system
that we have made in \refb{e1alt}. If we had chosen a different coordinate system,
then the
result would not have vanished. However the final result for any physical amplitude
will be independent of the choice of local coordinates. In particular $A_f$ will be
independent of the choice of local coordinate system since it can be extracted from the
scattering amplitude of $n$ closed string tachyons. We have illustrated this in appendix
\ref{sd} by working with a different choice of local coordinate system for the
C-O interaction vertex. $B_f$ by itself need not be independent of the
choice of local coordinate system, but the combination $2B_g-B_f$ will be independent
of the choice of local coordinates.

\sectiono{Annulus one point function in two dimensional string theory} \label{s7}

We now turn to the second building block for the amplitudes in two dimensional
string theory -- the one point function of closed string tachyon of energy $\omega$ on
the annulus. As discussed in \S\ref{sintro}, this has the form:
\be \label{eblock7}
2\, g_s\, \sinh(\pi|\omega|) \, g(\omega)\, ,
\ee
where $g(\omega)$ is given by  the sum of a finite piece and a divergent piece:
\be \label{e5.5}
g(\omega) = g_{\rm finite}(\omega) + g_{\rm div}(\omega)\, .
\ee
The expression for $g_{\rm finite}(\omega)$ has been given in \refb{egfinite} 
and will not be
repeated here. $g_{\rm div}(\omega)$ has the form:
\be\label{egdivor}
g_{\rm div}(\omega)=
  \int_0^1 dv \,  \int_0^{1/4} dx \left\{ {v^{-2} - v^{-1} \over \sin^2(2\pi x)}
+ 2\, \omega^2 \, v^{-1}\right\}\, .
\ee

As in \S\ref{s2}, we can analyze this by first relating the variables  $v$ and $x$
to the parameters $\{q_i\}$ associated with the propagators of string field theory,
and then using the replacement rule \refb{erp1}. We shall call this the world-sheet
contribution to $g_{\rm div}$. However there are several other contributions
to the annulus one point function that are not captured by the world-sheet expression
for $g_{\rm finite}$. We shall include these also in the definition of $g_{\rm div}$.
This includes the
contribution due to the $\psi^1$ field propagating in the internal lines of the Feynman
diagrams of string field theory. We also have two 
possible field redefinitions whose associated Jacobians give 
additional contributions to the one point function of the closed string tachyon. 
The first one is associated with the
redefinition that relates the zero mode in open string field theory to the collective 
coordinate associated with the translation of the D-instanton along the
time direction. The second redefinition involves the ghost field $\psi^2$ introduced
in \refb{efieldexpandfirst}. This can also be interpreted as the redefinition that relates the
gauge transformation
parameter in open string field theory associated with the vacuum state $|0\rangle$ to
the rigid 
$U(1)$ symmetry transformation parameter that gives a phase to any open string
stretching from the D-instanton to another D-brane. We shall analyze each of these
contributions in turn.

\subsection{World-sheet contribution}\label{s2.2}

We shall first analyze the contribution to \refb{egdivor} that we obtain by 
replacing the variables $x$ and/or $v$ near the singular points by appropriate
variables $\{q_i\}$ associated with the propagators of string field theory, and then
using the replacement rule \refb{erp1}. For this we need to know the world-sheet
interpretation of the variables $v$ and $x$. 
As in \refb{ecc1}, \refb{ecc2},  $-i\ln v$ is the periodicity of the 
$w$ 
coordinate on the annulus, with the real part of $w$ taking value in the interval
$(0,\pi)$, and $2\pi x$ denotes the real part of the location of the closed string
puncture in the $w$ plane\cite{1907.07688}. 
We shall divide the integration region over $v$ and $x$ in \refb{egdivor}
into four regions shown in Fig.~\ref{fignine}, representing the
contributions from different Feynman diagrams in Fig.~\ref{figfive},
and carry out the integration
over each region separately, using the replacement rule \refb{erp1} or
\refb{erp2}.
In this analysis we shall ignore terms that vanish
in the large $\alpha$ and large $\wt\lambda$ limit, including terms proportional to
$\alpha^2/\wt\lambda$ and $\wt\lambda^2/\alpha^2$, since these terms 
must cancel at the
end. The reader may be somewhat uncomfortable in ignoring terms of order 
$\wt\lambda^2/\alpha^2$, since this ratio, being given by $\lambda^2$, is large.
Nevertheless, it follows from our general arguments that such terms must cancel
at the end. As discussed at the end of this subsection, we have explicitly checked
that such terms do cancel.

We begin with the contribution from the region (d) given in \refb{e4104}:
\ben  \label{e4104repeat}
&& {\pi\over 2}\ge 2\pi x\ge  \wt\lambda^{-1} (1-\alpha^{-2}), 
\nonumber \\
&& {1\over \alpha^2 \wt\lambda^2\sin^2(2\pi x)}
\left(1 + {1\over 4\wt\lambda^2}\right)^{-2} \Bigg[1- 2 \left\{ \cot^2(2\pi x)- \wt\lambda^2 f^2\right\} \alpha^{-2}\wt\lambda^{-2}
\left(1 + {1\over 4\wt\lambda^2}\right)^{-2}\Bigg]\le v<1\, , \nonumber \\
\een
where $f$ stands for $f(\tan(\pi x))$
The integral takes the form:
\ben\label{eca2pre}
I_{(d)}&=&
\int_{(2\pi\wt\lambda)^{-1}(1-\alpha^{-2})}^{1/4} dx \, 
\int_{
{1\over \alpha^2 \wt\lambda^2\sin^2(2\pi x)}
\left(1 + {1\over 4\wt\lambda^2}\right)^{-2} \left[1- 2 \left\{ \cot^2(2\pi x)- \wt\lambda^2 f^2\right\} \alpha^{-2}\wt\lambda^{-2}
\left(1 + {1\over 4\wt\lambda^2}\right)^{-2}\right]
}^1\, dv \nonumber \\ &&
\hskip 1in \left[ {1\over \sin^2(2\pi x)} \left(v^{-2} - v^{-1}\right) 
+ 2\, \omega^2 \, v^{-1}\right] \nonumber \\
&=& \int_{(2\pi\wt\lambda)^{-1}(1-\alpha^{-2})}^{1/4} dx \Bigg[\alpha^2 \wt\lambda^2 
+{\alpha^2\over 2}+ 2 \, \{\cot^2(2\pi x) -\wt\lambda^2 f^2\} 
-{1\over \sin^2(2\pi x)}
\nonumber \\
&& - {1\over \sin^2(2\pi x)} \ln \{\alpha^2 \wt\lambda^2 \sin^2(2\pi x)\} + 2\,
\omega^2 \, \ln \{\alpha^2 \wt\lambda^2 \sin^2(2\pi x)\}
\Bigg]\, .
\een
The integrals in the third  line on the right hand side are 
elementary. 
For the terms in the last line, the integral can be simplified by writing 
$dx/\sin^2(2\pi x)$ as $- d\cot(2\pi x)/2\pi$ and then doing an integration by parts.
This gives, after dropping terms that have either powers of $\wt\lambda$ or of
$\alpha$ in the denominator: 
\ben\label{eca2}
I_{(d)}&=& \Bigg[{1\over 4} \alpha^2 \wt\lambda^2
- {\alpha^2 \wt\lambda\over 2\pi} + {\wt\lambda\over 2\pi}
+{\alpha^2\over 8}
+{\wt\lambda\over \pi}-{1\over 2} 
 - 2\, \wt\lambda^2\, 
 \int_{(2\pi\wt\lambda)^{-1}}^{1\over 4} dx \, f(\tan(\pi x))^2 -{\wt\lambda\over 2\pi}
\nonumber \\ &&
+{1\over 2\pi} \left\{-2\wt\lambda\, \ln\alpha + {\pi}-2\wt\lambda
\right\} + {1\over 2} \, \omega^2 \, \ln {\alpha^2 \wt\lambda^2 \over 4}
\Bigg]\, . 
\een

An alternative procedure for evaluating the $\omega$ independent part of the integral
is to express the integrand as a total derivative, {\it e.g.} as,
\be
d \AAA, \qquad \AAA = - v^{-1} \, {1\over \sin^2(2\pi x)} \, dx - {1\over 2\pi}\, v^{-1}\,
\cot(2\pi x) \, dv\, ,
\ee
and integrate the one form $\AAA$ along the boundary of region (d) in the anti-clockwise
direction. This has the advantage that we do not need to explicitly find the shape of
the boundary given in \refb{e4104}. For example, while evaluating the contribution from
the boundary separating regions (c) and (d), we can simply change variables from $(v,x)$
to $(u,\beta)$ using \refb{e2.90} that is appropriate in region (c), set $u=\alpha^{-2} 
(4\wt\lambda^2)^2 (1 + 4\wt\lambda^2)^{-2}$ to parametrize the boundary, and then integrate
over the variable $\beta$ in the range $(2\wt\lambda)^{-1}\le \beta\le 1$. 
This give the same
result, but this procedure is easier to use if we want to extend the analysis to higher
orders in the expansion in powers of $\wt\lambda^{-1}$ and $\alpha^{-2}$ and check that
these terms indeed cancel, as implied by our general arguments.

Next we turn to the contribution from region (c).
In order to evaluate the contribution from this region, we
need to 
change variables from $(v,x)$ to $(u,\beta)$ according to \refb{e2.90},
\ben\label{e2.90repeat}
&&\hskip -.3in  
2\pi x = 2\tan^{-1} \beta
+ {2\, u\, f} \, \wt\lambda^{-1} \,  - {1\over \beta} (1-\beta^2) \, \wt\lambda^{-2} \, u
\, , 
\nonumber \\ && 
\hskip -.3in v = u \,  {(1+\beta^2)^2\over 4\beta^2 \wt\lambda^{2}} 
\left\{ 1 + u\, \wt\lambda^{-2}\, {1\over 2\beta^2} (1-\beta^2 - 2\beta\wt\lambda f)^2\right\}
\, ,
\een
and then integrate
over the range $0\le u\le \alpha^{-2}\{1+(4\wt\lambda^2)^{-1}\}^{-2}$ and $(2\wt\lambda)^{-1}\le \beta\le 1$ using the
replacement rule \refb{erp2}. One finds:
\ben\label{eca3}
I_{(c)}&=&
{1\over \pi}\,
\int_{1/(2\wt\lambda)}^1 d\beta
\int_0^{\alpha^{-2}
(1+(4\wt\lambda^2)^{-1})^{-2}}
 du 
 \Bigg[\wt\lambda^2\,  u^{-2} {1\over 1+\beta^2} +\wt\lambda \, u^{-1} f'(\beta)
 +{1\over 4} u^{-1} \, {1+\beta^2\over \beta^2}\nonumber \\ && \hskip 1in
 +{2\over 1+\beta^2} \, \omega^2 \, u^{-1}
 \Bigg]
\nonumber \\
&\to & - {1\over \pi}\,
\alpha^2\wt\lambda^2\, \left(1+{1\over 4\wt\lambda^2}\right)^2
\int_{1/(2\wt\lambda)}^1 {d\beta\over 1+\beta^2}
= \left[-{1\over 4}
\alpha^2\wt\lambda^2 -{\alpha^2\over 8}
+ {\alpha^2\, \wt\lambda\over 2\pi}\right]\, \, .
\een

In region (b), $v$ is restricted to the range $\left(\alpha^{2}-{1\over 2}\right)^{-1}
\le v \le 1$, and
we use \refb{e2.81} to trade in $x$ for the variable $u$ via
\be 
2\pi x = u\, (1-\alpha^{-2}), \quad 0\le u\le \wt\lambda^{-1}\, .
\ee
This reduces the integral to
\ben\label{eca4}
I_{(b)}&=&
\int_{\left(\alpha^{2}-{1\over 2}\right)^{-1}}^1 dv \int_0^{\wt\lambda^{-1}} 
{du\over 2\pi} \, 
\Bigg[
(1+\alpha^{-2}) 
 (v^{-2}-v^{-1}) \{ u^{-2} +\OO(1)\}
+ (1-\alpha^{-2}) 
\, 2\, \omega^2 \, v^{-1}
\Bigg]\nonumber \\
&\to & 
{1\over 2\pi} \,\left[ -\wt\lambda\, (1+\alpha^{-2}) 
\int_{\left(\alpha^{2}-{1\over 2}\right)^{-1}}^{1} dv\, (v^{-2}-v^{-1}) 
\right] 
=\left[ -{1\over 2\pi} \wt\lambda\, \alpha^2 +{\wt\lambda\over 4\pi}
+{\wt\lambda\over \pi}
\, \ln\alpha
\right]\, . \nonumber \\
\een

Finally in region (a) the relevant change of variables is given in \refb{e171}:
\be\label{e171a}
v = u_2\left\{1 + {1\over 2} u_2\right\}, 
\qquad x = {1\over 2\pi} {u_1} (1-u_2), \qquad 0\le u_1\le \wt\lambda^{-1}, \qquad
0\le u_2\le \alpha^{-2} \, ,
\ee
ignoring fractional errors of order $u_1^2$ or $u_2^2$. This gives
\be
dx\wedge dv = {1\over 2\pi}\, du_1\wedge du_2 \, ,
\ee
and expresses the contribution to the integral from this region as
\ben \label{eia53}
I_{(a)}&=&
 {1\over 2\pi}\, \int_0^{\wt\lambda^{-1}} du_1 \, \int_0^{\alpha^{-2}} du_2 \ \Bigg[ 
u_1^{-2}\, (1+2\, u_2)\,
\left\{ u_2^{-2} (1- u_2) - u_2^{-1} \left(1-{1\over 2}u_2\right)\right\}
\nonumber \\ &&  \hskip 1in 
+2\, \omega^2 \, u_2^{-1}\left(1-{1\over 2} u_2\right)\Bigg] \nonumber \\
&=& {1\over 2\pi}\, \int_0^{\wt\lambda^{-1}} du_1 \, \int_0^{\alpha^{-2}} du_2 \ u_1^{-2}\,
\left\{ u_2^{-2} (1+ u_2) - u_2^{-1} \left(1+{3\over 2}u_2\right)
+2\, \omega^2 \, u_2^{-1}\left(1-{1\over 2} u_2\right)\right\}\, . \nonumber \\
\een
We now use the replacement rule \refb{erp2} to get:
\ben\label{eca5}
I_{(a)}&=&
{\wt\lambda \, \alpha^2\over 2\, \pi} \, .
\een

Adding \refb{eca2}, \refb{eca3}, \refb{eca4} and \refb{eca5}, we get the total 
world-sheet contribution to $g_{\rm div}$:
\ben \label{eidiv}
g_{\rm world} &=& 
\bigg[ - 2\, \wt\lambda^2\, 
 \int_{(2\pi\wt\lambda)^{-1}}^{1\over 4} dx \, f(\tan(\pi x))^2 +
 {\wt\lambda\over 4\pi} 
 + {1\over 2} \omega^2 \, \ln {\alpha^2 \wt\lambda^2 \over 4}\bigg]\nonumber \\
&=& 
\bigg[ - {2\over \pi}\, \wt\lambda^2\, 
 \int_{(2\wt\lambda)^{-1}}^{1} d\beta \, (1+\beta^2)^{-1}\,  f(\beta)^2 +
 {\wt\lambda\over 4\pi} 
 + {1\over 2} \omega^2 \, \ln {\alpha^2 \wt\lambda^2 \over 4}\bigg] \, ,
 \een
where in the last step we have changed variable from $x$ to $\beta=\tan(\pi x)$
and dropped terms containing inverse powers of $\wt\lambda$.

We have carried out two further checks on the result:
\begin{enumerate}
\item Terms of order $\wt\lambda^2/\alpha^2$
have been dropped in our analysis, despite the ratio being large, since we have argued
that such terms must cancel at the end. To confirm this,
we have checked that such terms appear in the
expressions for $I_{(c)}$ and $I_{(d)}$, and are given respectively by,
\be
-{1\over \pi}\, {\wt\lambda^2\over \alpha^2}\,  
\int_{(2\wt\lambda)^{-1}}^1 {d\beta\over (1+\beta^2)} \, f(\beta)^4\, ,
\ee
and
\be
{1\over \pi}\, {\wt\lambda^2\over \alpha^2}\,  
\int_{(2\wt\lambda)^{-1}}^1 {d\beta\over (1+\beta^2)} \, f(\beta)^4\, .
\ee
This shows that these terms do cancel at the end.
\item
The $\omega$ independent part of the integral
in \refb{egdivor} can also be computed using the 
alternative procedure described in \S\ref{salternative}.
For this we note that this part of the integrand may be expressed as a
total derivative, {\it e.g.} as,
\be
d \AAA, \qquad \AAA = - v^{-1} \, {1\over \sin^2(2\pi x)} \, dx - {1\over 2\pi}\, v^{-1}\,
\cot(2\pi x) \, dv\, .
\ee
Therefore the integral can be expressed as an anti-clockwise 
boundary integral of $\AAA$ in Fig.~\ref{fignine}, 
with the boundaries containing the lines $v=1$ and $x=1/4$, and the 
regularized boundaries at $v=0$ and $x=0$. The latter two may 
be viewed as the collection 
of boundaries
at $q_2=\delta_2$ in Figs.~\ref{figfive}(a) and (c)
and at $q_1=\delta_1$ in Figs.~\ref{figfive}(a) and (b).
We have checked that this gives the same
result as \refb{eidiv} after
dropping all terms
proportional to $\delta_1^{-1}$, $\delta_2^{-1}$, $\ln\delta_1$ and $\ln\delta_2$ in the
final expression, as described in \S\ref{salternative}. 
\end{enumerate}

\subsection{$\psi^1$ exchange contribution} \label{spsi}

We shall now evaluate the $\psi^1$ exchange contributions. For this we need the
$\psi^1$ propagator.  In order to demonstrate that
the final results are independent of the choice of normalization conventions, we shall
normalize the vacuum as
\be \label{econvention}
\langle 0|c_{-1} c_0 c_1 |0\rangle =K_0\, ,
\ee
where $K_0$ is some non-zero constant. 
This corresponds to normalizing the ghost
correlation function on the
upper half plane as:
\be\label{e615}
\langle c(z_1) c(z_2) c(z_3)\rangle= K_0 \, (z_1-z_2) (z_2-z_3) (z_1-z_3)\, .
\ee
If
we expand the string field as
\be\label{e2.3}
|\Psi\rangle= \psi^0 \, c_1|0\rangle + \psi^1 \, c_0|0\rangle +\cdots\, ,
\ee
then with the convention \refb{econvention} the quadratic term in the string field theory action takes the form:
\be \label{ekinprop}
{1\over 2}\langle\Psi|Q_B|\Psi\rangle = - {K_0\over 2} (\psi^0)^2 -K_0 \,  (\psi^1)^2\, .
\ee
Therefore the tachyon propagator is given by $K_0^{-1}$ while the $\psi^1$ propagator
is given by $K_0^{-1}/2$. 
Although the propagators are convention dependent, their ratio is convention independent.
Our final results will depend only on the convention 
independent ratios of this type.

\begin{figure}
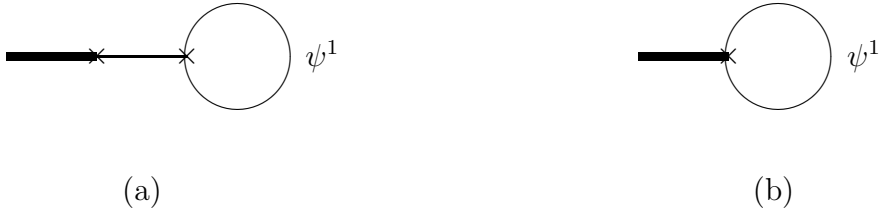

\begin{center}

\hbox{\figpsione}

\vskip -1.5in

\caption{Diagrams involving $\psi^1$ propagator that contribute to the annulus one
point function of the closed string tachyon. 
\label{figpsione}
}
\end{center}
\end{figure}

The C-O amplitude,
where the closed string is an on-shell tachyon state and the open string represents
$\psi^1$, vanishes by \refb{eocvanish}. 
Therefore the relevant diagrams are the ones
shown in Fig.~\ref{figpsione}.  Both these diagrams involve computing a disk amplitude
with one external C and two external O's, with the
O's corresponding to the $\psi^1$ state,
and then multiplying it by the $\psi^1$ propagator. One subtlety in this
computation is that the $\psi^1$ vertex operator $\p c$
is not a dimension
zero primary of the form $cW$ where $W$ is a dimension one matter primary. Therefore
we need to be careful in evaluating the integrand that we shall eventually integrate over
the modulus of this diagram -- the variable $q$ appearing in \refb{e4.9a} 
for Fig.\ref{figpsione}(a)
and the variable $\beta$ appearing in \refb{e5} for Fig.~\ref{figpsione}(b).
Defining $\beta=q/(2\wt\lambda)$ in \refb{e4.9a}, we can analyze the two diagrams together.
Let $z$ be the coordinate on the upper half plane
in which the closed string puncture is situated at $i$ and the open string punctures are
situated at $z_1=-\beta$ and $z_2=\beta$. Let
$w_1$ and $w_2$ be the local coordinates at the open string punctures, related to $z$
via the relations
\be \label{e6.18new}
z = F_a(w_a, \beta)\, ,
\ee
so that $z_a=F_a(0,\beta)$.
Then 
the integrand multiplying $d\beta$ is given by the correlation function of 
the vertex operators of the closed string tachyon 
at $i$ and of the
two $\psi^1$'s at $z_1$, $z_2$ in the $w_1,w_2$ coordinate system, and the
operator\cite{9206084,9705241,1703.06410}
\be\label{e7.14rep}
- K_1\sum_{a=1}^2 \ointop_a {\p F_a \over \p \beta} \, b(z) \, dz\, ,
\ee
where $\ointop_a$ is the anti-clockwise contour around $z_a=F_a(0;\beta)$ and
$K_1$ is another normalization constant. 
Note that $\p F_a/\p\beta$ is computed at fixed $w_a$, but after taking the derivative
we need to express this back in the $z$ coordinate system using \refb{e6.18new}.
Therefore the amplitude takes the form:
\be \label{epsiamp}
- K_1\int d\beta \, \sum_{a=1}^2 \ointop_a {\p F_a \over \p \beta} \, dz \, \left
\langle b(z) \, F_1\circ \p c(0) \, F_2\circ \p c(0)\, c\bar c V_T(i)\right\rangle\, ,
\ee
where $c\bar c V_T$ denotes the vertex operator of the closed string tachyon of energy 
$\omega$ and $F_a\circ \p c(0)$ is the conformal transform of $\p c(0)$ by the map $F_a$.
After computing the amplitude we can divide the result by $2\, g_s\, \sinh(\pi|\omega|)$ to
compute its contribution to $g(\omega)$.
We describe this computation in detail in appendix \ref{sc}, quoting here only the
final results for the contribution $g_{\psi^1}$ to $g_{\rm div}$:
\be \label{epsitot}
g_{\psi^1}(\omega) =
\Bigg[{2\over \pi}\, \wt\lambda^2 \, \int_{(2\wt\lambda)^{-1}}^1 \, d\beta\, (1+\beta^2)^{-1}
\, f(\beta)^2 + {\wt\lambda\over4\pi}\Bigg]\, .
\ee
The first term inside the square bracket is the contribution from Fig.~\ref{figpsione}(b)
and the second term is the contribution from Fig.~\ref{figpsione}(a). The constants
$K_0$ and $K_1$ drop out of the final computation since we compare the result with
the contributions where the $\psi^1$ in Fig.~\ref{figpsione} is replaced by the open
string tachyon. 
The latter contributions can be read out from \refb{eia53} and \refb{eca3} respectively.

\subsection{C-C-O  disk amplitude, zero mode field redefinition and the Jacobian} \label{scco2}

In this subsection we shall explore whether 
we need a field redefinition to relate the zero mode field $\phi$
of the open string field theory associated with the vertex operator $c\p X$ and the
collective coordinate $\wt\phi$ 
that translates the D-instanton along the time direction. This will be
done by studying the C-C-O disk amplitude with a pair of closed string tachyon vertex
operators carrying total energy $\omega$ and the open string vertex operator $c\p X$.
If the open string zero mode had been the same as the collective coordinate, then this
amplitude would be given by $-i\omega$ times the C-C disk amplitude, since the
dependence of the amplitude on the collective coordinate $\wt\phi$ is expected to
be of the form $e^{-i\omega\wt\phi}$. Therefore any
deviation from this form would have to be removed by a field redefinition and the
associated Jacobian would have to be taken into account in computing amplitudes.

Therefore we 
consider the C-C-O amplitude, with the closed string states 
corresponding to a pair of tachyons as in \S\ref{s2} and the
open string state corresponding to the zero mode $\phi$ associated with the
vertex operator $c\p X$. As in the previous cases, here by amplitude we shall
refer to the building block, obtained before getting the energy conserving delta
function or the $\NN \, e^{-1/g_s}$ factor. We shall insert the closed string vertex
operators at $i$ and $i\, y$ as in \S\ref{s2}, 
and the open string vertex operator at a point $x$ on the real axis, and 
express the amplitude as an integral over $x$ and $y$. The integrand is given by the integrand of the C-C
amplitude, multiplied by\footnote{The overall normalization has been adjusted to be $1/2\pi$ by choosing
suitable normalization of the open string zero mode $\phi$.}
\be\label{exmult}
{1\over 2\pi}\left[-{\omega_1\over x-i} + {\omega_1\over x+i} -{\omega_2\over x-i\, y}+{\omega_2\over x+i\, y}\right]=-{i\over \pi} \left[{\omega_1\over 1+x^2} + {\omega_2\, 
y\over x^2+y^2}
\right]\, .
\ee
Therefore if we integrate over $x$ for fixed $y$, we get the result:
\be
-i (\omega_1 +\omega_2) \,,
\ee
multiplying the integrand of the C-C amplitude. 
Since the coupling of the collective coordinate $\wt\phi$ to this order is
also supposed to be given by $-i(\omega_1+
\omega_2)$ times the
C-C amplitude, this would be consistent with the hypothesis that $\phi$ and $\wt\phi$ 
can be identified.

This analysis, however is only formal, since the C-C-O amplitude has divergences from different regions
of integration and they need to be treated following the replacement rule \refb{erp1}. 
Therefore we need to carefully
evaluate the 
contributions from the different regions shown in Fig.~\ref{figeight},
associated with the Feynman diagrams of Fig.~\ref{figfour}. 
Possible divergences are associated with the $q$ parameters of one or more internal
propagators in Fig.~\ref{figfour} approaching zero. 
To analyze this, we shall split the integrand into the sum of two parts -- one where
we multiply \refb{exmult} by the integrand of $f_{\rm finite}$ defined in \refb{effinite},
and the other where we multiply \refb{exmult} by the integrand of $f_{\rm div}$
defined in \refb{efdivexp}.
Now from
Fig.~\ref{figeight} one can easily 
see that all the potentially divergent regions I, I$'$,
II, II$'$, III and III$'$, associated with Feynman diagrams with at least one propagator,
correspond to the small $y$ region.   
On the other hand,
$f_{\rm finite}$ defined in \refb{effinite} has the property that it has no divergence from
the small $y$ region. We also see from \refb{exmult} that integration over $x$ for fixed
$y$ does not produce any additional singularity as $y\to 0$.
Therefore if we multiply the integrand of $f_{\rm finite}$ by \refb{exmult} and carry
out the integration over $x$ and $y$, we can  
exclude the region $y\le \eps$ from the $y$ integral at the cost of
making an error that will be suppressed by a power of $\eps$ and will
vanish in the $\eps\to 0$ limit. 
On the other hand, by the argument outlined in the
previous paragraph, 
the contribution to the $x$
integral for $y\ge \eps$ will just produce a multiplicative factor of 
$-i(\omega_1+\omega_2)$, reproducing the expected coupling of the collective
coordinate. 
Therefore the
problematic terms come from the second integral where we multiply the integrand
of $f_{\rm div}$ by \refb{exmult} and integrate over $x$ and $y$. 
This contribution to the C-C-O amplitude is given by:
\ben\label{emastersec}
&& \int_{-\infty}^\infty dx\int_0^1 dy\, 
{1\over 2\pi}\left[-{\omega_1\over x-i} +{\omega_1\over x+i} -{\omega_2\over x-i\, y}+{\omega_2\over x+i\, y}\right] \nonumber \\ && \hskip 1in \times\ 
4\,  g_s \,  \sinh(\pi|\omega_1|)\, \sinh(\pi|\omega_2|)\, \Bigg[
{1\over 2}
\, y^{-2}\,  (1+2\, \omega_1 \omega_2 y)
\Bigg]\, .
\een
A reader unconvinced by this argument can verify this
explicitly by replacing the second line of \refb{emastersec} by $y^\beta$ for 
$\beta\ge 0$,
and then carrying out the integration over $x$ and $y$ from different regions of
Fig.~\ref{figeight} following the same procedure as described in appendix \ref{sa}.
For $\beta\ge 0$, the result will always be given by $-i(\omega_1+\omega_2)$ times the
integral of $y^\beta$.

In appendix \ref{sa} we have analyzed the
contributions to \refb{emastersec} from the different regions shown in Fig.~\ref{figeight},
associated with the Feynman diagrams of Fig.~\ref{figfour}.  The result is:
\be \label{ecco5}
2 \, i \, (\omega_1+\omega_2) \, g_s \,  \sinh(\pi|\omega_1|)\, \sinh(\pi|\omega_2|) \left\{ 1 -\, {2\over \pi}
\, \wt\lambda 
- 2\, \omega_1\, 
\omega_2 \ln \wt\lambda^2\right\}\, .
\ee
The desired result for this is $-i(\omega_1+\omega_2)$ times the divergent part of
the C-C amplitude given in \refb{efdiv}:
\be\label{eunexpect}
i(\omega_1+\omega_2) \times 4 \, g_s \,  \sinh(\pi|\omega_1|)\, \sinh(\pi|\omega_2|) 
\, {1\over 2} \left\{ 1 - 2\, \omega_1\, 
\omega_2 \ln \lambda^2\right\}\, .
\ee
Note that we have put back the overall normalization factor
$4 \, g_s \,  \sinh(\pi|\omega_1|)\, \sinh(\pi|\omega_2|)$ that multiplies 
$f(\omega_1,\omega_2)$ in the disk two point amplitude.
We now
see that \refb{ecco5} differs from \refb{eunexpect} by the term:
\be \label{ecco5extra}
-2 \, i \, (\omega_1+\omega_2) \,  g_s \,  
\sinh(\pi|\omega_1|)\, \sinh(\pi|\omega_2|) \left\{ {2\over \pi}
\, \wt\lambda 
+ 2\, \omega_1\, 
\omega_2 \ln {\wt\lambda^2\over \lambda^2} \right\}\, .
\ee
This means that the zero mode dependence of the effective action
is not of the form $e^{-i(\omega_1+\omega_2)\phi}$ times the amplitude
without the zero mode. 
Let us denote by $\Phi_C(\omega,P)$   the closed string tachyon field of energy $\omega$
and Liouville momentum $P$,
with positive $\omega$ describing outgoing states and negative $\omega$ describing
incoming states.
Then the extra term in the effective action
to order $\phi$ is of the form\footnote{We are using the convention that 
if $S$ is the action then the weight factor in the path integral
is $e^S$. 
In that case a term in the action contributes to the amplitude without any
additional minus sign or factors of $i$.
Also
note that when we compute the amplitude from \refb{eextra}, we shall get an extra factor
of 2 due to two possible ways of contracting $\Phi_C$ with the
external closed string states. This would account for the factor of 2 in \refb{ecco5extra}.}
\ben \label{eextra}
S_{\rm extra} &=&  -i \,  g_s \, \phi\, \int {d\omega_1}\, dP_1\, {d\omega_2} \, d P_2\,
(\omega_1+\omega_2)  \,  \sinh(\pi|\omega_1|)\, \sinh(\pi|\omega_2|) \left\{{2\over \pi}
\, \wt\lambda 
+ 2\, \omega_1\, 
\omega_2 \ln {\wt\lambda^2\over \lambda^2} \right\} \nonumber \\ && \hskip 2in
\times\, \Phi_C(\omega_1,P_1) \, \Phi_C(\omega_2,P_2)\, .
\een
Note that the integrand could be multiplied by some function of $\{\omega_i\}$ and
$\{P_i\}$ that reduces to 1 for on-shell closed string fields. But since eventually
we shall use this to compute on-shell amplitudes, this factor will not be important and
we shall drop it.
We shall now look for an appropriate relation between the open string field $\phi$ and
the collective coordinate $\wt\phi$ so that the effective 
action has $\wt\phi$ dependence proportional to $e^{i\omega\wt\phi}$. 
This will allow us to recover the standard energy conserving delta function after integration over $\wt\phi$.

For closed string tachyon of energy $\omega$, the
one point function on the disk is given by,
\be\label{edisk}
2 \, \sinh(\pi |\omega|)\, .
\ee
The C-O amplitude with a closed string tachyon and an open string zero mode $\phi$
will be given by $-i\omega$ times  this 
amplitude. This shows the presence of a term in the
effective action of the form
\be \label{edefact}
-2 \,  i\,  \phi \int\,  {d\omega} \, dP  \, \omega \, \sinh(\pi|\omega|) \, 
 \Phi_C(\omega,P)\, .
\ee
Since there is no divergence in this amplitude, this result is not modified. 
We now make a field redefinition
\be\label{efred}
\phi=\wt\phi +  2 \, g_s\,\wt\phi\, \int {d\omega'} \, dP'\, \sinh(\pi|\omega'|) \, a(\omega') \, \Phi_C(\omega', P')\, ,
\ee
for some function $a(\omega)$ that will be determined shortly.
Upon substitution into \refb{edefact}, this produces \refb{edefact} with
$\phi$ replaced by $\wt\phi$, plus an extra term:
\ben\label{enew}
&&  - 4 \, i\, g_s\, \wt\phi \int {d\omega}\, dP\, \int {d\omega'} \,  dP' \, 
\, \omega\,
\sinh(\pi|\omega|) \Phi_C(\omega,P)
\, \sinh(\pi|\omega'|) \, a(\omega') \, \Phi_C(\omega',P') \nonumber \\
&=&   -2\, i\, g_s\, \wt\phi \int {d\omega} \, dP\, 
\int {d\omega'}  \, dP'\, 
\sinh(\pi|\omega|)  \, \sinh(\pi|\omega'|) \, \left\{\omega\, a(\omega') 
+ \omega'\, a(\omega)\right\} \nonumber \\ && \hskip 2in \times
\, \Phi_C(\omega,P)
\, \Phi_C(\omega',P')\, , 
\een
where in the second step we have symmetrized the integrand in $\omega$ and $\omega'$.
On the other hand, to order $g_s$, \refb{eextra} retains the same form with
$\phi$ replaced by $\wt\phi$. If we choose
\be \label{efchoice}
a(\omega) = - \left[{1\over \pi}\, \wt\lambda 
+ \omega^2\ln {\wt\lambda^2\over \lambda^2}
\right]\, ,
\ee
then \refb{enew} cancels the unwanted extra term \refb{eextra} in the effective
action. Therefore we can identify  $\wt\phi$ as the collective coordinate to this
order.

When we express the path integral over $\phi$ in terms of path integral over
$\wt\phi$ using the field redefinition \refb{efred}, we get a
Jacobian
\ben\label{ejac}
&& 1+2 \, g_s\, \int {d\omega'} \, dP'\, 
\sinh(\pi|\omega'|) \, a(\omega') \, \Phi_C(\omega',P') \nonumber \\
&=& \exp\left[2 \, g_s \,
\int {d\omega'} \, dP'\, \sinh(\pi|\omega'|) \, a(\omega') \, \Phi_C(\omega',P')+\OO(g_s^2)
\right]\, .
\een
This can be interpreted as a 
new term in the effective action, given by:
\ben
&& 2 \,g_s\, \int {d\omega'} \, dP'\,  \sinh(\pi|\omega'|) \, a(\omega') \, \Phi_C(\omega',P')
\nonumber \\
&=&  -2 \, g_s \, \int {d\omega} \, dP\, \sinh(\pi|\omega|) \, \left[{1\over \pi}\, 
\wt\lambda 
+ \omega^2\ln {\wt\lambda^2\over \lambda^2}
\right] \, \Phi_C(\omega,P)\, .
\een
This will give an additional contribution to the one point function of closed string
tachyon besides the ones given by Feynman diagrams:
\be \label{eca1pre}
 -2 \,   g_s \,  \sinh(\pi|\omega|) \, \left[{1\over \pi}\, 
\wt\lambda 
+ \omega^2\ln {\wt\lambda^2\over \lambda^2}
\right] \, .
\ee
Comparing this with \refb{eblock7} we see that 
\refb{eca1pre} can be interpreted as an additional
contribution to $g(\omega)$, given by:
\be \label{eca1}
g_{\rm jac}(\omega) =-  \left[{1\over \pi}\, 
\wt\lambda 
+ \omega^2\ln {\wt\lambda^2\over \lambda^2}
\right] \, .
\ee

We end this section with two remarks: 
\begin{enumerate}
\item The field redefinition described here
is only one of the many field redefinitions that $\phi$ undergoes. In general 
\refb{efred} will contain  
terms involving other closed string fields and higher powers of closed string
fields. These will affect higher order results but will not affect the closed string tachyon
one point function to the order that we are doing our computation.
There will also be terms that involve
redefinition of $\phi$ involving higher powers of $\phi$\cite{1702.06489}.  
Such terms will lead
to Jacobian factor involving non-linear terms in $\wt\phi$, which can be regarded as 
a contribution to the effective action for $\wt\phi$. However all such terms must cancel 
with explicit
loop contribution to the effective action for $\phi$, so that when the dependence of the
amplitude on $\wt\phi$ takes the form $e^{-i\omega\wt\phi}$, the integration measure 
becomes $\wt\phi$ independent. In this case the integration over $\wt\phi$ will produce 
the $\delta(\omega)$ factor.
\item The notion of effective action used here is different from the notion of closed string
effective action discussed in \cite{2012.00041}. Here the effective action refers to the sum over
contributions from connected world-sheet before integrating over the collective modes,
while in the analysis of \cite{2012.00041} 
the effective action includes disconnected world-sheets
and also include the final integration over the collective modes. Therefore in the  analysis of
\cite{2012.00041} the Jacobian given in \refb{ejac} will appear as a multiplicative 
contribution to the integrand that, after integration over the collective coordinates,
gives the effective action. The difference between
the two papers can be traced to the fact that here we are trying to analyze the building
blocks of the amplitude instead of the full amplitude.
\end{enumerate}

\subsection{C-O-O-O disk amplitude, ghost field redefinition and the Jacobian}
\label{sghost}

In this subsection we shall find  the
relation between the infinitesimal gauge transformation parameter $\theta$
corresponding to the state 
$|0\rangle$ in open string field theory and 
the rigid $U(1)$ transformation parameter
$\wt\theta$.
This will be done as follows. Let us imagine 
bringing a second D-instanton, which we shall call the spectator, on top of the original
D-instanton, and consider an open string 
that stretches from the original D-instanton to the spectator D-instanton. 
We shall compare the transformation laws of this state under the gauge 
transformation of string field theory with the expected transformation laws under
rigid $U(1)$ to find the relation between the two gauge transformation parameters.
For definiteness we shall work with a particular field $\xi$ that multiplies the vacuum state
$|0\rangle$
of the open string stretched
between the two D-instantons. 
The same analysis may be carried out with
any other physical state of the open string stretched between the two D-instantons 
{\it e.g.} the field described by the
vertex operator $c\p X$.
We have verified that this gives the same relation between $\theta$ and $\wt\theta$.

The conjugate 
anti-field $\xi^*$ of $\xi$
will have vertex operator $c\p c \p^2 c/2$ and will stretch from the spectator 
D-instanton to the original D-instanton. The states associated with the fields
$\xi$ and $\xi^*$ may be assigned Chan-Paton 
factors $\pmatrix{0 & 1\cr 0 & 0}$ and $\pmatrix{0 & 0\cr 1 & 0}$ respectively, while
the open string states with both ends lying on the original D-instanton will have 
Chan-Paton factor $\pmatrix{1 & 0\cr 0 & 0}$. 
Now from the gauge transformation laws of the string fields discussed
below \refb{esftgauge}, it follows that the gauge transformation of $\xi$ under the
gauge transformation generated by $\theta$ 
is given by the second derivative of the Wilsonian effective action with
respect to $\xi^*$ and $\psi^2$, where $\psi^2$ is the string field multiplying 
$|0\rangle\otimes \pmatrix{1&0\cr0&0}$.
The lowest order contribution to this gauge transformation law comes from the
$\xi$-$\xi^*$-$\psi^2$ coupling in the action, whose coefficient is given by:
\be \label{eKapp}
\langle I(z_1) \{c\p c \p^2 c(z_2)/2\} I(z_3) \rangle \, Tr\left[
\pmatrix{0 & 1\cr 0 & 0}
\pmatrix{0 & 0\cr 1 & 0}\pmatrix{1 & 0\cr 0 & 0}
\right] = - K_0 \, ,
\ee
where we have normalized the ghost correlator as in \refb{e615}
\be\label{e780a}
\langle c(z_1) c(z_2) c(z_3) \rangle = K_0 (z_1-z_2) (z_2-z_3) (z_1-z_3)
\, .
\ee 
\refb{eKapp} corresponds to the presence of a term:
\be \label{eghost0}
- K_0   \, \xi\, \xi^*\, \psi^2\, ,
\ee
in the action.
This in turn implies that under the gauge transformation generated by the parameter
$\theta$, $\delta\xi = -K_0 \theta\xi$.
On the other hand, under a rigid $U(1)$ gauge transformation by an infinitesimal
parameter $\wt\theta$, we expect $\delta\xi=i\wt\theta \xi$. 
Therefore, to this order $\theta$
and $\wt\theta$ 
are related by simple multiplicative constant: $\theta=-i\wt\theta/(K_0 )$. 
In this analysis we have
not been very careful in keeping track of the overall signs, which depend on the 
precise arrangement of the different fields in the action and the order in which we take
the derivatives, since some of the fields are grassmann odd. However such
sign errors will be eventually cancelled since the important quantity is the ratio of
the coupling described in \refb{eghost0} and the coupling in the presence of closed
string tachyon, as given in \refb{eghost1} below.

When we take into account other terms in the action containing more fields, 
the rigid $U(1)$ gauge transformation law of $\xi$ remains unchanged,
but the  string field theory
gauge transformation law given by the second derivative of the action
with respect to $\psi^2$ and $\xi^*$, gets modified. This causes $\theta$
and $\wt\theta$ to be related by a field dependent transformation. We shall now
study a
particular
example of this.
Let us denote by $A(\omega)$ the disk amplitude (with contribution from
massless internal open string states removed) with one closed string tachyon of
energy $\omega$ and three open string states described by the vertex operators
$I$, $c\p c \p^2c/2$ and $I$, carrying the same Chan-Paton factors as in \refb{eKapp}. 
This translates to a term in the effective action
of the form:
\be\label{eghost1}
\int {d\omega}\, dP\,  A(\omega) \Phi_C(\omega,P)  \, \xi \, \xi^*\, \psi^2\, ,
\ee
where $\Phi_C(\omega,P)$ is the closed string tachyon field of energy $\omega$ and
Liouville momentum $P$. 
As in \refb{eextra}, the integrand in \refb{eghost1} can contain an extra factor that reduces
to 1 when the closed string field is on-shell, but we shall not include this since we shall
use this formula for on-shell closed string states.
Adding this to \refb{eghost0} we get the following contribution to the effective action:
\be 
- K_0 \, \xi \, \xi^*\,  \psi^2 \, \left[1 - 
K_0^{-1} \int {d\omega} \, dP\, A(\omega) \Phi_C(\omega,P)\right]\, .
\ee
Taking derivative with respect to $\psi^2$ and $\xi^*$, we get the modified gauge
transformation law of $\xi$ under the gauge transformation parameter $\theta$:
\be \label{etrxyyy}
\delta \xi =- K_0 \, \theta \,\xi \left[1 -K_0^{-1}  
\int {d\omega} \, dP\, A(\omega) \Phi_C(\omega,P) 
\right]\, .
\ee
Comparing this with $\delta \xi=i\wt\theta \xi$ we see that $\theta$ and $\wt\theta$
are related by:
\be\label{e6.50a}
\theta = -{ i} \, K_0^{-1}\,  \wt\theta \left[1 + K_0^{-1}
 \int {d\omega} \, dP\, A(\omega) \Phi_C(\omega,P)  +\cdots
\right]\, ,
\ee
where $\cdots$ denote higher order corrections that we shall ignore.
Therefore we have
\be 
\int d\theta = \hbox{constant} \, \times \, \left[1 + K_0^{-1}
 \int {d\omega}\, dP A(\omega) \Phi_C(\omega,P) 
\right] \, \int d\wt\theta\, .
\ee
Hence dividing the path integral over the open string fields
by $\int d\theta$ can be regarded as a division
by $\int d\wt\theta$ and a multiplication by 
$\left[1 -K_0^{-1}
 \int {d\omega}\, dP\, A(\omega) \Phi_C(\omega,P) 
\right] =\exp[- K_0^{-1}
 \int {d\omega}\, dP\, A(\omega) \Phi_C(\omega,P) ]$ up to higher order terms. 
 The effect of division by
 $\int d\wt\theta$ generates a constant factor since $\int d\wt\theta$
 gives the volume of the rigid U(1) gauge group. This can be absorbed into a
 redefinition of the overall normalization constant $\NN$ that accompanies all
 amplitudes. Therefore we can ignore this term. This 
 leaves us with a correction to the action of the form:
 \be \label{eactionghost}
 - K_0^{-1}
 \int {d\omega}\, dP\,  A(\omega) \Phi_C(\omega,P)\, .
 \ee

 $A(\omega)$ has been evaluated in appendix \ref{sb}, leading to the result:
\be\label{eappfina}
-K_0^{-1} \, A = -{2\over \pi}\, g_s\, \sinh(\pi|\omega|)
\, \left[{\pi\over 2} - {\wt\lambda\over 2}\right] \, .
\ee
This leads to an additional contribution to the closed string tachyon one point function
due to the extra term \refb{eactionghost} in the effective action.
Comparing this with \refb{eblock7} we get the contribution to $g(\omega)$ from the
ghost jacobian:
\be \label{eghosttot}
g_{\rm ghost}(\omega) = -
\left[{1\over 2} - {\wt\lambda\over 2\pi}\right]\, .
\ee

We note in passing that $\xi$ is the ghost field associated with the gauge 
transformation parameter $I\otimes \pmatrix{0 & 1\cr 0 & 0}$. Therefore, 
our calculation can also
be interpreted as computing the algebra of the gauge transformations generated by
$\theta$ and $\xi$, and finding appropriate redefinition of parameters that maps the
algebra to the standard SU(2) algebra expected on a pair of D-instantons.

Before concluding this section, let us discuss one subtlety that could have affected our
analysis. Usually when one tries to relate the standard gauge transformation laws of the
form $\delta\xi=i\wt\theta \xi$ to the string field theory gauge transformation laws as given in
\refb{etrxyyy}, we could add to \refb{etrxyyy} trivial gauge transformation proportional to the
equations of motion\cite{9110038}. Therefore one could wonder if it is possible to change the
coefficient of the second term inside the square bracket in \refb{etrxyyy} by adding such
trivial gauge transformation, and modify \refb{e6.50a}. For this to happen, there must be
a field $\chi$ whose equation of motion is proportional to $\xi\Phi_C(\omega,P)$. Using 
various conservation laws one can see that
 the only candidate for $\chi$ is the coefficient of the open string state $c_1c_{-1}|0\rangle\otimes
 \pmatrix{0&0\cr 1&0}$. However explicit computation along the line of appendix 
 \ref{sc} shows that the $\chi$-$\xi$-$\Phi_C$ coupling on the disk vanishes, 
 and therefore we cannot modify \refb{etrxyyy} by adding to it trivial gauge transformation.
 The vanishing of the $\chi$-$\xi$-$\Phi_C$ coupling can in turn be traced to the fact that
 $\xi$ and on-shell $\Phi_C$ represent BRST invariant states, whereas $\chi$ represents
 a BRST trivial state.

\subsection{Total contribution} \label{stotal}

We now collect the various contributions to $g_{\rm div}(\omega)$ as given in
\refb{eidiv},  \refb{epsitot},  \refb{eca1} and 
\refb{eghosttot}:
\ben \label{egdivtotal}
g_{\rm world} &=& 
\bigg[ - {2\over \pi}\, \wt\lambda^2\, 
 \int_{(2\wt\lambda)^{-1}}^{1} d\beta \, (1+\beta^2)^{-1}\,  f(\beta)^2 +
 {\wt\lambda\over 4\pi} 
 + {1\over 2} \omega^2 \, \ln {\alpha^2 \wt\lambda^2 \over 4}\bigg] \, , \nonumber\\
 g_{\psi^1}(\omega) &=&
\Bigg[{2\over \pi}\, \wt\lambda^2 \, \int_{(2\wt\lambda)^{-1}}^1 \, d\beta\, (1+\beta^2)^{-1}
\, f(\beta)^2 + {\wt\lambda\over4\pi}\Bigg]\nonumber\\
g_{\rm jac}(\omega) &=& -  \left[{1\over \pi}\, 
\wt\lambda 
+ \omega^2\ln {\wt\lambda^2\over \lambda^2}
\right]\, , \nonumber\\
g_{\rm ghost}(\omega) &=&  -
\left[{1\over 2} - {\wt\lambda\over 2\pi}\right]\, .
 \een
Adding these, we get the following
total contribution to $g_{\rm div}(\omega)$: 
\be\label{egdiv}
g_{\rm div}(\omega)=-{1\over 2}+ {1\over 2}\, \omega^2 \, \ln{\lambda^2\over 4}\, .
\ee
Note that $g_{\rm div}$ stands for any contribution to the annulus one point function that is
not already included in $g_{\rm finite}$ defined in \refb{egfinite}. 
Comparing this with \refb{efgpar} we get
\be \label{egconstants}
A_g= -{1\over 2}, \qquad B_g=  {1\over 2}
\ln {\lambda^2\over 4}\, .
\ee

\sectiono{Strategy for computing the correction to the D-instanton action} \label{s5.4} \label{ecompC}

In this section we shall outline the strategy for computing 
the remaining building block that contributes to the one instanton correction 
to closed
string tachyon scattering to the first subleading order. This involves
a zero point function on a surface of Euler number $-1$. There
are two different surfaces of the latter kind -- a disk with two holes and a 
torus with a hole. If we denote the total contribution to this building block 
by $C\, g_s$, then the net contribution to the
$(n+1)$-point tachyon amplitude from the product of $(n+1)$ disk one point
functions and a zero point function on surfaces of Euler number $-1$ 
will be given by:
\be\label{eCfirst}
\NN\, e^{-1/g_s}\, C\, g_s\, \prod_{i=1}^{n+1} \left\{ 2\sinh(\pi|\omega_i|)\right\}\, .
\ee
Adding \refb{eCfirst} to the leading order contribution 
$\NN\, e^{-1/g_s}\, \prod_i \left\{ 2\sinh(\pi|\omega_i|)\right\}$,  we see that
the effect of \refb{eCfirst} is to make the replacement,
\be\label{ecomp1}
e^{-1/g_s}\to e^{-1/g_s} (1+C\, g_s) = \exp\left[-{1\over g_s}\left\{1 - C\, g_s^2 
+\OO(g_s^3)\right\}
\right]\, .
\ee
Therefore the instanton action is modified to
\be \label{einst1}
-{1\over g_s} \left\{1 - C\, g_s^2 +\OO(g_s^3)\right\}\, .
\ee
Furthermore, this modification is independent of which amplitude we consider
-- this is the overall factor that multiplies the one instanton contribution to
any amplitude.

Now \refb{einst1} shows that the value of $C$ can be changed by a redefinition
of $g_s$ of the form $g_s\to g_s(1 + a g_s^2+\OO(g_s^4))$ for some constant $a$. 
In a critical string theory, 
$g_s$ is the expectation value of the dilaton field.
Since different versions of
open-closed string field theory are related by field redefinition,  the string coupling
$g_s$ used in different versions of string field theory are also related by redefinition
of the kind described above\cite{1209.5461,1411.7478}.
If the same story holds in two dimensional string theory, then it would seem that
$C$ takes
different values in different versions of string field theory. 
On the other hand, the matrix model yields a definite
result for the tachyon scattering amplitudes, and determines $C$ via \refb{emulti1int} 
without any ambiguity. To resolve this apparent contradiction, let us
recall that while comparing the matrix model results with the string theory results,
\cite{1907.07688} made a specific choice of the relation between the matrix 
model parameter
$\mu$ and the string coupling $g_s$:
\be\label{egmureln}
2\pi \mu = {1\over g_s}\, .
\ee
Let us take \refb{egmureln} as the definition of $g_s$, and denote by $\tilde g_s$ the string
coupling in some version of string field theory. 
Now in the matrix model one can compute perturbative contribution to the 
closed string tachyon 
scattering amplitude with three or more tachyons in the external state. 
This will have an expansion in powers of $g_s^2$.
In the dual string theory the same amplitude will have an expansion in powers of
$\tilde g_s^{2}$. These expansion coefficients may or may not depend on the 
version of string field theory we use, depending on whether or not $\tilde g_s$ in
different versions of string field theory are different. 
In either case, in any given version
of string field theory, we can determine the relation between $\tilde g_s$ and $g_s$
by comparing any particular perturbative amplitude in the matrix model and 
string theory. On general ground we expect this relation to take the 
form\footnote{In principle there may also be corrections of order $e^{-1/g_s}$ on
the right hand side of \refb{eredef}. Such a term, if present, would have
given rise to an additional contribution to the three point function 
of order $e^{-1/g_s}$ multiplying the
tree level 3-point function, and spoil the agreement between the matrix model and
world-sheet results at order $g_s\, e^{-1/g_s}$.
Therefore they do not seem to be present.
}
\be \label{eredef}
\tilde g_s=g_s(1+\OO(g_s^2))\, .
\ee 
Once this is done, we can use the same string field theory to
compute the zero point function on surfaces of Euler number $-1$, use this to determine
the quantum corrected instanton action as a function of
$\tilde g_s$ and then express this in terms of $g_s$ using the relation between $g_s$
and $\tilde g_s$ determined from perturbative amplitudes. This will generate an
unambiguous expansion of the quantum corrected D-instanton action in powers of
$g_s$, and determine $C$
unambiguously. Note also that the difference between $g_s$ and $\tilde g_s$ 
will not affect the functions $f$ and
$g$ appearing in \refb{emulti1int} since they are 
corrections of order $g_s$ to the leading order
instanton induced amplitudes.

We have not computed $C$ using this procedure. However, we shall now argue that
if we assume the equality of the quantum corrected D-instanton action in two dimensional
string theory and the quantum corrected instanton action in the matrix model, then
$C$ must vanish. 
In the matrix model, the dynamics is that of free fermions moving in an inverted
harmonic oscillator potential. Therefore the fluctuations around the instanton solution
have purely quadratic action and the instanton action does not get renormalized 
(except for the one loop determinant that is related to the overall normalization 
constant $\NN$). 
Therefore 
the overall $\mu$ dependent
multiplicative factor associated with the instanton will be given by the leading order
result:
\be\label{ecomp2}
e^{-2\pi \mu} \, .
\ee
Comparing this with the right hand side of \refb{ecomp1}, we get
\be\label{ecomp3}
2\, \pi\, \mu = {1\over g_s} \left\{1 - C\, g_s^2 +\OO(g_s^4)\right\}\, .
\ee
On the other hand, $g_s$ is defined by \refb{egmureln} without any correction.
By comparing this with \refb{ecomp3} we get
\be\label{ecz}
C=0\, .
\ee

It should be noted however that the quantum corrected instanton action, by itself, is
not an observable, and therefore this argument is not watertight. Nevertheless, agreement
of \refb{ecz} with the numerical results based on the comparison of the
S-matrix elements in the matrix model and the world-sheet theory\cite{private} provides
a verification of
the argument given
above.

\sectiono{Some open problems} \label{sopen}

Clearly, the most urgent issue is to understand the origin of the discrepancy between the
value of $A_g$ obtained from the string field theory computation and the result obtained by
comparison with the matrix model. We discuss various possible sources of this discrepancy,
in order of their likelihood.
\begin{enumerate}
\item The most likely explanation is that there is an error in the string field theory
computation of $A_g$. However we have not been able to identify the origin of this
error. Furthermore the string field theory results pass various 
internal consistency checks
-- the final result is independent of the string field theory parameters $\wt\lambda$,
$\alpha$ and the function $f(\beta)$. 
Some further consistency checks have been described in
appendix \ref{se}. Although not reported here, we have also performed the computation with
a different choice of local coordinates for the O-O-O vertex, where we take the local
coordinates at the punctures at 0, 1 and $\infty$ to be 
$\alpha z$, $\alpha(z-1)$ and $-\alpha/z$ respectively and then average over cyclic permutations
generated by $z\to 1/(1-z)$. Consequently the C-O-O vertex will also have to be
defined by averaging over permutations. However, the final result for $A_g$ does
not change. We have also used the procedure of \S\ref{salternative} to reproduce our results.
In fact this method can also be used to get the result without assuming $\alpha$ and / or
$\lambda$ to be large, and the final result remains unchanged.
\item Another possibility is that for the D-instanton, the world-sheet theory 
has an additional contribution that is not related to divergences from the boundary
of the moduli space. This could give an 
additive constant of 1/2 in the
computation of $g_{\rm finite}$ and cancel the value $-1/2$
of $A_g$ that we find, 
restoring the equality between the string field theory results and the matrix model results.
If this is the case, then it will be important to understand what the missing contribution
is, since it is likely to affect the D-instanton contribution to other string theories as well.
\item A third possibility is that the matrix model result for the instanton induced scattering
amplitude may have an additional term that makes the result agree with the choice
$A_g=-1/2$ in 
\refb{emulti1intfinite}. This would restore the agreement between 
the results of the
world-sheet theory and the matrix model.
\item The final (unlikely) possibility is that the duality between the bosonic string theory and
the matrix model does not hold in the form suggested in \cite{1907.07688}.
\end{enumerate}

We end this paper by listing some other open problems.

\begin{enumerate}
\item The divergences we discussed here also exist for D-instantons in superstring 
theories\cite{2002.04043} and the method described here can be used to 
get finite, unambiguous results.
Therefore we could use this technique to find systematic $g_s$ expansion of
D-instanton induced amplitudes. On the other hand, using non-perturbative duality
symmetries, many such corrections have already been 
predicted\cite{9701093,9704145,9706175,9707018,9710078,9903113,9910055,
0510027,0708.2950,0712.1252,1001.3647,
1001.2535,1004.0163,1111.2983,1308.1673,1404.2192,1502.03377,1510.07859,
1712.02793}. 
It will be
interesting to verify these predictions using direct string theory computation.
\item Matrix model dual to two dimensional string theory is described by free fermions
in inverted harmonic oscillator potential. The analysis of 
\cite{1907.07688,1912.07170} identified the two dimensional bosonic string theory with
the particular vacuum where the exact energy eigenstates representing the scattering
states incident from the left are filled up to the fermi level, 
but the scattering states incident from the right
are empty. Refs.\cite{0307083,0307195} 
proposed a different correspondence where they identified
two dimensional type 0B theory to the system where both sides are filled up to the
Fermi level. It will be interesting to compute D-instanton induced amplitudes in type 0B 
theory to verify this. This would also establish that the bosonic string theory and type
0B string theory are different vacua of the same underlying theory, at least in two
dimensions.
\item In our analysis we have not fixed the overall normalization constant $\NN$ that
appears in \refb{emulti1int}. This measures the relative normalization between the
path integral measure in the D-instanton background and the path integral measure
around the vacuum. One way to fix this will be to begin with a configuration of unstable
D-branes so that both the vacuum and the D-instanton are solutions in this theory. 
In this case the path integral over the fields in the original theory with unstable D-branes
will in principle fix the relative normalization in the path integral measure in the
instanton sector and the vacuum sector. For two dimensional string theory we can 
in fact start with the D-instanton and regard the perturbative vacuum as the
minimum of the open string 
tachyon potential. Therefore the integration measure of the open-closed
string field theory on the D-instanton should in principle 
be able to determine the integration 
measure for closed string field theory around
the perturbative vacuum. This in turn would 
determine the normalization constant $\NN$.
\item Perturbative string theory is expected to be non-perturbatively incomplete.
This is clearly demonstrated by the example of two dimensional bosonic and type 0B
string theories, which should have the same perturbative amplitudes but different
non-perturbative amplitudes. Since we now also have a fully systematic 
perturbation expansion for D-instanton induced amplitudes, one might ask: 
does this help make
string theory have unique non-perturbative extension, possibly with the help of
resurgence\cite{1206.6272,1511.05977,1802.10441}? 
If not, what else is needed?
\end{enumerate}

\bigskip

\noindent {\bf Acknowledgement:} I wish to thank Bruno Balthazar, Nathan Berkovits, 
Carlo Maccaferri, Victor Rodriguez, Xi Yin and
Barton Zwiebach for useful discussions. I specially thank Bruno Balthazar, 
Victor Rodriguez and Xi Yin for sending me the unpublished
numerical results for $A_f$, $A_g$ and
$C$.
Part of these results was reported at the online `Workshop on Fundamental Aspects of 
String Theory'  organized by ICTP-SAIFR and `Strings 2020' organized by the University of
Cape Town. I thank the organizers of these conferences for 
giving me an opportunity to present the results there.
This work was
supported in part by the 
J. C. Bose fellowship of 
the Department of Science and Technology, India and the Infosys chair professorship. 

\appendix

\sectiono{Disk two point function in general coordinate system} \label{sd}

In this appendix we shall repeat the analysis of \S\ref{s2} 
using a more general local coordinate
system at the open string puncture of the C-O interaction vertex 
than what was given in \refb{e1alt}. 

We insert the closed string puncture at $i$ and
the open string puncture at 0 on the
upper half plane as usual, but choose the local coordinate at the open string puncture
to be:
\be \label{elocalwz}
w = {\lambda z \over 1 + \gamma z}\, ,
\ee
for some constant $\gamma $. 
The sewing relation $ww'=-q$ now gives
\be 
\lambda^2 {z \over 1 + \gamma  z} {z' \over 1 + \gamma  z'}=-q\, .
\ee
In the $z$ plane the closed string 
punctures at $z=i$ and $z'=i$ are located at, respectively,
\be
z_1=i, \qquad z_2 = i\, u \, {1 + i\, \gamma \over 1-iu\gamma +u\gamma ^2}, \qquad u={q\over \lambda^2},
\qquad 0\le u\le \lambda^{-2}\, .
\ee
We now introduce a new coordinate $\hat z=(z-a)/ (1+a\, z)$ for some real
constant $a$. This maps the closed string punctures to
\be
\hat z_1=i, \qquad \hat z_2= {z_2-a\over 1+a\, z_2}\, .
\ee
Our goal is to adjust $a$ so that $\hat z_2$ is purely imaginary and then identify $\hat z_2$
as $i\, y$. Since it follows from the analysis of \S\ref{serror} that we need the relation
between $u$ and $y$ to the first subleading order in the expansion in powers of $u$,
we can work only to this order.
We get
\be
a=-\gamma  u - \gamma  u^2 + \gamma ^3 u^2 +\OO(u^3), \qquad
y=u - 2 \gamma ^2 u^2 +\OO(u^3)\, .
\ee
Therefore the range $0\le u\le\lambda^{-2}$ translates to $0\le y\le \lambda^{-2}
-2\, \gamma ^2 \lambda^{-4}$.
In this range we change variable from $y$ to $q=u\lambda^{2}$ in \refb{efdivor} 
and apply the
replacement rule \refb{erp1}, and in the rest of the range of integration we carry out the
$y$ integration explicitly. This gives the following result for \refb{efdivor}:
\ben\label{egenpsi}
{1\over 2}  \int_0^1 dy\, y^{-2} (1+2\, \omega_1\omega_2 y)  \hskip -.1in
&=&   \hskip -.1in {1\over 2}  \int_0^1 {dq} \, \left\{\lambda^2 q^{-2} +2\, \omega_1\omega_2 q^{-1}\right\}+
{1\over 2} \, 
\int_{\lambda^{-2} -2 \gamma ^2 \lambda^{-4}}^1 dy\, y^{-2} (1+2\, \omega_1 \omega_2 y)
\nonumber \\   \hskip -.1in &\to&   \hskip -.1in
-{1\over 2} \lambda^2 + {1\over 2} {\lambda^2 \over 1-2\, \gamma ^2 \lambda^{-2}} -{1\over 2}
+\omega_1 \omega_2 \ln\lambda^2 \nonumber \\   \hskip -.1in
&=&    \gamma ^2-{1\over 2} \left\{ 1 - 2\, \omega_1\, 
\omega_2 \ln \lambda^2\right\}\, ,
\een
where in all the steps we have ignored terms containing negative powers of $\lambda$.
This differs from \refb{enetcc} by the additive term $\gamma ^2$. We shall now show that this is
cancelled by the $\psi^1$ exchange contribution to Fig.~\ref{figthree}(a).

As usual, in order to avoid normalization issues, it will be simplest to compare the
$\psi^1$ exchange contribution 
with the tachyon exchange contribution, given by the first term in the second
line of \refb{egenpsi}. Since it follows from \refb{ekinprop} that 
the $\psi^1$ propagator is 1/2
times the tachyon propagator, and we see from
the first term in the second line of \refb{egenpsi} that the tachyon 
exchange contribution
is $-\lambda^2/2$, 
we can express the $\psi^1$ exchange contribution as:
\be\label{epsirat}
-{1\over 4} \, \lambda^2 \, \left( {C_{\Phi \psi^1}\over C_{\Phi \psi^0}}\right)^2\, ,
\ee
where $C_{\Phi\psi^0}$ and $C_{\Phi\psi^1}$ are respectively the disk two point function 
of a closed string tachyon and an open string tachyon associated with the
vertex operator $c$, and the disk two point 
function of a closed string tachyon  
and the open string $\psi^1$ field associated with the vertex operator $\p c$. 
To compute these, let
us express \refb{elocalwz} as:
\be
z=F(w), \quad F(w) = {w\over \lambda-\gamma  w}\, .
\ee
This gives the conformal transforms of the open string tachyon and $\psi^1$
vertex operators:
\be\label{eration11}
F\circ c(0) = (F'(0))^{-1} c(F(0))= \lambda \, c(0), \quad 
F\circ \p c(0) = \p c (F(0)) - {F''(0)\over F'(0)^2} c(F(0))= \p c(0) - 2\ \gamma \, c(0)\, .
\ee
Now as shown in \refb{eocvanish}, the correlation function of a closed string tachyon
vertex operator at $i$ and $\p c(0)$ vanishes. Therefore we see from 
\refb{eration11} that $C_{\Phi \psi^0}/ C_{\Phi \psi^1}= -2 \gamma  \lambda^{-1}$.
Substituting this into \refb{epsirat} we arrive at the $\psi^1$ exchange 
contribution:
\be
-{1\over 4} \, \lambda^2 \, \left( {2 \gamma \over \lambda}\right)^2 = - \gamma ^2\, .
\ee
This exactly cancels the first term on the right hand side of \refb{egenpsi}, giving
back the original result \refb{efdiv}.

\sectiono{$\psi^1$ exchange contribution to the annulus one point
function} \label{sc}

In this appendix we shall compute the 
$\psi^1$ exchange contribution to the annulus one point
function discussed in \S\ref{spsi}.
The relevant diagrams are the ones
shown in Fig.~\ref{figpsione}. 

Let us begin with Fig.~\ref{figpsione}(b). This involves a
C-O-O interaction vertex with both O's representing the field $\psi^1$. 
Let $z$ be the coordinate on the upper half plane
in which the closed string puncture is situated at $i$ and the open string punctures are
situated at $z_1=-\beta$ and $z_2=\beta$. Let
$w_1$ and $w_2$ be the local coordinates at the open string punctures, related to $z$
via the relations
\be \label{ezFrel}
z = F_a(w_a, \beta)\, .
\ee
According to \refb{epsiamp}, the amplitude is given by:
\be \label{e7.14}
- K_1\int d\beta \, \sum_{a=1}^2 \ointop_a {\p F_a \over \p \beta} \, dz \, \left
\langle b(z) \, F_1\circ \p c(0) \, F_2\circ \p c(0)\, c\bar c V_T(i)\right\rangle\, ,
\ee
where $c\bar c V_T$ denotes the vertex operator of the closed string tachyon of energy 
$\omega$, 
$\ointop_a$ is the anti-clockwise contour around $z_a=F_a(0;\beta)$ and
$K_1$ is a normalization constant. We shall assume that $\ointop$ includes a
$1/(2\pi i)$ factor. $F_a\circ \p c(0)$ denotes the conformal transform of $\p c(0)$ by
$F_a$.

Let $F_a(w_a,\beta)$ have an expansion of the form:
\be\label{eFexpand}
F_a(w_a,\beta)= f_a(\beta) + g_a(\beta)\, w_a + {1\over 2} h_a(\beta) \, 
w_a^2 +\OO(w_a^3)
\, .
\ee
Then the locations of the punctures are given by,
\be
z_a = F_a(0,\beta)=f_a(\beta)\, .
\ee
Furthermore, $F_a\circ c(0)$ and 
$F_a\circ \p c(0)$  can be found from \refb{eration11} and \refb{eFexpand}:
\be\label{eA.8}
F_a\circ c(0) = g_a(\beta)^{-1} c(z_a), \qquad F_a\circ \p c(0) = \p c(z_a)
- {h_a(\beta)\over g_a(\beta)^2} c(z_a)\, .
\ee
Let us introduce another constant $K_2$ via the relation:
\be\label{e5.72}
\langle c(z)\ c\bar c V_T(i)\rangle = K_2 \, (z^2+1), \qquad
\langle \p c(z)\  c\bar c V_T(i)\rangle=2\, K_2\, z\, .
\ee
Then \refb{e7.14} takes the form,
\ben\label{eAmpli}
&& \hskip -.3in 
- K_1\, \int d\beta\, \sum_{a=1}^2 \ointop_a 
\Bigg\langle\left[{\p f_a\over \p\beta} + {1\over g_a} {\p g_a\over \p\beta} (z-z_a)
+\OO\left((z-z_a)^2\right)\right] b(z) \, dz\, 
\left(\p c(z_1) - g_1^{-2} h_1\, c(z_1)\right)
\nonumber \\ && \hskip 1in
\left(\p c(z_2) - g_2^{-2} h_2 \, c(z_2)\right)
c\bar c V_T(i)\Bigg\rangle\nonumber \\ 
&=& - K_1 \, K_2 \, \int d\beta\,\left[\left\{{1\over g_1} {\p g_1\over \p \beta}
 - {h_1\over g_1^2} {\p f_1\over
\p\beta} \right\}\left\{ 2f_2 - g_2^{-2} h_2 (f_2^2+1) \right\} \right. \nonumber \\ &&
\hskip 1in \left. -
\left\{{1\over g_2} {\p g_2\over \p \beta} - {h_2\over g_2^2} {\p f_2\over
\p\beta} \right\}\left\{ 2f_1 - g_1^{-2} h_1 (f_1^2+1) \right\}
\right]\, .
\een
In arriving at the right hand side of \refb{eAmpli} we have used the operator
product expansion $b(z) c(z_a) \sim (z-z_a)^{-1}$, $b(z) \p c(z_a)\sim
(z-z_a)^{-2}$, where $\sim$ denotes equality up to non-singular terms.

Now we have, from \refb{e5},
\ben\label{epsi1}
w_1 &=& {\alpha\wt\lambda} \, {4\wt\lambda^2+1\over 4\wt\lambda^2} \,
{z+\beta\over (1-\beta\, z) - \wt\lambda\, f(\beta) (z+\beta)},
\nonumber \\
w_2 &=& {\alpha\wt\lambda} \, {4\wt\lambda^2+1\over 4\wt\lambda^2} \,
{z-\beta\over (1+\beta\, z) + \wt\lambda\, f(\beta) (z-\beta)}\, .
\een
Inverting these relations and expanding them in powers of $w_a$, we get
\ben
z \equiv F_1(w_1,\beta) 
&=& -\beta + {4\wt\lambda^2\over 4\wt\lambda^2+1} {1+\beta^2\over
\alpha\wt\lambda} w_1 - \left({4\wt\lambda^2\over 4\wt\lambda^2+1}\right)^2
{\beta+\wt\lambda f\over (\alpha\wt\lambda)^2} \, (1+\beta^2)\, w_1^2 +\OO(w_1^3), \nonumber \\
z\equiv F_2(w_2,\beta) 
&=& \beta + {4\wt\lambda^2\over 4\wt\lambda^2+1} {1+\beta^2\over
\alpha\wt\lambda} w_2 + \left({4\wt\lambda^2\over 4\wt\lambda^2+1}\right)^2
{\beta+\wt\lambda f\over (\alpha\wt\lambda)^2} \, (1+\beta^2)\, w_2^2 +\OO(w_2^3)\, .
\nonumber \\
\een
Comparing these with \refb{eFexpand} we get,
\ben\label{efglist}
&& z_1 =f_1(\beta)=-\beta, \quad g_1(\beta)={4\wt\lambda^2\over 4\wt\lambda^2+1} {1+\beta^2\over
\alpha\wt\lambda}, \quad h_1(\beta) = -2\, \left({4\wt\lambda^2\over 4\wt\lambda^2+1}\right)^2
{\beta+\wt\lambda f\over (\alpha\wt\lambda)^2}\, (1+\beta^2)\, , \nonumber \\ 
&& z_2 =f_2(\beta)=\beta, \quad g_2(\beta)={4\wt\lambda^2\over 4\wt\lambda^2+1} {1+\beta^2\over
\alpha\wt\lambda}, \quad h_2(\beta) = 2\, \left({4\wt\lambda^2\over 4\wt\lambda^2+1}\right)^2
{\beta+\wt\lambda f\over (\alpha\wt\lambda)^2}\, (1+\beta^2)\, .\nonumber \\
\een
Substituting these into \refb{eAmpli} and multiplying the result by the $\psi^1$ propagator
$K_0^{-1}/2$ obtained from 
\refb{ekinprop}, we get the contribution
to Fig.~\ref{figpsione}(b):
\be \label{epsiexchange}
-4 \, K_1 K_2 K_0^{-1}\, \int_{1/(2\wt\lambda)}^1 {d\beta\over 1+\beta^2} \wt\lambda^2 
f(\beta)^2\, .
\ee

We shall now determine the constant $K_1 K_2 K_0^{-1}$. For this let us consider
the effect of replacing $\psi^1$ by the open string tachyon in Fig.~\ref{figpsione}(b).
The analysis proceeds as above, except that instead of the vertex operator of $\psi^1$,
we insert the tachyon vertex operators as given in the first equation of \refb{eA.8}. The
analog of \refb{eAmpli} now takes the form:
\ben\label{eAmplitach}
&&\hskip -.3in - K_1 \, \int \, d\beta\, \sum_{a=1}^2 \ointop_a 
\left\langle\left[{\p f_a\over \p\beta} + {1\over g_a} {\p g_a\over \p\beta} (z-z_a)
+\OO(z-z_a)^2\right] b(z) \, dz\, 
g_1^{-1} \, c(z_1) \, g_2^{-1} c(z_2) \, c\bar c V_T(i)\right\rangle
\nonumber \\ 
&=& - K_1 \, K_2 \, \int \, d\beta\, \left[ {\p f_1\over \p\beta}\, (1+z_2^2)- {\p f_2\over \p\beta}\, 
(1+z_1^2) \right]\, g_1^{-1} \, g_2^{-1}
\, .
\een
Multiplying this by the tachyon propagator $K_0^{-1}$ and using \refb{efglist}, we get 
the contribution to Fig.~\ref{figpsione}(b) with $\psi^1$ replaced by the open string
tachyon:
\be\label{etachcon}
2\, K_1 \, K_2 \, K_0^{-1} \, \alpha^2 \wt\lambda^2 \, \left( {4\wt\lambda^2+1\over 
4\wt\lambda^2}\right)^2 \int_{1/(2\wt\lambda)}^1 {d\beta\over 1+\beta^2} \, .
\ee
On the other hand, the tachyon contribution, computed from the first term in the
last line of \refb{eca3}, is given by:
\be \label{eca3rep}
- {2\over \pi}\, g_s\, \sinh(\pi|\omega|)\, 
\alpha^2\wt\lambda^2\, \left(1+{1\over 4\wt\lambda^2}\right)^2
\int_{1/(2\wt\lambda)}^1 {d\beta\over 1+\beta^2} \, .
\ee
Note that we have put back a factor of $2\, g_s\, \sinh(\pi|\omega|)$ that was factored out from
\refb{eblock7} in defining $g(\omega)$.
Comparison of
\refb{etachcon} and \refb{eca3rep} now gives
\be\label{ekvalue}
K_1\, K_2 \, K_0^{-1}= -{g_s\over \pi}\, \, \sinh(\pi|\omega|)\, .
\ee
Substituting this into \refb{epsiexchange} we get the $\psi^1$ exchange contribution to
Fig.~\ref{figpsione}(b):
\be\label{e5.79}
{4\over \pi}\, \, g_s\, \sinh(\pi|\omega|)
\int_{(2\wt\lambda)^{-1}}^1 \, d\beta\, (1+\beta^2)^{-1}
\, \wt\lambda^2\, f(\beta)^2\, .
\ee

The contribution from Fig.~\ref{figpsione}(a) may be analyzed similarly. The main difference
is that the local coordinates $w_a$ now have a different relation to the coordinate $z$
in the upper half plane. 
We have, from \refb{e4.9a}, \refb{e5a}, 
\be\label{epsi11}
w_1 = -2\, \alpha \, {z+\beta \over z-3\,\beta}, \qquad 
w_2 = 2\, \alpha \, {z-\beta \over z+3\,\beta}, \qquad \beta\equiv {q\over 2\wt\lambda}\, .
\ee
Inverting these relations we can identify the functions $F_1$ and $F_2$:
\ben
z &\equiv& F_1(w_1,\beta) = -\beta + {2\beta\over \alpha} w_1 - {\beta\over \alpha^2}
w_1^2 +\OO(w_1^3)\, , \nonumber \\
z &\equiv& F_2(w_2,\beta) = \beta + {2\beta\over \alpha} w_2 + {\beta\over \alpha^2}
w_2^2 +\OO(w_1^3)\, . 
\een
Comparison with \refb{eFexpand} now gives:
\be
f_1=-\beta, \quad g_1 =  {2\beta\over \alpha} , \quad 
h_1=- {2\beta\over \alpha^2}, \quad f_2=\beta, \quad 
g_2 =  {2\beta\over \alpha} , \quad 
h_2={2\beta\over \alpha^2}\, .
\ee
Substituting this into \refb{eAmpli}, multiplying this by the $\psi^1$
propagator $K_0^{-1}/2$ and integrating over $\beta$, we get the
contribution to Fig.~\refb{figpsione}(a):
\be\label{e5.94}
-K_1 K_2 K_0^{-1} \int_0^{1/(2\wt\lambda)}\, d\beta\, 
\left[{3\over 4} - {1\over 4\beta^2}\right]
= {g_s\over 2\pi}\, \, \sinh(\pi|\omega|) \, \wt\lambda \, ,
\ee
where in the last step we have used \refb{ekvalue},  
replaced $\beta$ by $q/(2\wt\lambda)$ according to \refb{e4.9a} and
used the replacement rule \refb{erp1}, throwing away terms of order $1/\wt\lambda$.

Adding \refb{e5.79} and \refb{e5.94}, and dividing the result by 
$2\, g_s\, \sinh(\pi|\omega|)$, we get the total $\psi^1$ exchange contribution
to $g(\omega)$:
\be \label{epsitotapp}
g_{\psi^1}(\omega) =
\Bigg[{2\over \pi}\int_{(2\wt\lambda)^{-1}}^1 \, d\beta\, (1+\beta^2)^{-1}
\, \wt\lambda^2\, f(\beta)^2 + {\wt\lambda\over4\pi}\Bigg]\, .
\ee

\sectiono{Computation related to collective mode redefinition} \label{sa}

As discussed in \S\ref{scco2}, in order to find the field redefinition that relates
the open string zero mode associated with the vertex operator $c\p X$ to the
collective mode associated with time translation of the D-instanton, we need to
evaluate the `divergent' part of the contribution to the disk amplitude
with a pair of closed string tachyons of energies $\omega_1$ and $\omega_2$ 
and an open string zero mode field described by the vertex operator $c\p X$.
According to \refb{emastersec}, the relevant integral is:
\ben\label{emaster}
&& \int_{-\infty}^\infty dx\int_0^1 dy\, 
{1\over 2\pi}\left[-{\omega_1\over x-i} +{\omega_1\over x+i} -{\omega_2\over x-i\, y}+{\omega_2\over x+i\, y}\right] \nonumber \\ && \hskip 1in \times\ 
4\,  g_s \,  \sinh(\pi|\omega_1|)\, \sinh(\pi|\omega_2|)\, \Bigg[
{1\over 2}
\, y^{-2}\,  (1+2\, \omega_1 \omega_2 y)
\Bigg]\, ,
\een
and we have to evaluate the contribution to this integral 
from the different regions shown in Fig.~\ref{figeight},
associated with the Feynman diagrams of Fig.~\ref{figfour}. 
In this appendix we shall carry out this computation.

First we evaluate  the contribution to \refb{emaster} 
from region IV in Fig.~\ref{figeight}. There are no singularities from this region and we need to evaluate the integral explicitly.
We see from \refb{eivrange} that the $x$ integration is unrestricted and the $y$
integration range may be expressed as:
\be
1\ge y \ge \wt\lambda^{-2} - \wt\lambda^{-4} \Delta(x)\, ,
\ee
where
\be \label{edefDelta}
\Delta(x) = F(x)^2 +{1\over 4} \, .
\ee
In the above we have ignored fractional corrections of order $\wt\lambda^{-4}$.
Our strategy for evaluating the integral is to divide region IV into two parts: 
$y\ge \wt\lambda^{-2}$, and $y$ in the range $(\wt\lambda^{-2}- \wt\lambda^{-4} \Delta(x),
\wt\lambda^{-2})$.
Since the latter range is small, we can evaluate the $y$ integral
by replacing $y$ by $\wt\lambda^{-2}$ in the integrand and then multiplying 
the integrand
by $\wt\lambda^{-4}\Delta(x)$. If we denote these two integrals by $I_1$ and $I_2$,
then we have:
\ben \label{ecco1}
\hskip -.5in
I_1&=& \int_{-\infty}^\infty dx \int_{\wt\lambda^{-2}}^1 dy\, {1\over 2\pi}\left[-{\omega_1\over x-i} +
{\omega_1\over x+i} -{\omega_2\over x-i\, y}+{\omega_2\over x+i\, y}\right] \nonumber \\ && \times
2\, g_s \,  
\sinh(\pi|\omega_1|)\, \sinh(\pi|\omega_2|)\,
\, y^{-2} (1+2\, \omega_1 \omega_2 y)
\nonumber \\
&=& -i(\omega_1+\omega_2) \, 2\, g_s \,   \sinh(\pi|\omega_1|)\, \sinh(\pi|\omega_2|)\,
\int_{\wt\lambda^{-2}}^1 dy
\, y^{-2} (1+2\, \omega_1 \omega_2 y)
\nonumber \\
&=& -2\,  i\, (\omega_1+\omega_2) \,  g_s \,  
\sinh(\pi|\omega_1|)\, \sinh(\pi|\omega_2|)\, \left\{
\wt\lambda^2  - 1 + 2\, \omega_1\, \omega_2 \, \ln \wt\lambda^2\right\}
\, . 
\een
On the other hand, we get,
\ben \label{ecco1ext}
I_2 &=& \int_{-\infty}^\infty dx \, \wt\lambda^{-4} \Delta(x) \,
{1\over 2\pi}\left[-{\omega_1\over x-i} +{\omega_1\over x+i} -{\omega_2\over x-i\, 
\wt\lambda^{-2}}+{\omega_2\over x+i\, \wt\lambda^{-2}}\right] \nonumber \\ && 
\hskip 1in \times\, 
2\, g_s \, \sinh(\pi|\omega_1|)\, \sinh(\pi|\omega_2|)\,
\, \wt\lambda^4 (1+2\, \omega_1 \omega_2 \wt\lambda^{-2})
\nonumber \\
&& \hskip -.5in
=4 \, i\,  g_s \,   \sinh(\pi|\omega_1|)\, \sinh(\pi|\omega_2|)\,
(1+2\, \omega_1 \omega_2 \wt\lambda^{-2})
\, \int_{-\infty}^\infty dx \,  \Delta(x) \,
{1\over 2\pi}\left[-{\omega_1\over x^2+1}  -{\omega_2\, \wt\lambda^{-2}\over x^2
+ \wt\lambda^{-4}}\right]\, . \nonumber \\
\een
Making a change of variable $x\to \wt\lambda^{-2}/x$ in the second term, and
using \refb{eFid}, we can express this as
\be \label{ei2mid}
I_2 = -{2\, i\over \pi}\, (\omega_1+\omega_2)\,
g_s \,   \sinh(\pi|\omega_1|)\, 
\sinh(\pi|\omega_2|)\,
\, \int_{-\infty}^\infty dx \,  \Delta(x) \,
\left[{1\over x^2+1} \right]\, ,
\ee
where we have ignored fractional errors of order $\wt\lambda^{-2}$. This can be justified
as follows. 
$F(x)$ is given by \refb{edefFx1}, \refb{edefFx2}, \refb{edefFx3}.
Since $f(\beta)$ in these expressions
are bounded from above by 1, $F(x)$ and hence $\Delta(x)$ defined in \refb{edefDelta}
are also bounded. This shows that $I_2$ is a finite integral, and justifies working to
leading order in the expansion in powers of $\wt\lambda^{-2}$ for this integral.
Using \refb{edefFx1}, \refb{edefFx2}, \refb{edefFx3}, 
\refb{edefDelta}, and ignoring terms containing negative powers
of $\wt\lambda$, \refb{ei2mid} may be expressed as:
\be\label{ei2fin}
I_2 = -{8\, i\over \pi}\, (\omega_1+\omega_2)\,
g_s \,   \sinh(\pi|\omega_1|)\, \sinh(\pi|\omega_2|)\, 
\int_{1\over 2\wt\lambda}^1 d\beta\, (1+\beta^2)^{-1} \, \left\{ f(\beta)^2+{1\over 4}\right\}\, .
\ee

Next we consider regions III and III$'$. In these regions, 
we need to change variable from $(x,y)$ to $(u,\beta)$ according to 
\refb{em7}, \refb{em6}, and then 
integrate
over the range
\refb{em4a}. It will be more convenient to trade in the variable $\beta$ for $\xi$,
and introduce the function $\phi(\xi)$ via the relations:
\be \label{exk1}
\xi = \pm {2\beta\over 1-\beta^2}, \qquad \phi(\xi)=f(\beta)\, ,
\ee
so that \refb{em7}, \refb{em6} may be combined into the equation:
\be\label{exk2}
y= {u\over \wt\lambda}\,   \{1 + u^2\, \phi(\xi)^2\}^{-1}, 
\qquad x= 
\xi - u^2 \, \wt\lambda^{-1}\,   \phi(\xi) \, .
\ee
In these variables the range \refb{em4a} translates to:
\be\label{exk3}
0\le u\le {4\, \wt\lambda\over 4\, \wt\lambda^2+1}, \qquad {4\wt\lambda\over 4\wt\lambda^2-1}\le |\xi| < \infty\, .
\ee
We also have
\be\label{exk4}
dx\wedge dy = \wt\lambda^{-1} d\xi\wedge du \, \left\{1 - 3\, u^2 \, \phi(\xi)^2 
\right\}\, ,
\ee
ignoring fractional corrections of order higher than $\wt\lambda^{-2}$.
This may be justified by noting that after using the replacement rule \refb{erp2},
the leading divergent term in the integral \refb{emaster} is of order $\wt\lambda^2$.
Therefore fractional corrections of order higher than $\wt\lambda^{-2}$ will be 
suppressed in the large $\wt\lambda$ limit. 
Using \refb{exk2}-\refb{exk4} we may now express
the contribution to the integral \refb{emaster} from regions III and III$'$ as:
\ben \label{ei3first}
I_3 &=& -2\, i\, \int\int_{{\rm III+III}'} dx  \, dy\, 
 {1\over 2\pi}\left[{\omega_1\over x^2+1} +{\omega_2\, y\over x^2+ y^2}
 \right] \nonumber \\ && \hskip 1in \times\,
4\,  g_s \,  \Bigg[
{1\over 2} \sinh(\pi|\omega_1|)\, \sinh(\pi|\omega_2|)\,
\, y^{-2} (1+2\, \omega_1 \omega_2 y)
\Bigg]\nonumber \\
&=& -{2\over \pi}\, i\, \wt\lambda\, 
g_s \,   \sinh(\pi|\omega_1|)\, \sinh(\pi|\omega_2|)
\int_{|\xi|\ge 4\wt\lambda/(4\wt\lambda^2-1)} \, d\xi \, \int_0^{4\wt\lambda/
(4\wt\lambda^2+1)}\, du \nonumber \\ &&
\left\{ {\omega_1\over \{\xi - u^2 \, \wt\lambda^{-1}\,   \phi(\xi) \}^2 + 1}
+{\omega_2\, u \, \wt\lambda^{-1} \, \{1+u^2\, \phi(\xi)^2\}^{-1} \over \{\xi - u^2 \, \wt\lambda^{-1}\,   \phi(\xi) \}^2
+ u^2\, \wt\lambda^{-2} \, \{1+u^2\, \phi(\xi)^2\}^{-2}}
\right\} \nonumber \\ &&
u^{-2} \, \{1+u^2\, \phi(\xi)^2\}^{2}\, \left[1 + 2\, \omega_1\, \omega_2\, u\, \wt\lambda^{-1}\,
\{1+u^2\, \phi(\xi)^2\}^{-1}
\right] \left\{1 - 3\, u^2 \, \phi(\xi)^2 
\right\}\, . \nonumber \\
\een
Note that the $\xi$ integration range includes both positive and negative $\xi$.
We now expand the integrand in a Laurent series expansion in $u$ and use the
replacement rules:
\be
\int_0^{4\wt\lambda/
(4\wt\lambda^2+1)}\, du \, u^{-2} \to - {4\wt\lambda^2+1\over 4\wt\lambda}, 
\qquad \int_0^{4\wt\lambda/
(4\wt\lambda^2+1)}\, du \, u^{-1} \to 0\, .
\ee
Integrals with non-negative powers of $u$ in the integrand may be carried out explicitly.
This gives
\be\label{ei3fina}
I_3 = {2\over \pi}\, i\, \wt\lambda\, 
g_s \,   \sinh(\pi|\omega_1|)\, \sinh(\pi|\omega_2|)
\int_{|\xi|\ge 4\wt\lambda/(4\wt\lambda^2-1)} \, d\xi 
 {\omega_1\over \xi^2+1}
\left[{4\wt\lambda^2+1\over 4\wt\lambda} + {4\wt\lambda\over 4\wt\lambda^2+1}
\phi(\xi)^2 
\right]\, ,
\ee
where we have thrown away the terms that would vanish in the large $\wt\lambda$
limit. Using \refb{exk1} this may be written as:
\be\label{ei3fin}
I_3 = {8\over \pi}\, i\, \omega_1
g_s \,   \sinh(\pi|\omega_1|)\, \sinh(\pi|\omega_2|)
\int_{1\over 2\wt\lambda}^1 d\beta \, 
 {1\over \beta^2+1}
\left[{4\wt\lambda^2+1\over 4} +f(\beta)^2 \right]\, .
\ee

The change of variables in region II can be found in \refb{em2} and \refb{em4}. 
Let us first define
\be\label{exyx}
\tilde x = -y/x\, ,
\ee
and express \refb{em2} and \refb{em4} as:
\be \label{em2new}
y= {u\over \wt\lambda}\,   \{1 + u^2\, \phi(\xi)^2\}^{-1}, 
\quad \tilde x= \xi  - u^2 \, \wt\lambda^{-1}\, \phi(\xi)\, ,
\ee
where $\xi$ and $\phi(\xi)$ have 
been defined in \refb{exk1}. As already discussed in \S\ref{s1}, under
\refb{exyx} the regions II and II$'$ get mapped to III and III$'$. Therefore the
original integral \refb{emaster} 
over regions II and II$'$ can be rewritten as follows:
\ben\label{ei4first}
I_4 &=& -2\, i\, \int\int_{{\rm II+II}'} dx  \, dy\, 
 {1\over 2\pi}\left[{\omega_1\over x^2+1} +{\omega_2\, y\over x^2+ y^2}
 \right] \nonumber \\ &&\hskip 1in \times \
4\, g_s \,  \Bigg[
{1\over 2} \sinh(\pi|\omega_1|)\, \sinh(\pi|\omega_2|)\,
\, y^{-2} (1+2\, \omega_1 \omega_2 y)
\Bigg]\nonumber \\
&=& -2\, i\, \int\int_{{\rm III+III}'} d\tilde x  \, dy\, 
 {1\over 2\pi}\left[{\omega_1\, y\over \tilde x^2+y^2} +{\omega_2\over \tilde x^2+ 1}
 \right] \nonumber \\ && \hskip 1in \times \
2\, g_s \,  \sinh(\pi|\omega_1|)\, \sinh(\pi|\omega_2|)\,
\, y^{-2} (1+2\, \omega_1 \omega_2 y)
\, .
\een
\refb{em2new}, \refb{ei4first} have 
the same form as  \refb{exk2}, \refb{ei3first}, with 
$\omega_1\leftrightarrow \omega_2$ and $x$ replaced by $\tilde x$. 
Therefore this integral
can be analyzed in the same way, leading to the analog of \refb{ei3fin}:
\be\label{ei4fin}
I_4 = {8\over \pi}\, i\, \omega_2
 g_s \,   \sinh(\pi|\omega_1|)\, \sinh(\pi|\omega_2|)
\int_{1\over 2\wt\lambda}^1 d\beta\,  
 {1\over \beta^2+1}
\left[{4\wt\lambda^2+1\over 4} 
+f(\beta)^2 \right]\, .
\ee

Finally let us turn to regions I and I$'$. 
The change of variables is given in \refb{ech1p}, \refb{ech2p}  
for 
region I 
\be \label{ech1prepeat}
y = u_1 u_2 \left\{1 -{u_1^2\over 4} - {u_2^2\over 4} + 
\OO(u_1^4, u_2^4, u_1^2 u_2^2)\right\}, \qquad
x = -u_1 \left\{1 +{u_1^2\over 4} - {u_2^2\over 2} + \OO(u_1^4, u_2^4, u_1^2 u_2^2)
\right\} \, ,
\ee
and \refb{ech3p}, \refb{ech4p} in region I$'$,
\be \label{ech3prepeat}
y = u_1 u_2 \left\{1 - {u_1^2\over 4} - {u_2^2\over 4} 
+ \OO(u_1^4, u_2^4, u_1^2 u_2^2)\right\}, \qquad
x = u_1 \left\{1 +{u_1^2\over 4} - {u_2^2\over 2}+ \OO(u_1^4, u_2^4, u_1^2 u_2^2)
\right\}\, .
\ee  
This gives:
\be
dx dy = u_1 \, du_1\, du_2\, \left\{1 +{u_1^2\over 2} - {u_2^2\over 4}\right\}\, .
\ee
Using this, the contribution to the integral from this region may be expressed as:
\ben\label{ei5first}
\hskip-.2in I_5 &=& -2\, i\, \int\int_{{\rm I+I}'} dx  \, dy\, 
 {1\over 2\pi}\left[{\omega_1\over x^2+1} +{\omega_2\, y\over x^2+ y^2}
 \right] \nonumber \\ && \times
4\, g_s \,  \Bigg[
{1\over 2} \sinh(\pi|\omega_1|)\, \sinh(\pi|\omega_2|)\,
\, y^{-2} (1+2\, \omega_1 \omega_2 y)
\Bigg]\nonumber \\
&=& -{4\over \pi}\, i\, 
g_s \,   \sinh(\pi|\omega_1|)\, \sinh(\pi|\omega_2|)\, 
\int_0^{\wt\lambda^{-1}} du_1 \, \int_0^{\wt\lambda^{-1}} du_2\, 
u_1^{-2} \, u_2^{-2} \, du_1\, du_2
\nonumber \\ &&
\left[\omega_1\, u_1\, \left\{1 + {u_2^2 \over 4}\right\}
+ \omega_2 \, u_2\, \left\{1 + {u_1^2\over 4}\right\}
\right] \left[ 1 + 2\, \omega_1\, \omega_2 \, u_1 \, u_2 \, \left\{ 1 - {u_1^2\over 4}
-{u_2^2\over 4}\right\}
\right]\, .
\een
We evaluate this by using the replacement rules:
\be
\int_0^{\wt\lambda^{-1}} du_i\, u_i^{-2} \to -\wt\lambda, \qquad
\int_0^{\wt\lambda^{-1}} du_i\, u_i^{-1} \to 0\, .
\ee
This leads to
\be \label{ei5fin}
I_5 = 0\, .
\ee

Finally we note that the $\psi^1$ exchange contribution to any of the 
diagrams in Fig.~\ref{figfour}(a), (b) or (c) vanishes, since each internal
open string propagator is connected at least at one end to an open-closed 
two point interaction vertex, and for such vertices, the contribution vanishes according to
\refb{eocvanish} when the open string represents $\psi^1$ and the 
closed string represents
the tachyon.

Adding \refb{ecco1}, \refb{ei2fin}, \refb{ei3fin}, \refb{ei4fin} and \refb{ei5fin}, 
and carrying out the $\beta$ integrals, we get the
total C-C-O amplitude to be
\be \label{ecco5app}
2 \, i \, (\omega_1+\omega_2) \, 
g_s \,  \sinh(\pi|\omega_1|)\, \sinh(\pi|\omega_2|) \left\{ 1 -\, {2\over \pi}
\, \wt\lambda 
- 2\, \omega_1\, 
\omega_2 \ln \wt\lambda^2\right\}\, .
\ee
This concludes our derivation of \refb{ecco5} which was used to find the field redefinition
that relates the open string zero mode $\phi$ to the collective coordinate $\wt\phi$.

We end this appendix by noting that we can also use the procedure described in
\S\ref{salternative} to evaluate the integral given in \refb{emaster}. For example, the 
integrand of
\refb{emaster} may be expressed as:
\be
- \, dx\wedge dy\, {i\over \pi}\left[ {\omega_1\over x^2+1}+ {\omega_2 y\over x^2+y^2}
\right]
4\,  g_s \,  \sinh(\pi|\omega_1|)\, \sinh(\pi|\omega_2|)\, \Bigg[
{1\over 2}
\, y^{-2}\,  (1+2\, \omega_1 \omega_2 y)
\Bigg] = d\, \BBB\, ,
\ee
where,
\be 
\BB = -{i\over\pi} \, dy\, \left[\omega_1 \, \tan^{-1} x + \omega_2 \, \tan^{-1} {x\over y} \right]\,
4\,  g_s \,  \sinh(\pi|\omega_1|)\, \sinh(\pi|\omega_2|)\, \Bigg[
{1\over 2}
\, y^{-2}\,  (1+2\, \omega_1 \omega_2 y)
\Bigg]\, .
\ee
Therefore, \refb{emaster} may be represented as a boundary integral of $\BB$ in
Fig.~\ref{figeight}, with the boundaries including the segments $x=\infty$, $y=1$, $x=-\infty$,
and the regularized boundaries at $y=0$, consisting of the segments $q_1=\delta_1$ in
Fig.~\ref{figfour}(a) and (b) and  $q_2=\delta_2$ in
Fig.~\ref{figfour}(a) and (c). String field theory instructs us to drop all terms
proportional to $\delta_1^{-1}$, $\delta_2^{-1}$, $\ln\delta_1$ and $\ln\delta_2$ in the
final expression. We have checked that this gives the same result 
as \refb{ecco5app}.

\sectiono{Computation related to gauge parameter redefinition} \label{sb}

As described in \S\ref{sghost}, in order to find the relation between the string field
theory gauge transformation parameter accompanying the state $|0\rangle$, and
the rigid $U(1)$ gauge transformation parameter, we need to compute the disk 
amplitude $A(\omega)$ 
of the closed string tachyon vertex operator of energy $\omega$, 
and the open string vertex
operators $I$, $c\p c \p^2c/2$ and $I$. 
Furthermore, the open string vertex operators carry Chan-Paton factors so that only
one cyclic ordering contributes.
In this appendix 
we shall 
compute $A(\omega)$.

We shall use the upper half plane
representation of the disk with the closed string tachyon vertex operator $c\bar c V_T$ 
inserted at
$i$ and the vertex operators $I$, $c\p c \p^2c/2$ and $I$ inserted at three points
$z_1$, $z_2$ and $z_3$ respectively on the real axis. The upper half plane with three punctures
on the boundary and one puncture in the bulk has two real moduli. Let us denote them
by $\vec\beta =(\beta_1,\beta_2)$. Then we have
$z_1=f_1(\vec\beta)$,
$z_2=f_2(\vec\beta)$ and $z_3=f_3(\vec\beta)$ where 
$f_1,f_2,f_3$ are three functions of these moduli.  As in \refb{eFexpand},
we describe the choice of
local coordinates $w_a$ at the punctures via the relations:
\be\label{e7.87}
z= F_a(w_a, \vec \beta) = f_a(\vec\beta) + g_a(\vec\beta) \, w_a +{1\over 2} h_a(\beta) 
\, w_a^2
+ \OO\left( w_a^3\right)\, .
\ee
We can now
use the generalization of \refb{e7.14} to 
express the C-O-O-O amplitude $A(\omega)$ as:
\ben\label{edefAnew}
A(\omega)&=&K_3\, \int d\beta_1\wedge d\beta_2 \Bigg\langle
\Bigg\{\sum_{a=1}^3 \ointop_a {\p F_a\over \p \beta_1} b(z) dz
\Bigg\} \Bigg\{\sum_{a=1}^3 \ointop_{a} {\p F_a\over \p \beta_2} b(z) dz
\Bigg\} \nonumber \\ &&
\left\{ c\p c \p^2 c (z_2)/2\right\}  c\bar c V_T(i) \Bigg\rangle\, ,
\een
where $K_3$ is a normalization constant and $\ointop_a$ is the anti-clockwise
contour surrounding $z_a$. In writing \refb{edefAnew} we have made use of the
fact that all the vertex operators are dimension zero primary operators, so that
conformal transformations by the $F_a$'s leave them unchanged.
Using \refb{e7.87},  \refb{e5.72}, and evaluating the contour integrals using the 
$b$-$c$ operator product expansion, we can express  \refb{edefAnew} as:
\ben \label{eAint}
A(\omega) &=& {1\over 2} \, K_2\, K_3\, \int \Bigg[ 
(1+f_2^2) g_2^{-3} dh_2\wedge dg_2 + 2 f_2 h_2 g_2^{-3} dg_2\wedge df_2
\nonumber \\ && \hskip 1in - 2 f_2 g_2^{-2} dh_2\wedge df_2 + 2 g_2^{-1} dg_2\wedge df_2\Bigg] \nonumber \\
&=&  K_2 \, K_3\, 
\int d \JJ \, ,
\een
where
\be \label{edefJJ}
\JJ\equiv  - f_2 g_2^{-1} dg_2 -{1\over 2} (1+f_2^2) h_2 g_2^{-3} dg_2 + 
{1\over 2} (1+f_2^2) g_2^{-2} dh_2 + df_2\, .
\ee
In this form $\JJ$ is manifestly invariant under the $SL(2,R)$ transformation 
$z\to (z-c)/(1+c\, z)$ that keeps the point at $i$ fixed.  This can be seen by
examining the transformation laws of $f_i$, $g_i$ and $h_i$ 
from \refb{e7.87} and then
substituting them into \refb{edefJJ}. This in turn ensures that when we evaluate 
the integral in \refb{eAint} by expressing it as a boundary integral of $\JJ$, we do not
need to worry about possible boundary contributions from regions where some
open string puncture approaches infinity, since $\infty$ can
be brought to a finite point on the real axis using the $SL(2,R)$ transformation 
$z\to (z-c)/(1+c\, z)$.

Next we shall discuss how to fix the orientation of the integral, i.e.\  given an integral
of the form $\int_R d\beta_1\wedge d\beta_2$ over some finite region $R$, 
how to decide whether 
it gives positive or negative result. If we were using the same coordinate system
everywhere in the moduli space then this would not be necessary, since an overall
sign can always be absorbed inside $K_3$. 
However we shall use different coordinate systems
in different regions, and relative signs will be important.\footnote{If we were 
evaluating 
the amplitude in string field theory from
first principles, then the orientation would have been automatically fixed. Here we are
combining string field theory insights with world-sheet methods, and therefore we need
to make extra effort to fix the orientation.}
For this we shall
first make an $SL(2,R)$ transformation 
\be
z'={z-f_3\over 1+z f_3}\, ,
\ee
to bring the third puncture to 0, keeping the closed string puncture fixed at $i$. This
gives:
\be
z_1'= {f_1-f_3\over 1+f_1f_3}, \qquad z_2'={f_2-f_3\over 1+f_2f_3},
\ee
and
\ben\label{efixorient}
dz_1'\wedge dz_2' &=& (1+f_1f_3)^{-2} (1+f_2f_3)^{-2} (1+f_3^2)
\nonumber \\ && \left[ (1+f_3^2)\, df_1\wedge df_2
+  (1+f_1^2)\, df_2\wedge df_3 +  (1+f_2^2)\, df_3\wedge df_1
\right]\, .
\een
Since $dz_1'\wedge dz_2'$ fixes the global orientation of the moduli space of a 
disk with one closed string puncture and three open string punctures, we need to ensure
that the orientations we use in different segments coming from different Feynman 
diagrams agree with this orientation. We shall take $dz_1'\wedge dz_2'$ to be the 
positive integration measure, i.e.\ assume that $\int_R dz_1'\wedge dz_2'$ for some finite
region $R$ gives positive result. The other convention where $\int_R dz_2'\wedge dz_1'$
gives positive result is related to the one used here by a change in the sign of $K_3$
introduced in \refb{edefAnew}. Once we decide that $\int_R dz'_1\wedge dz'_2$
gives positive result, \refb{efixorient} 
can be used to fix the orientation of the moduli space integration 
coming from different Feynman diagrams. This will be illustrated  below.

\begin{figure}
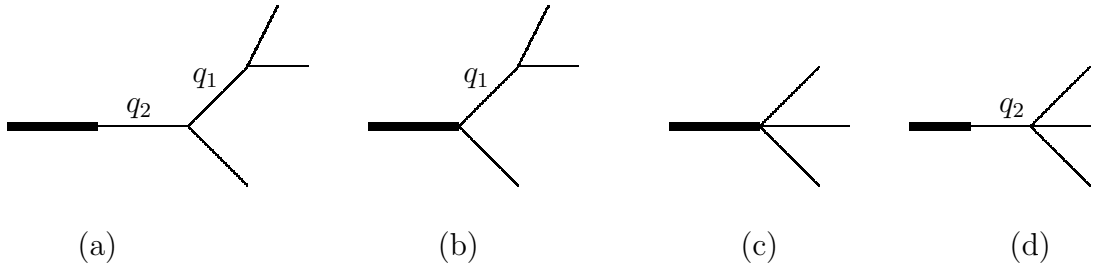

\begin{center}

\hbox{\figcooo}

\vskip -.8in

\caption{This figure shows four Feynman diagrams contributing to the disk amplitude with 
one external
closed string and three external open strings. We have not displayed diagrams that differ
from each other by permutation of external states. As usual, $q_1$ and $q_2$ represent
the sewing parameters of the corresponding open string propagators.
\label{figcooo}
}
\end{center}
\end{figure}

We shall now analyze the contribution to $A(\omega)$ by expressing this as a sum of four Feynman
diagrams shown in Fig.~\ref{figcooo}. First we shall argue that the contributions from
Fig.~\ref{figcooo}(a) and \ref{figcooo}(b) vanish. 
To see this, note that since we are constructing the interaction terms of the
Wilsonian effective action in which we
remove all the $L_0=0$ states from the internal propagators, and since the
contributions from internal states with $L_0>0$ are suppressed in the limit of
large $\alpha$ and/or $\lambda$, we only need to focus on
the contributions where both internal states are tachyons. 
Now, using the fact that the external open strings are
described by vertex operators $I$, $c\p c \p^2c /2$ and $I$, 
it is easy to see that ghost number conservation  
prevents the internal open string that connects to the pair of external
open strings in Fig.~\ref{figcooo}(a) or Fig.~\ref{figcooo}(b)
to be tachyonic.\footnote{This can be seen more explicitly by 
expressing the integral
in \refb{eAint} as integral over the variables $q_i$ associated to the propagators in 
Fig.~\ref{figcooo}(a) and (b) and the internal variable $\beta$ of the C-O-O vertex for
Fig.~\ref{figcooo}(a), and then showing that there are no $q_1^{-2}$ singularity in
the integrands where $q_1$ is the sewing 
parameter associated with the open string propagator
connected to the pair of external open strings. However we shall not describe this analysis
here.} For example, if the external open string states are $I$ and $c\p c \p^2 c/2$, 
then the
lowest $L_0$ state that has a three point function with this pair of states is $I$, 
whereas
if the external open string states are $I$ and $I$, then the
lowest $L_0$ state that has a three point function with this pair of states  
is $c\p c \p^2 c/2$. All of these
are $L_0=0$ states and are prevented from propagating in the internal lines.

Next we shall analyze the
contribution from Fig.~\ref{figcooo}(d). Let us suppose that the local
coordinates at the four punctures at the four point open string interaction vertex
are given by $w_i=G_i(\hat z, \tau)$, where $\hat z$ is the coordinate on the upper
half plane representing this interaction vertex, $\tau$ is the modular parameter
of a plane with four boundary punctures and $G_i$'s are some given functions. 
We can find the range of $\tau$ and the constraints on $G_i$ by requiring that the
four point interaction vertex correctly interpolates between the s, t and  u channel
diagrams constructed using the O-O-O vertex given in \refb{eooo}, but we
shall not need this for our analysis. The only information on $G_i(\tau)$ that we shall
use is that while it depends on the parameter $\alpha$ that appears in the O-O-O
interaction vertex \refb{eooo}, it does not depend on the parameter 
$\lambda$ that appears in the
C-O interaction vertex \refb{e1alt}.
Denoting by $z$ the upper half plane coordinate associated
with the C-O interaction
vertex, and using \refb{e1alt}, we get the sewing relation associated with
Fig.~\ref{figcooo}(d) to be:
\be
\lambda \, z \, G_4(\hat z, \tau)= -q_2\, , \qquad 0\le q_2\le 1\, ,
\ee
where we have assumed that the sewing takes place between the external open 
string of the C-O vertex and the 4-th external open string of the O-O-O-O vertex.
We can rewrite this as
\be \label{eC10}
G_i(\hat z, \tau) = - u/z, \quad u\equiv q_2/\lambda, \quad 0\le u\le \lambda^{-1}\, .
\ee
Note now that $z$ and $u$ appear in this equation in the combination $z/u$. As a 
result, the locations of the other three open string punctures in the $z$ plane must
have the form $u \, \phi_i(\tau)$ for some functions $\phi_i(\tau)$. Furthermore, the
local coordinates $w_a$ at the open string punctures must be functions of $z/u$ and
$\tau$. Therefore we have the expansion:
\be
z = u\, \phi_a(\tau) + u\, \gamma_a(\tau) \, w_a+{u\over 2} \eta_a(\tau) \,
w_a^2 + \OO(w_a^3), \qquad 1\le a\le 3
\, .
\ee
Comparing this with \refb{e7.87} we see that we have
\be\label{eutau}
f_a = u\, \phi_a(\tau), \quad g_a= u \, \gamma_a(\tau), \quad h_a(\tau) = u\, \eta_a(\tau)\, .
\ee
Substituting this into the first expression in
\refb{eAint} we get the form of the integral inside the
region covered by Fig.~\ref{figcooo}(d) to be:
\ben \label{ecoood}
I_{(d)}&=& {1\over 2} \, K_2\, K_3\, \int du\wedge d\tau \, \Bigg[u^{-2}
\Bigg\{ \gamma_2^{-3} \eta_2 \gamma_2' - \gamma_2^{-2} \eta_2' 
\Bigg\} + \OO(u^0)\Bigg]\, ,
\een
where $'$ denotes derivative with respect to $\tau$.

Let us define $\ve$ such that it can take values 1 or $-1$ and  
that $\ve du\wedge d\tau$ gives the integration measure with the correct sign, i.e.\
$\ve \int du\wedge d\tau$ gives positive result when integrated over a finite
region in $u$-$\tau$ space. $\ve$ can be determined from \refb{efixorient}, but we
shall not need this information.
We now use the replacement rule \refb{erp1} to  
replace $\int_0^{1/\lambda} du u^{-2}$ by
$-\lambda$.  Terms in the integrand with non-negative powers of $u$ can be made
zero by taking the large $\lambda$ limit.
This gives the contribution to $A(\omega)$ from Fig.~\ref{figcooo}(d)
to be
\be \label{eid2}
I_{(d)}=-{1\over 2} \, K_2 \, K_3\, \lambda \,  \ve 
\int_{\BB_2} d\tau \Bigg[ \gamma_2^{-3} \eta_2 \gamma_2' - \gamma_2^{-2} \eta_2' 
\Bigg] \, ,
\ee
where $\BB_2$ is the boundary at $u=\lambda$, with its normal along the direction
of increasing $u$.

The contribution from Fig.~\ref{figcooo}(c) may be analyzed by using the fact that 
the integrand in \refb{eAint} is a total derivative and therefore the integral may be 
expressed as boundary integrals. Now the region of the moduli space covered by
Fig.~\ref{figcooo}(c) has two boundaries: $\BB_1$ where the moduli space connects to
the $q_1=1$ boundary of the region covered by Fig.~\ref{figcooo}(b), and  
$\BB_2$ where the moduli space connects to
the $q_2=1$ boundary of the region covered by Fig.~\ref{figcooo}(d).
Therefore we have
\be
I_{(c)}= I_{(c),1} + I_{(c),2}\, ,
\ee
where
\be \label{eici}
I_{(c),i} = K_2\, K_3\, 
\int_{\BB_i}   \left\{ - f_2 g_2^{-1} dg_2 -{1\over 2} (1+f_2^2) h_2 g_2^{-3} dg_2 + 
{1\over 2} (1+f_2^2) g_2^{-2} dh_2 + df_2\right\}\, .
\ee
First let us evaluate the contribution from the boundary $\BB_2$. For this we can
use the $u,\tau$ coordinate system introduced in \refb{eC10}, and, describing the
boundary as the curve $u=1/\lambda$, express the boundary integral as:
\be\label{eic2}
I_{(c),2} =  -  K_2\, K_3\, \lambda\, \ve\, 
\int_{\BB_2} d\tau \left[ -{1\over 2} \eta_2\gamma_2^{-3}\gamma_2'+{1\over 2} 
\gamma_2^{-2}\eta_2'
\right]\, ,
\ee
where we have dropped terms with negative powers of $\lambda$. An extra negative
sign in \refb{eic2} comes from the fact that $u=1/\lambda$ is the lower limit of integration
in the region covered by the diagram \ref{figcooo}(c). Adding this to \refb{eid2} we get,
\be\label{sumid2ic2}
I_{(d)}+I_{(c),2} = 0\, .
\ee

It remains to evaluate the contribution from the boundary $\BB_1$.
Since this boundary is shared with the boundary of the region associated with 
Fig.~\ref{figcooo}(b), we can use the moduli space parameters
inherited from this diagram and
describe the boundary to be at $q_1=1$, where $q_1$ is the sewing parameter 
associated with the internal open string propagator of this diagram.\footnote{Actually,
$\BB_1$ consists of six components. A factor of three comes from the
possibility of cyclically permuting the external open strings. Another factor of two
comes from the freedom of inserting the external open string that attaches to the
C-O-O vertex to be either at $\beta$ or at $-\beta$ in the convention of \S\ref{scoo}.}
Therefore we shall now study the moduli space parameters inherited from
Fig.~\ref{figcooo}(b). Let  $z$ denote the upper half plane coordinate associated with the
C-O-O vertex and $\hat z$ denote the upper half plane coordinate associated with the
O-O-O vertex. Let us
first sew the puncture at $z=\beta$ from the C-O-O 
vertex with the puncture at $\hat z=0$
from the O-O-O vertex, and  
call the puncture at $z=-\beta$ as puncture 1, the puncture at
$\hat z=1$ as puncture 2 and the puncture at $\hat z=\infty$ as puncture 3.
In this case the sewing relation takes the form
\be \label{ed18}
w_1 \hat w_1 = -q_1\, ,
\ee
where $w_1$ is the local coordinate at the puncture at $\beta$ for the C-O-O vertex,
and $\hat w_1$ is the local coordinate at the puncture at 0 for the O-O-O vertex. Using
\refb{eooo} and \refb{e5}, \refb{ed18} may be written as:
 \be\label{ed20}
  {\alpha\wt\lambda} \, {4\wt\lambda^2+1\over 4\wt\lambda^2} \,
{z-\beta\over (1+\beta\, z) + \wt\lambda\, f(\beta) (z-\beta)} 
\, \alpha \, {2\hat z\over 2-\hat z} = -q_1\, .
 \ee
 The open string punctures at $z=-\beta$, $\hat z=1$ and $\hat z=\infty$ are located in
 the $z$ plane at
 \ben\label{ed21}
&& z_1= -\beta, \quad z_2=\beta - u\, {1+\beta^2\over 1 + u(\beta+\wt\lambda f)}
= {\beta+u (\beta\wt\lambda f-1)\over 1 + u(\beta+\wt\lambda f)}, \nonumber \\ &&
 z_3=\beta + u\, {1+\beta^2\over 1 - u(\beta+\wt\lambda f)}
 = {\beta-u (\beta\wt\lambda f-1)\over 1 - u(\beta+\wt\lambda f)}\, ,
 \een
 where
 \be
 u\equiv {q_1\over 2\alpha^2\wt\lambda}\, {4\wt\lambda^2\over 4\wt\lambda^2+1}\, ,
 \qquad 0\le u\le u_0, \qquad u_0\equiv {1\over 2\alpha^2\wt\lambda}\, 
 {4\wt\lambda^2\over 4\wt\lambda^2+1}
 \, .
 \ee
Furthermore, the local coordinates at the open string punctures are given by,
\ben\label{ed23}
 w_1 &=&   {\alpha\wt\lambda} \, {4\wt\lambda^2+1\over 4\wt\lambda^2} \,
{z+\beta\over (1-\beta\, z) - \wt\lambda\, f(\beta) (z+\beta)}
\,, \nonumber \\ 
w_2 &=& -2\, \alpha\, {1-\hat z\over 1+\hat z}= 2\, \alpha \,{ (z-\beta)\{1 + u(\beta+\wt\lambda f)\} + u(1+\beta^2)\over 3u(1+\beta^2) - (z-\beta) + 
3u(\wt\lambda f+\beta)(z-\beta) 
}, \nonumber \\
w_3 &=& \alpha\, {2\over 1-2\hat z}= 2\, \alpha \,{ (z-\beta)\{1 - u(\beta+\wt\lambda f)\} -
u(1+\beta^2)\over 3u(1+\beta^2) + (z-\beta) + 
3u(\wt\lambda f+\beta)(z-\beta) 
}\, .
\een
Inverting these relations, we get:
\ben
z&\equiv& F_1(w_1,\beta) =-\beta + 
{4\wt\lambda^2\over 4\wt\lambda^2+1}\, {1+\beta^2\over
\alpha\wt\lambda} w_1 - \left({4\wt\lambda^2\over 4\wt\lambda^2+1} {1\over \alpha\wt\lambda}\right)^2 (1+\beta^2) (\beta+\wt\lambda\, f)w_1^2
+\OO(w_1^3)\, , \nonumber \\
z&\equiv& F_2(w_2,\beta)
= {\beta+ u(\beta\wt\lambda f-1)\over 
1 + u(\beta+\wt\lambda f)} +{2\, u\over \alpha} 
{(1+\beta^2) \over \{1+ u(\beta+\wt\lambda f)\}^2} \, w_2
\nonumber \\ &&
- {u\over \alpha^2}\, 
{(1+\beta^2) \{1- 3 u(\beta+\wt\lambda f)\}\over  \{1 + u (\beta+\wt\lambda f)\}^3}
w_2^2
\,
+\OO(w_2^3) \, , \nonumber \\
z&\equiv& F_3(w_3,\beta) = {\beta- u(\beta\wt\lambda f-1)\over 
1 - u(\beta+\wt\lambda f)} +{2\, u\over \alpha} 
{(1+\beta^2) \over \{1- u(\beta+\wt\lambda f)\}^2} \, w_3 
\nonumber \\ &&
+ {u\over \alpha^2}\, 
{(1+\beta^2) \{1+ 3 u(\beta+\wt\lambda f)\}\over  \{1 - u (\beta+\wt\lambda f)\}^3}
w_3^2
\,
+\OO(w_3^3)\, . 
\een
Comparing this with the expansion given in \refb{e7.87} we get:
\ben \label{effgv1}
&& f_1= -\beta, \quad f_2 = \beta - u (1+\beta^2)+u^2 (1+\beta^2) (\beta+\wt\lambda f)
+\OO(u^3), \nonumber \\ &&
f_3 = \beta + u(1+\beta^2) +u^2 (1+\beta^2) (\beta+\wt\lambda f)
+\OO(u^3)\, , \nonumber \\ &&
g_1 = {4\wt\lambda^2\over 4\wt\lambda^2+1}\, {1+\beta^2\over
\alpha\wt\lambda},
\quad g_2 = 2\, \alpha^{-1} (1+\beta^2) \, u  \, \left\{1 - 2\, u\, (\beta+\wt\lambda f)
+ \OO(u^2) \right\}, \nonumber \\ &&
g_3 =2\, \alpha^{-1} (1+\beta^2) \, u  \, \left\{1 + 2\, u\, (\beta+\wt\lambda f)
+ \OO(u^2) \right\}, \nonumber \\ &&
h_1 = - 2\left({4\wt\lambda^2\over 4\wt\lambda^2+1} {1\over \alpha\wt\lambda}\right)^2 (1+\beta^2) (\beta+\wt\lambda\, f), \nonumber \\ && 
h_2 = -2\, {u\over \alpha^2} (1+\beta^2)
\{1 - 6u(\beta+\wt\lambda f) +\OO(u^2)\}\,, \nonumber \\ &&
h_3 = 2\, {u\over \alpha^2} (1+\beta^2)
\{1 + 6u(\beta+\wt\lambda f) +\OO(u^2)\} \, .
\een
There are two other configurations related to the one described above by cyclic 
permutation of $1,2,3$. Since \refb{eici} involves only  $f_2$, $g_2$ and $h_2$ and 
fixing the orientation via \refb{efixorient} involves also $f_1$ and $f_3$, we shall
quote only these values:
\ben\label{effgv2}
&&  f_1= \beta + u(1+\beta^2) +u^2 (1+\beta^2) (\beta+\wt\lambda f)
+\OO(u^3), \quad f_2= -\beta,  \quad g_2={1\over \alpha\wt\lambda} \, 
{4\wt\lambda^2\over 4\wt\lambda^2+1}  {(1+\beta^2)}\, ,\nonumber \\ &&
h_2=- 2\left({4\wt\lambda^2\over 4\wt\lambda^2+1} {1\over \alpha\wt\lambda}\right)^2 (1+\beta^2) (\beta+\wt\lambda\, f), \nonumber \\ && 
f_3=  \beta - u (1+\beta^2)+u^2 (1+\beta^2) (\beta+\wt\lambda f)
+\OO(u^3),
\een
and
\ben\label{effgv3}
&& f_1 = \beta - u (1+\beta^2)+u^2 (1+\beta^2) (\beta+\wt\lambda f)
+\OO(u^3), \nonumber \\ &&
f_2 = \beta + u(1+\beta^2) +u^2 (1+\beta^2) (\beta+\wt\lambda f)
+\OO(u^3)\, , \nonumber \\ &&
g_2 =2\, \alpha^{-1} (1+\beta^2) \, u  \, \left\{1 + 2\, u\, (\beta+\wt\lambda f)
+ \OO(u^2) \right\}\, , \nonumber \\
&& h_2 = 2\, {u\over \alpha^2} (1+\beta^2)
\{1 + 6u(\beta+\wt\lambda f) +\OO(u^2)\}\, , \nonumber \\
&& f_3=-\beta\, .
\een

We shall now fix the orientation of the integral. For all three cases, the integration 
measure \refb{efixorient} takes the form, for small $u$,
\be \label{eorientexpl}
dz'_1\wedge dz'_2 = F(\beta) \, d\beta \wedge du\, ,
\ee
where $F(\beta)$ is a positive function of $\beta$. Therefore, $d\beta\wedge du$ is
the positive integration measure.

We can now evaluate the contribution to $I_{(c)1}$ given in \refb{eici} 
from \refb{effgv1}, \refb{effgv2} and \refb{effgv3}, by identifying the boundary $\BB_1$ as
the subspace
$u=u_0$, $(2\wt\lambda)^{-1}\le\beta\le 1$ in each case. The results are given,
respectively, by:
\be\label{ers1}
 \, K_2\, K_3 \,\int_{(2\wt\lambda)^{-1}}^1 d\beta\, \left[{2\over 1+\beta^2} 
+ \wt\lambda \, f'(\beta)\right]\, ,
\ee
\be\label{ers3}
 \, K_2\, K_3\, \int_{(2\wt\lambda)^{-1}}^1 d\beta\, \left[-{2\over 1+\beta^2} 
- \wt\lambda \, f'(\beta)\right]\, ,
\ee
and
\be\label{ers2}
 \, K_2\, K_3\, 
\int_{(2\wt\lambda)^{-1}}^1 d\beta\, \left[{2\over 1+\beta^2} + \wt\lambda \, f'(\beta)\right]\, .
\ee
In arriving at the above equations, two minus signs have cancelled.
First we see from \refb{eorientexpl} that $-du\wedge d\beta$ is the positive integration
measure. Second, the outward  normal to the boundary of the region associated with
Fig.~\ref{figcooo}(c) at $u=u_0$ is directed towards the
direction of decreasing $u$.

We also need to consider the case where we sew the 
puncture at $z=-\beta$ from the C-O-O vertex with the puncture at $\hat z=0$
from the O-O-O vertex.\footnote{Alternatively, as discussed near the end
of \S\ref{scoo},
we could simply extend the range
of $\beta$ in  \refb{ers1}-\refb{ers2} all the way to $\beta=2\wt\lambda$. This gives the
same result after we use the result $f(1/\beta)=f(-\beta)=-f(\beta)$.}
Since the C-O-O vertex is symmetric under $\beta\to -\beta$, the results in this case
can be read out from those in the previous case by making a $\beta\to -\beta$,
$f\to -f$ change in the results. 
This would produce an extra minus sign in \refb{eorientexpl} and \refb{ers1}-\refb{ers2}
which compensate each other. Therefore the results are identical to those given in
\refb{ers1}-\refb{ers2}.
The sum of all the terms gives:
\be \label{ersf}
A(\omega)=2\,  \, K_2 \, K_3\, 
\int_{(2\wt\lambda)^{-1}}^1 d\beta\, \left[{2\over 1+\beta^2} + \wt\lambda \, f'(\beta)\right]
=2\,  \, K_2 \, K_3\, 
\left[{\pi\over 2} - {\wt\lambda\over 2}\right]\, ,
\ee
where we have used $f((2\wt\lambda)^{-1})=1/2+\OO(\wt\lambda^{-2})$ and $f(1)=0$, and ignored terms
containing inverse power of $\wt\lambda$.
Using \refb{eactionghost} we now see that this 
gives an additional contribution to the closed string tachyon one
point function of the from:
\be \label{e6132}
- K_0^{-1} A(\omega)=-2\, K_2\, K_3\, K_0^{-1} \, 
\left[{\pi\over 2} - {\wt\lambda\over 2}\right]\, .
\ee

We shall now determine the ratio $K_2K_3/K_0$.
For this consider a C-O-O-O amplitude where all external open string states are tachyons,
described by the vertex operator $c$, but they carry the same Chan-Paton factors as the
original vertex operators so that only one cyclic ordering contributes.. 
Since $c$ is a primary operator of dimension $-1$, we have
$F_a\circ c(0)=(F_a'(w_a=0))^{-1} c(z_a)= g_a^{-1} c(z_a)$.
In that case,  starting from the analog of
\refb{edefAnew}:
\ben \label{edeftA}
\wt A(\omega)&=&K_3\, \int d\beta_1\wedge d\beta_2 \Bigg\langle
\Bigg\{\sum_a \ointop_{a} {\p F_a\over \p \beta_1} b(z) dz
\Bigg\} \Bigg\{\sum_a \ointop_{a} {\p F_a\over \p \beta_2} b(z) dz
\Bigg\} \nonumber \\ &&
(g_1g_2g_3)^{-1} \, c(z_1)  c(z_2)  c(z_3) c\bar c V_T(i) \Bigg\rangle\, ,
\een
and following the same steps as outlined below \refb{edefAnew}, we arrive at the
analog of \refb{eAint}:
\be
\wt A(\omega)= K_2\, K_3\,  \int\,  (g_1 g_2 g_3)^{-1} 
\left[(1+f_3^2) df_2\wedge df_1 + (1+f_1^2) df_3\wedge df_2 + (1+f_2^2)
df_1\wedge df_3\right]\, .
\ee
For Fig.~\ref{figcooo}(b), we can use \refb{effgv1} to express this as:
\be
\wt A_{(b)}(\omega)=- K_2\, K_3 \, {4\wt\lambda^2+1\over 4\wt\lambda^2} \int
\, {\alpha^3\wt\lambda\over (1+\beta^2) u^2}
d\beta\wedge du\, ,
\ee
where we have kept the leading term in the expansion in a power series in $u$.
Note that this is the contribution from one Feynman diagram and we have not summed
over cyclic permutations.
Furthermore, using \refb{eorientexpl} we see that $d\beta\wedge du$ is the positive
integration measure. Now using $u= \{1+(4\wt\lambda^2)^{-1}\}^{-1} q/(2\alpha^2\wt\lambda)$ and
the replacement rule \refb{erp1}, we can carry out integration over $q$ and express
the leading contribution from Fig.~\ref{figcooo}(b) to the amplitude with one closed
string tachyon and three open string tachyons to be:
\be \label{ecomp1new}
\wt A_{(b)}(\omega) =  K_2\, K_3 \, \left({4\wt\lambda^2+1\over 4\wt\lambda^2}\right)^2\,
2\, \alpha^2 \, \wt\lambda\, 
\int_{1/2\wt\lambda}^1 d\beta {\alpha^3\wt\lambda\over (1+\beta^2)}\,
\, 
\, .
\ee
On the other hand, it follows from Fig.~\ref{figcooo}(b) that this contribution should be
proportional to the product of a C-O-O vertex with two open string tachyons and a
closed string tachyon as external
states, an open string tachyon propagator and an O-O-O vertex with three open
string tachyons as external states. Now we see from \refb{eca3rep}  
that the product of the C-O-O vertex and the open
string tachyon propagator is given by:
\be \label{eca3reptwo}
- 2\,  g_s\, \sinh(\pi|\omega|)\, 
\alpha^2\wt\lambda^2\, {1\over \pi} 
\int_{1/2\wt\lambda}^1 d\beta (1+\beta^2)^{-1} 
\left(1+{1\over 4\wt\lambda^2}\right)^2
\, .
\ee
On the other hand, the O-O-O vertex, with all external tachyon states, can be
computed using \refb{eooo} and \refb{e780a}. 
This is given by
\be \label{eooofin}
- K_0\, \alpha^3\, .
\ee
Multiplying \refb{eca3reptwo} by \refb{eooofin} we get the expected contribution to
Fig.~\ref{figcooo}(b):
\be\label{ecomp2new}
\wt A_{(b)}(\omega)= 2\, K_0\,  \alpha^3 \, g_s\, \sinh(\pi|\omega|)\, 
\alpha^2\wt\lambda^2\, {1\over \pi} 
\int_{1/2\wt\lambda}^1 d\beta (1+\beta^2)^{-1} 
\left(1+{1\over 4\wt\lambda^2}\right)^2\, .
\ee
Comparing \refb{ecomp1new} and \refb{ecomp2new}, we get
\be\label{ek2k3k0}
{K_2 K_3\over K_0} = {1\over \pi}\, g_s\, \sinh(\pi|\omega|)\, .
\ee

Substituting this into \refb{e6132}, we get,
\be\label{eappfin}
-K_0^{-1} A(\omega)= -{2\over \pi}\, g_s\, 
\sinh(\pi|\omega|)\, \left[{\pi\over 2} - {\wt\lambda\over 2}\right]\, .
\ee
This concludes the derivation of \refb{eappfina}.

Finally we would like to note that the integral \refb{eAint}, having a total derivative as the
integrand, can also be evaluated using the method described in \S\ref{salternative}. This
is given as an integral of $\JJ$ over the boundaries of the moduli space, corresponding 
to the segments $q_1=\delta_1$ in Fig.~\ref{figcooo} (a) and (b) and $q_2=\delta_2$
in Fig.~\ref{figcooo} (a) and (c). Each of these boundary segments actually represent
three separate terms, obtained by cyclic permutation of the three external lines. We have
checked that the integral, evaluated this way, gives the same final result \refb{ersf} after
dropping all terms
proportional to $\delta_1^{-1}$, $\delta_2^{-1}$, $\ln\delta_1$ and $\ln\delta_2$ in the
final expression. 

\newcommand{\YY}{\wp}

\sectiono{Closed string
tachyon one point function on the annulus
with a more general C-O-O vertex} \label{se}

In the analysis of the C-O-O amplitude in
\S\ref{scoo}, we could include on the right hand side of
\refb{e5} a multiplicative function of $\beta$ that takes
value 1 at $\beta=(2\wt\lambda)^{-1}$, 
but is otherwise arbitrary. In this appendix we shall
show that the final result is independent of the choice of this function. 
For simplicity
we shall assume that this function differs from 1 by infinitesimal amount, so that
we can express this as $1+\YY(\beta)$ and work to first order in $\YY(\beta)$. 
Therefore we replace \refb{e5} by
\be\label{replace}
w_a = (1+\YY(\beta))\, {\alpha\wt\lambda} \, {4\wt\lambda^2+1\over 4\wt\lambda^2} \,
{z-z_a\over (1+z_a\, z) + \wt\lambda\, f(z_a) (z-z_a)}, \qquad a=1,2, \qquad z_1=-\beta, \quad z_2=\beta\, .
\ee
The analysis may be easily generalized to the case of finite multiplicative 
function.

We shall first outline the general strategy for dealing with this problem. Let us suppose 
that we have an amplitude where one of the O legs of the C-O-O vertex gets 
connected by a propagator to some other vertex. Let
$w$ be the local coordinate of the O of the C-O-O vertex and $w'$ be the local
coordinate of the other O to which it gets connected. Then the sewing relation 
takes the form:
\be \label{eE1}
w\, w' = -\wt q, \quad 0\le \wt q\le 1\, ,
\ee
where we have labelled the sewing parameter by $\wt q$ to distinguish this from the
original sewing parameter $q$ that appeared for $\YY(\beta)=0$. When we express
this in terms of the upper half plane coordinates associated with the two vertices, then
the $\YY(\beta)$ dependence appears in the form of a multiplicative factor $(1+\YY)$
on the left hand side of \refb{eE1}. Therefore the sewing relations with the modified vertex
may be read out from the original sewing relations by making the identification:
\be\label{eE2}
\wt q= (1+\YY(\beta))\, q\, .
\ee
The effect of this can be analyzed from the results of \S\ref{serror}. As shown there,
a scaling of this form does not affect integrals of the form $\int dq\, q^{-2}$ -- if we
replace $q$ in terms of $\wt q$ using \refb{eE2} and then use the replacement rule
\refb{erp1}, we get the same result. However a rescaling of this type does affect
integrals of the form $\int dq\, q^{-1}$. This is because according to \refb{erp1}, we are
instructed to drop terms of the form $\int_0^1 dq \, q^{-1}$ 
from the amplitudes. For the original choice of the
C-O-O vertex, we drop such terms in the range $0\le q\le 1$, whereas with the new 
choice of the C-O-O vertex, we drop such terms in the range $0\le \wt q\le 1$. Since the
latter translates to the range $0\le q\le (1-\YY(\beta))$, we see that effectively with the
new choice of the C-O-O vertex, we shall have an extra contribution:
\be\label{eE3}
\int_{1-\YY(\beta)}^1 dq\, q^{-1} = \YY(\beta)\, .
\ee
This leads to the following simple prescription: Identify all diagrams where a 
C-O-O vertex appears, and if $q$ denotes the sewing parameter of the propagator
to which one of the O legs of the C-O-O vertex connects, then for every $\int_0^1
dq q^{-1}$ in the original expression, we have an extra contribution where we 
replace $\int_0^1
dq q^{-1}$ by $\YY(\beta)$, leaving the rest of the expression unchanged. If both
ends of the propagator connects to the O legs of the C-O-O vertex, then the
contribution is doubled. 

Therefore our task is to identify all diagrams  in our analysis
that contain a C-O-O vertex. We list them below:
\begin{enumerate}
\item Contributions to the C-C-O amplitude from Fig.~\ref{figfour} (b) and (c) involve the
C-O-O vertex. This will modify the contributions given in \refb{ei3first} and \refb{ei4first}.
\item Contribution to the annulus one point function given in Fig.~\ref{figfive}(c) 
involves the
C-O-O vertex. This, in turn, will modify the contribution given in \refb{eca3}.
\end{enumerate}

Besides these, there are two further effects that need to be taken into account:
\begin{enumerate}
\item  The modification of the C-O-O vertex also affects 
the $\psi^1$ exchange 
diagrams where the internal $\psi^1$ connects to the O of the C-O-O vertex. This has
to be analyzed separately. Fig.~\ref{figpsione}(b) is the only diagram of this type in 
our analysis. This will modify the analysis of \S\ref{spsi} and appendix \ref{sc}.
\item Contribution to the C-O-O-O amplitude shown in Fig.~\ref{figcooo}(b) involves
the C-O-O vertex. In this case, besides the effect mentioned above, there will be
additional dependence on $\YY(\beta)$ since one of the external vertex operators 
involve $c\p c\p^2 c/2$ that can connect to one of the O's of the C-O-O vertex. 
Since this vertex operator
is not of the form $cW$ where $W$ is a dimension one matter primary,
the integrand will have a complicated dependence on the local coordinates of the
C-O-O vertex, and hence on $\YY(\beta)$. This
needs to be taken into account separately.
\end{enumerate}

\subsection{Modification of the annulus one point function}

The off-shell C-O-O vertex appears in Fig.~\ref{figfive}(c). Therefore it modifies 
the contribution from this diagram as given in \refb{eca3}. Since the integration
variable $u$ in \refb{eca3} is proportional to the sewing parameter $q$, the
additional term is obtained by replacing $\int du u^{-1}$ term by $2\YY(\beta)$.
The extra factor of 2 comes from the fact that the propagator now connects two
O's of the C-O-O vertex.
This gives the following additional contribution to $g(\omega)$:
\be \label{eE5}
{2\over \pi}\,
\int_{1/(2\wt\lambda)}^1 d\beta\, \YY(\beta)\, 
\Bigg[\wt\lambda \,  f'(\beta) +{1\over 4}  \, 
{1+\beta^2\over \beta^2}+{2\over 1+\beta^2} \, \omega^2
\Bigg]\, .
\ee

\subsection{Modification of the C-C-O disk 
amplitude and its consequences}

Contributions to the C-C-O amplitude from Fig.~\ref{figfour} (b) and (c) involve the
C-O-O vertex, and their expressions, for C's given by a pair of closed string tachyons of
energies $\omega_1$ and $\omega_2$ and O given by an external open string state
with vertex operator $c\p X$, have been given in  \refb{ei3first} and \refb{ei4first}.
Since the contribution to \refb{ei4first} is related to the one in \refb{ei3first} by 
$\omega_1\leftrightarrow\omega_2$ exchange, let us focus on \refb{ei3first}. Here the
integration variable $u$ is proportional to $q$, so according to the algorithm given earlier
we simply need to replace $\int du\, u^{-1}$ by $\YY(\beta)$ to get the extra term. 
Adding the contribution from \refb{ei4first} by $\omega_1\leftrightarrow\omega_2$ 
exchange, we get the net extra contribution to be:
\ben
&& -{2\, i\over \pi} \, g_s \,   
 \sinh(\pi|\omega_1|)\, \sinh(\pi|\omega_2|) \,  (\omega_1+\omega_2)\,
 \int_{|\xi|\ge 4\wt\lambda/(4\wt\lambda^2-1)} 
 d\xi\, \YY(\beta)\, \left\{ {1\over \xi^2} + 2\omega_1 \, \omega_2\, {1\over 1+\xi^2} 
 \right\}\, , \nonumber \\ && \hskip 1in \xi = \pm {2\beta\over 1-\beta^2}\, .
 \een
Converting the integration variable from $\xi$ to $\beta$, and ignoring terms with
inverse powers of $\wt\lambda$, we get
\be
-{2 \, i\over \pi} \, g_s \,   
  \sinh(\pi|\omega_1|)\, \sinh(\pi|\omega_2|)  \,  (\omega_1+\omega_2)\,
 \int_{1/(2\wt\lambda)}^1 
 d\beta\, \YY(\beta)\, \left\{ {1+\beta^2\over  \beta^2} 
 + 8\, \omega_1 \, \omega_2\, {1\over 1+\beta^2} 
 \right\}\, .
 \ee
This term needs to be added to \refb{ecco5extra}. This in turn requires us to 
add a new term to the effective action \refb{eextra} and therefore a new term to the
required field redefinition \refb{efred}, \refb{efchoice}. This will lead to an additional contribution to $g(\omega)$
as in \refb{eca1}. Since the analysis is identical to that
leading from \refb{ecco5extra} to \refb{eca1}, we only quote the final result 
for the additional contribution to $g(\omega)$:
\be\label{eE8}
-{1\over \pi} \,    \int_{1/(2\wt\lambda)}^1 d\beta\, \YY(\beta)\, 
 \left\{ {1+\beta^2\over 2\beta^2} + 4\, \omega^2 {1\over 1+\beta^2}\right\}\, .
 \ee

\subsection{Modification of the $\psi^1$ exchange contribution} 

We shall now study the effect of modifying the C-O-O vertex on the $\psi^1$ 
exchange contribution shown in Fig.~\ref{figpsione}(b) and analyzed in appendix
\ref{sc}. In particular our modification of the C-O-O vertex has the effect of replacing
$w_a$ by $w_a(1-\YY(\beta))$ in \refb{eFexpand}. Following the subsequent analysis
in appendix \ref{sc}, it is easy to see the the net effect of this change is to 
replace
\be 
{1\over g_a}{\p g_a\over \p\beta} \to {1\over g_a}{\p g_a\over \p\beta} - \YY'(\beta)\, ,
\ee
in \refb{eAmpli}. $h_a/g_a^2$ and $f_a$ are not affected by $\YY(\beta)$.
This gives a net extra contribution to \refb{eAmpli} of the form
\be
-4\, K_1\, K_2\,\int_{1/(2\wt\lambda)}^1 d\beta\,  \YY'(\beta) \, \wt\lambda\, f(\beta)\,.
\ee
Multiplying this by the $\psi^1$ propagator $K_0^{-1}/2$,
we get the net extra contribution to the annulus one point function of the closed
string tachyon from the $\psi^1$ exchange diagram:
\be
-2\, K_1\, K_2\, K_0^{-1}\, \int_{1/(2\wt\lambda)}^1 d\beta \, 
\YY'(\beta) \, \wt\lambda\, f(\beta)
=-2\, {g_s\over \pi}\, \sinh(\pi|\omega|)\, \int_{1/(2\wt\lambda)}^1 d\beta \,
\YY(\beta) \, \wt\lambda\, f'(\beta)
\,,
\ee
where in the second step we have used \refb{ekvalue} and carried out integration by parts
over $\beta$, ignoring the boundary terms since $\YY(\beta) f(\beta)$ vanishes at
the boundaries. This translates to the following additional contribution to 
$g(\omega)$:
\be \label{eE12}
- {1\over \pi} \, \int_{1/(2\wt\lambda)}^1 d\beta \, \YY(\beta) \, \wt\lambda\, f'(\beta)
\, .
\ee

\subsection{Modification of the C-O-O-O amplitude and its consequences}

We shall now examine how the modification of the C-O-O vertex affects the results
of \S\ref{sghost} and appendix \ref{sb}. The C-O-O vertex appears in Fig.~\ref{figcooo}(b).
Even though we have argued that the contribution from this diagram vanishes, this
affects the final result indirectly, since the contribution from the boundary $\BB_1$
of the region of the moduli space associated with Fig.~\ref{figcooo}(c) was 
evaluated by parametrizing this boundary by the coordinate system inherited from
Fig.~\ref{figcooo}(b). In particular the modification of the C-O-O vertex changes
\refb{ed20} to
 \be\label{ed20new}
(1+\YY(\beta)) \, {\alpha\wt\lambda} \, {4\wt\lambda^2+1\over 4\wt\lambda^2} \,
{z-\beta\over (1+\beta\, z) + \wt\lambda\, f(\beta) (z-\beta)} 
\, \alpha \, {2\hat z\over 2-\hat z} = -\wt q\, ,
 \ee
where $\wt q$ is the new sewing parameter. Comparing this with \refb{ed20} we get 
the relation:
\be
q=\wt q\, (1-\YY(\beta))\, .
\ee
This has the effect of replacing $u$ by $\wt u(1-\YY(\beta))$ in \refb{ed21}, \refb{ed23}, with
\be
\wt u={1\over 2\alpha^2\wt\lambda}\, {4\wt\lambda^2\over 4\wt\lambda^2+1} \, \wt q,
\qquad 0\le \wt u\le u_0, \qquad 
u_0\equiv {1\over 2\alpha^2\wt\lambda}\, {4\wt\lambda^2\over 4\wt\lambda^2+1}
\, .
\ee
$\wt q=1$ defines the modified boundary $\BB_1$. Furthermore the expression for
$w_1$ in \refb{ed23} is multiplied by a factor of $(1+\YY(\beta))$ since $w_1$
is inherited from the local coordinate on the C-O-O vertex. The net effect of these
changes may be summarized as follows:
\begin{enumerate}
\item In \refb{effgv1}, $u$ is replaced by $\wt u(1-\YY(\beta))$, the expression
for $g_1$ is multiplied
by $(1-\YY(\beta))$ and the expression
for $h_1$ is multiplied
by $(1-2\YY(\beta))$.
\item In \refb{effgv2}, $u$ is replaced by $\wt u(1-\YY(\beta))$, the expression
for $g_2$ is multiplied
by $(1-\YY(\beta))$ and the expression for $h_2$ is multiplied by
$(1-2\YY(\beta))$.
\item In \refb{effgv3}, $u$ is replaced by $\wt u(1-\YY(\beta))$.
\end{enumerate}
The effect of these changes is to modify \refb{ers1}, \refb{ers3} and 
\refb{ers2} by additive terms of the form:
\be
-K_2 \, K_3\, \wt\lambda \int_{1/(2\wt\lambda)}^1 d\beta \YY'(\beta) f(\beta),
\quad 
K_2 \, K_3\, \wt\lambda \int_{1/(2\wt\lambda)}^1 d\beta \YY'(\beta) f(\beta),
\quad -K_2 \, K_3\, \wt\lambda \int_{1/(2\wt\lambda)}^1 d\beta \YY'(\beta) f(\beta)\, ,
\ee 
respectively.
The second set of contributions, obtained by exchanging the punctures at $\beta$
and $-\beta$ of the C-O-O vertex, gives identical contribution. Therefore the net
modification of the amplitude $A(\omega)$ is:
\be
-2\, K_2 \, K_3\, \wt\lambda \int_{1/(2\wt\lambda)}^1 d\beta \YY'(\beta) f(\beta) 
=2\, K_2\, K_3 \,  \wt\lambda \int_{1/(2\wt\lambda)}^1 d\beta \YY(\beta) f'(\beta)\, ,
\ee 
where we have performed an integration by parts. Since the contribution to
the annulus one point 
function is given by $-2 K_0^{-1} A(\omega)$, the modification of the annulus one point
function is given by:
\be 
-4\, K_2\, K_3 \, K_0^{-1}\, 
\wt\lambda \int_{1/(2\wt\lambda)}^1 d\beta \YY(\beta) f'(\beta)
= -{2\over \pi} \, g_s\, \sinh(\pi|\omega|) \, \wt\lambda \int_{1/(2\wt\lambda)}^1 
d\beta \YY(\beta) f'(\beta)\, ,
\ee 
where in the second step we have used \refb{ek2k3k0}. This, in turn, gives an 
additional contribution to $g(\omega)$ of the form:
\be\label{eE19}
-{1\over \pi} \, \wt\lambda \int_{1/(2\wt\lambda)}^1 
d\beta \YY(\beta) f'(\beta)\, .
\ee 

We now see that the sum of \refb{eE5}, \refb{eE8}, \refb{eE12} and \refb{eE19}
vanishes. Therefore the modification of the C-O-O vertex has no effect on 
$g(\omega)$.

\end{document}